\documentclass[twocolumn,ip]{jpsj3}
\usepackage{txfonts}

\usepackage{graphicx}
\usepackage{dcolumn}
\usepackage{bm}
\usepackage{amsmath}

\title{Topological Insulator Materials}

\author{\name{Yoichi \surname{Ando}}\thanks{E-mail: y{\_}ando@sanken.osaka-u.ac.jp}}
\inst{Institute of Scientific and Industrial Research, Osaka University, Ibaraki, Osaka 567-0047, Japan} 

\abst{Topological insulators represent a new quantum state of matter
which is characterized by peculiar edge or surface states that show up
due to a topological character of the bulk wave functions. This review presents a
pedagogical account on topological insulator materials with an emphasis
on basic theory and materials properties. After presenting a
historical perspective and basic theories of topological
insulators, it discusses all the topological insulator materials
discovered as of May 2013, with some illustrative descriptions of the developments in
materials discoveries in which the author was involved. A summary is given for
possible ways to confirm the topological nature in a candidate material.
Various synthesis 
techniques as well as the defect chemistry that are important for
realizing bulk-insulating samples are discussed. 
Characteristic properties of topological insulators are discussed with
an emphasis on transport properties. In particular, the Dirac fermion physics and the
resulting peculiar quantum oscillation patterns are discussed in detail. It is
emphasized that proper analyses of quantum oscillations make it possible
to unambiguously identify surface Dirac fermions through transport
measurements. The prospects of topological insulator materials for elucidating novel
quantum phenomena that await discovery conclude the review.}

\kword{topological insulator, Dirac fermions, surface state, quantum oscillations}

\begin{document}
\maketitle

\section{Introduction}

The progress in condensed matter physics is often driven by discoveries
of novel materials. In this regard, materials presenting unique
quantum-mechanical properties are of particular importance. Topological
insulators (TIs) are a class of such materials and they are currently
creating a surge of research activities \cite{HasanKane, Moore, QiZhang}. 
Because TIs concern a
qualitatively new aspect of quantum mechanics, i.e. the topology of the
Hilbert space, they opened a new window for understanding the elaborate
workings of nature. 

TIs are called ``topological" because the wave functions describing
their electronic states span a Hilbert space that has a nontrivial
topology. Remember, quantum-mechanical wave functions are 
described by linear combinations of orthonormal vectors forming a 
basis set, and the abstract space spanned by this orthonormal basis is
called Hilbert space. In crystalline solids, where the wave vector $\mathbf{k}$
becomes a good quantum number, the wave function can be viewed as a mapping
from the $\mathbf{k}$-space to a {\it manifold} in the Hilbert space
(or in its projection), and 
hence the topology becomes relevant to electronic states in solids.
Depending on the way the Hilbert-space topology becomes
nontrivial, there can be various different kinds of TIs \cite{Schnyder}.
An important consequence of a nontrivial topology associated with the
wave functions of an insulator is that a gapless interface state
necessarily shows up when the insulator is physically terminated and
faces an ordinary insulator (including the vacuum). This is because the
nontrivial topology is a discrete characteristic of gapped energy
states, and as long as the energy gap remains open, the topology cannot
change; hence, in order for the topology to change across the interface
into a trivial one, the gap must close at the interface. Therefore,
three-dimensional (3D) TIs are always associated with gapless surface
states, and so are two-dimensional (2D) TIs with gapless edge states.
This principle for the necessary occurrence of gapless interface states
is called bulk-boundary correspondence in topological phases. 

\begin{figure}[b]
\begin{center}
\includegraphics[width=8.2cm]{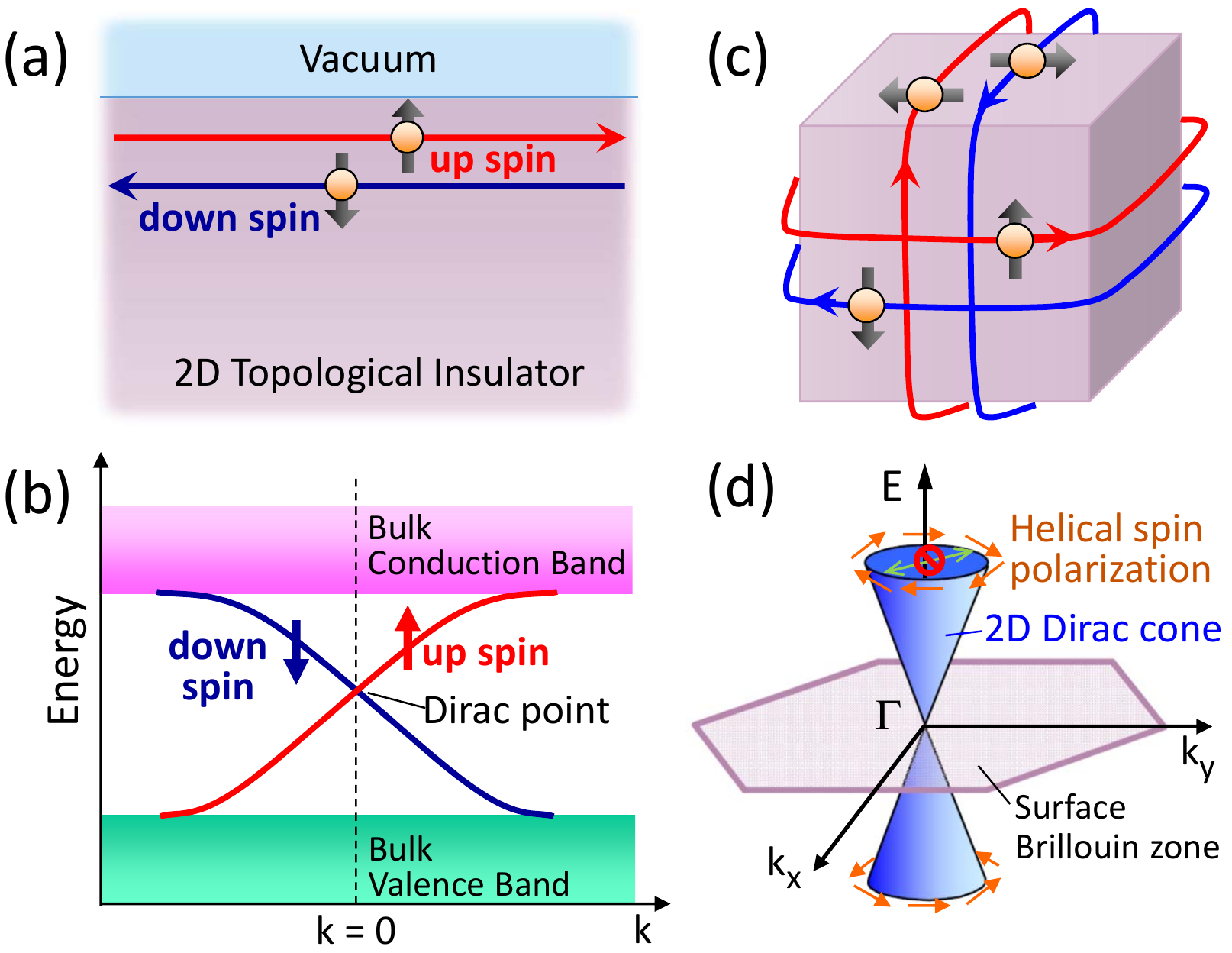}
\caption{(Color online) 
Edge and surface states of topological insulators with Dirac dispersions.
(a) Schematic real-space picture of the 1D helical edge state of a 2D TI. 
(b) Energy dispersion of the spin non-degenerate edge state of a 2D TI forming
a 1D Dirac cone.
(c) Schematic real-space picture of the 2D helical surface state of a 3D TI.
(d) Energy dispersion of the spin non-degenerate surface state of a 3D TI forming
a 2D Dirac cone; due to the helical spin polarization, back scattering from 
$\mathbf{k}$ to $\mathbf{-k}$ is prohibited.}
\label{fig:HelicalSS}
\end{center}
\end{figure}

A large part of the unique quantum-mechanical properties of TIs come
from the peculiar characteristics of the edge/surface states. Currently,
the TI research is focused mostly on time-reversal (TR) invariant systems,
where the nontrivial topology is protected by time-reversal symmetry (TRS)
\cite{HasanKane, Moore, QiZhang}. 
In those systems, the edge/surface states present Dirac dispersions
(Fig. \ref{fig:HelicalSS}),
and hence the physics of relativistic Dirac fermions becomes relevant.
Furthermore, spin degeneracy is lifted in the Dirac
fermions residing in the edge/surface states of TR-invariant
TIs and their spin is locked to the momentum (Fig. \ref{fig:HelicalSS}). 
Such a spin state is
said to have ``helical spin polarization" and it provides an opportunity to
realize Majorana fermions \cite{Wilczek} in the presence of proximity-induced
superconductivity \cite{FuKaneMajorana}, not to mention its obvious implications 
for spintronics applications. An important objective of the experimental 
studies of TIs has been to elucidate the existence and the nature of such 
helically spin-polarized Dirac fermions in the topological surface state.

This review is aimed at providing a pedagogical introduction to the 
field of topological insulators, putting emphasis on the basic theory 
and materials properties. It also elaborates on the basics of the
characterizations of TI materials using transport measurements.

\section{Historical Perspective}

It is useful to understand the historical background in order to
appreciate the importance of TIs in condensed matter physics. In this
section, I will briefly discuss several relevant issues that preceded
the discovery of TIs and also describe how the TI research developed 
in its early days.

\subsection{Integer quantum Hall effect}

In 1980, von Klitzing {\it et al.} discovered the quantum Hall effect in
a high-mobility 2D semiconductor under high magnetic fields \cite{Klitzing}. 
The occurrence of this effect is usually limited to very low temperatures,
where localization of electrons and Landau quantization of their energy
spectrum lead to vanishing longitudinal conductivity $\sigma_{xx}$
together with quantization of the Hall conductivity $\sigma_{xy}$ to
integer multiples of $e^2/h$ when the chemical potential is located in
between Landau levels. Such a quantization of transport coefficients
obviously pointed to a macroscopic quantum phenomenon, as was
made clear by Laughlin's gauge argument \cite{Laughlin1981}.
It is prudent to mention that  this quantization phenomenon was  
theoretically anticipated already in 1974 \cite{Ando1974}.
In 1982, it
was recognized by Thouless, Kohmoto, Nightingale, and den Nijs (TKNN)
\cite{TKNN} that this phenomenon not only is
quantum mechanical but also is {\it topological}; namely, TKNN showed \cite{TKNN}
that in the quantum Hall
system the $\mathbf{k}$-space is mapped to a topologically-nontrivial 
Hilbert space, whose topology can be specified by an integer topological invariant 
called TKNN invariant $\nu$,
and that $\sigma_{xy}$ becomes equal to $\nu$ times $e^2/h$. The TKNN
invariant is also called the first Chern number or the winding number, and
it is equal to the Berry phase of the Bloch wave function calculated
around the Brillouin zone (BZ) boundary divided by $2\pi$ (actual calculations
are shown in the next section).

In hindsight, the quantum Hall system can be considered to be the first
topological insulator that became known to physicists, because when the
quantization is taking place, the energy spectrum is gapped due to the
Landau quantization and the chemical potential is located within the
gap, which is a situation akin to an insulator. In this case, the
nontrivial topology specified by the TKNN invariant is characteristic of
a 2D system with broken TRS. 
Also, as was shown by Halperin \cite{Halperin}, the integer
quantum Hall effect is always accompanied by chiral edge states, and
those gapless states residing at the interface to the vacuum can be
understood to be a result of the bulk-boundary correspondence due to the
topological 2D ``bulk" state. 

It is prudent to mention that the integer quantum Hall effect was a 
tip of an iceberg.  The fractional quantum Hall (FQH) effect discovered in 1982 by 
Tsui, Stormer, and Gossard \cite{TsuiFQHE} turned out to contain richer 
physics, because electron correlations play essential roles in the FQH effect and 
they lead to the appearance of 
fractionally-charged quasiparticles \cite{LaughlinFQHE}.  In terms of topology, 
however, FQH states do not have much relevance to topological insulators, because the 
former present ground-state degeneracy and their topological character is 
described by quite an abstract concept of {\it topological order} \cite{Wen-Niu}.

\subsection{Quantum spin Hall effect and $Z_2$ topology}

On a different front in condensed matter physics, generation and
manipulation of spin currents have been attracting a lot of interest,
since they will have a profound impact on future spintronics
\cite{Spintronics}. In this
regard, the spin Hall effect, the appearance of transverse spin current
in response to longitudinal electric field, has been discussed
theoretically since 1970s \cite{spinHall1, spinHall2, spinHall3,
Murakami2003, Sinova2004}, but its experimental confirmation by Kato
{\it et al.} \cite{Kato2004} in 2004 gave a big boost to the research of this
phenomenon. It was soon recognized that the spin Hall effect in
nonmagnetic systems is fundamentally related to the anomalous Hall
effect in ferromagnets \cite{AHE}, and similarly to the latter effect, there are
both intrinsic and extrinsic origins of the spin Hall effect. The
intrinsic mechanism of the spin Hall effect stems from the Berry
curvature of the valence-band Bloch wave functions integrated over the
Brillouin zone \cite{Murakami2003, Sinova2004}. 
Since such an integral can become finite even in an
insulator, Murakami, Nagaosa, and Zhang went on to propose the idea of
{\it spin Hall insulator} \cite{SHI}, 
which is a gapped insulator with zero charge
conductivity but has a finite spin Hall conductivity due to a finite
Berry phase of the occupied states.

Later it was shown \cite{Onoda2005} that the proposed spin Hall insulators cannot really
generate spin currents in the absence of any electrons at the Fermi
level, but this idea triggered subsequent proposals of its quantized
version, the quantum spin Hall (QSH) insulator, by Kane and Mele \cite{KM_QSH,
KM_Z2}, followed by an independent proposal by Bernevig and Zhang \cite{BZ2006}.
The QSH insulators are essentially two copies of the quantum Hall system, in
which the chiral edge state is spin polarized and the two states
form a time-reversed pair to recover the overall TRS. 
When current flows using the edge states of a QSH insulator, 
a quantized version of the spin Hall effect, the QSH effect, is predicted
to be observed. Since the predicted phenomenon is based
on the quantum Hall effect, it only exists in 2D. While it is not a priori
clear how one can achieve such a state with quantized edge
states in zero magnetic field, the ingenious proposal by Kane and Mele 
provided a concrete model to realize the QSH insulator \cite{KM_QSH}. 
Their model is
essentially a graphene model with spin-orbit coupling (SOC).

In graphene, the band structure near the Fermi level consists of two
linearly dispersing cones located at $K$ and $K'$ points in the BZ 
\cite{graphene}; 
since the low-energy physics on these cones is described
by employing the Dirac equation with the rest mass set to zero
\cite{CastroNeto}, this
dispersion is called {\it Dirac cone} and the electrons are said to
behave as {\it massless Dirac fermions}. Kane and Mele showed \cite{KM_QSH}
that a
finite SOC leads to an opening of a gap at the crossing point of the
cone (called {\it Dirac point}) and, furthermore, that a time-reversed
pair of spin-polarized one-dimensional (1D) states indeed show up at the
edge in some parameter range; in this model, the desired spin
polarization of the edge state is achieved due to the SOC which has an
inherent tendency to align spins in relation to the momentum direction.
This peculiar spin-non-degenerate state [Fig. \ref{fig:HelicalSS}(a,b)]
is often said to have {\it
helical spin polarization} or {\it spin-momentum locking}. Intriguingly,
those electrons in the gapless edge state behave as 1D massless Dirac
fermions within the gap opened in the 2D Dirac cone. In this case, the
2D ``bulk" electrons can be viewed as {\it massive} Dirac fermions
because of the finite energy gap at the Dirac point.

Most importantly, Kane and Mele recognized that the electronic states of
their QSH insulator is characterized by a novel topology specified by
a $Z_2$ index \cite{KM_Z2}, 
which expresses whether the number of times the 1D edge
state crosses the Fermi level between 0 and $\pi/a$ is even or odd ($a$
is the lattice constant). Remember, in mathematics the group of integer
numbers is called $Z$ and its quotient group classifying even and odd
numbers is called $Z_2$; hence, a $Z_2$ index generally gives a
topological classification based on parity. (A detailed description of
the $Z_2$ index for TR-invariant TIs is given in the
next section.) The theoretical discovery of the $Z_2$ topology in
insulators was a big step in our understanding of topological phases of
matter, because it indicated that nontrivial topology can be embedded in
the band structure of an ordinary insulator and that breaking of 
TRS by application of magnetic fields is not mandatory for realizing 
a topological phase. 

\begin{figure}[t]
\begin{center}
\includegraphics[width=8.2cm]{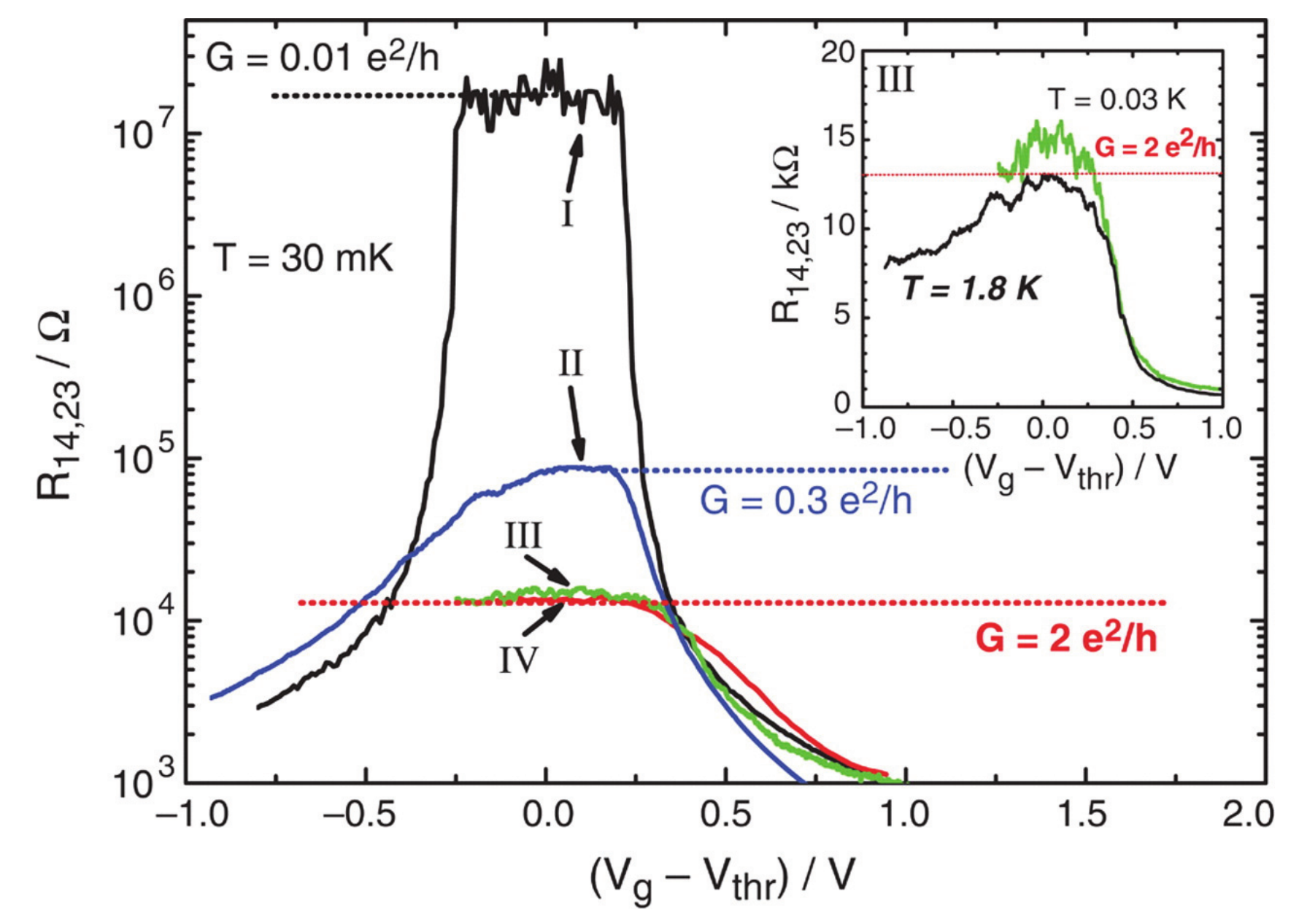}
\caption{(Color online) 
Longitudinal four-terminal resistance of
various CdTe/HgTe/CdTe quantum-well structures
as a function of the gate voltage
measured in zero magnetic field at 30 mK. 
Devices with the size of 1 $\times$ 1 $\mu$m$^2$
or less in the band-inverted regime (III and IV) show quantized conductance 
of $2e^2/h$, giving evidence for the 2D TR-invariant TI phase. 
Taken from Ref. \citen{Konig2007}; copyright
American Association for the Advancement of Science (2007).}
\label{fig:Koenig}
\end{center}
\end{figure}

Unfortunately, the SOC in graphene is very weak, and hence it is
difficult to experimentally observe the QSH effect predicted in the
Kane-Mele model. However, another theoretical breakthrough was soon made by
Bernevig, Hughes, and Zhang (BHZ) \cite{BHZ}, 
who constructed a 2D model to produce
a $Z_2$ topological phase based on the band structure of HgTe; based on
their model, BHZ predicted that a CdTe/HgTe/CdTe quantum well should
give rise to the QSH effect. This prediction was verified in 2007 by
K\"onig {\it et al.} \cite{Konig2007}, 
who observed $\sigma_{xx}$ to be quantized to
$2e^2/h$ in zero magnetic field when the chemical potential is tuned
into the bulk band gap (Fig. \ref{fig:Koenig}), giving evidence 
for the gapless edge states in
the insulting regime. This was the first experimental confirmation of
the TR-invariant TI characterized by the $Z_2$ topology. 

\begin{figure}[t]
\begin{center}
\includegraphics[width=8.2cm]{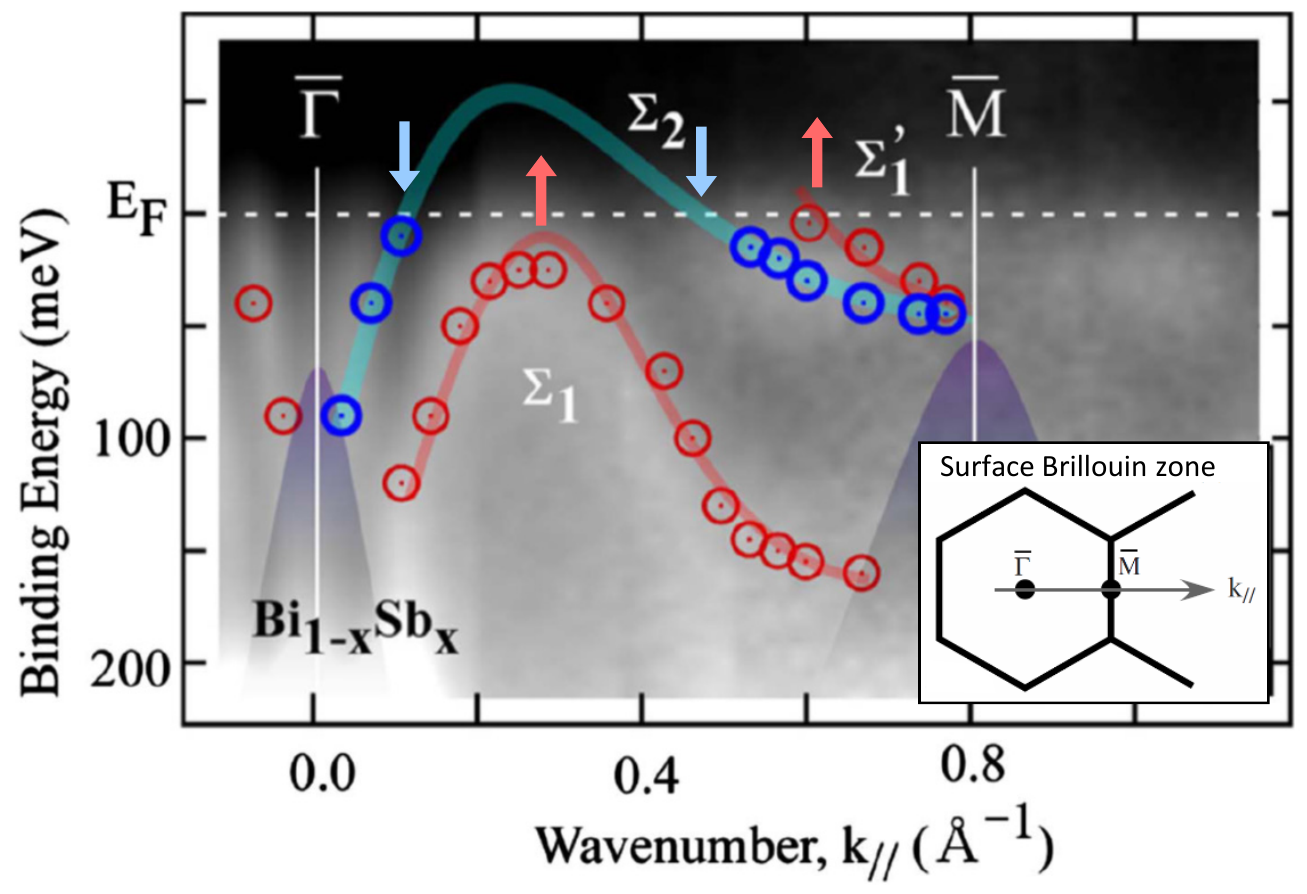}
\caption{(Color online) 
Spin-resolved surface band structure of Bi$_{1-x}$Sb$_x$ 
($x = 0.12-0.13$) on the (111) cleaved surface. Its surface BZ and the 
momentum direction of the data are shown in the inset.
Dispersions shown by symbols are determined from the peak positions 
of the spin-resolved 
energy dispersion curves and they are plotted on top of the spin-integrated
ARPES data shown in gray scale. There are three Fermi-level crossings 
of the surface states, which signifies the $Z_2$-nontrivial nature of this
system. Taken from Ref. \citen{Nishide}; copyright
American Physical Society (2009).}
\label{fig:BiSb}
\end{center}
\end{figure}

Without waiting for experimental verification of the $Z_2$ topology in
2D, theorists noticed that this topological classification of insulators
can be extended to 3D systems, where there are four $Z_2$ invariants to
fully characterize the topology \cite{3DTI1, 3DTI2, 3DTI3}. 
In fact, the term ``topological insulator" was coined by Moore and Balents
in their paper to propose the existence of TIs in 3D systems \cite{3DTI1}.
For 3D TIs, Fu and Kane made a concrete
prediction in 2006 that the Bi$_{1-x}$Sb$_x$ alloy in the insulating
composition should be a TI, and they further proposed that the
nontrivial topology can be verified by looking at the surface states
using the angle-resolved photoemission spectroscopy (ARPES) and counting
the number of times the surface states cross the Fermi energy between
two TR-invariant momenta \cite{FuKane3D}. The proposed experiment was
conducted by Hsieh {\it et al.} who reported in 2008 that
Bi$_{1-x}$Sb$_x$ is indeed a 3D TI \cite{Hsieh2008}. 
The experimental identification of
Bi$_{1-x}$Sb$_x$ as a TR-invariant TI opened a lot of new
experimental opportunities to address a topological phase of matter.
For example, the first transport study of Bi$_{1-x}$Sb$_x$ to detect 
topological 2D
transport channels was reported by Taskin and Ando \cite{Taskin2009}, 
and the first
scanning tunneling spectroscopy (STS) study that addressed the peculiar
spin polarization was reported by Roushan {\it et al.} \cite{Roushan}, 
both in 2009. 
Direct observation of the helical spin polarization of the surface states 
in Bi$_{1-x}$Sb$_x$ using
spin-resolved ARPES was first partially done by Hsieh {\it et al.} 
\cite{Hsieh2009} and
then fully accomplished by Nishide {\it et al.} (Fig. \ref{fig:BiSb}) 
\cite{Nishide}.

\subsection{Topological field theory}

While the notion of topological insulators became popular only after the 
discovery of the $Z_2$ topology 
by Kane and Mele \cite{KM_Z2}, there had been theoretical efforts to conceive topological
states of matter beyond the scope of the quantum Hall system. In this respect,
an important development was made in 2001 by Zhang and Hu, who generalized
the 2D quantum Hall state to a four-dimensional (4D) TR-invariant state 
possessing an integer topological invariant \cite{ZhangHu}. 
The effective field theory for this 4D topological system was 
constructed by Bernevig {\it et al.} \cite{Bernevig2002}. 
After the $Z_2$ topology was discovered
for TR-invariant systems in 2D and 3D
\cite{KM_Z2, 3DTI1, 3DTI2, 3DTI3}, it was shown by Qi, Hughes, 
and Zhang \cite{QHZ} that the
framework of topological field theory is useful for describing those systems
as well, and they further demonstrated that the $Z_2$ TIs in 2D and 3D 
can actually be deduced from the 4D effective field theory 
by using the dimensional reduction. 
From a practical point of view,
the topological field theory is suitable for describing the electromagnetic 
response of TIs and has been used for predicting novel topological 
magnetoelectric effects \cite{QHZ,monopole,MacDonald,Nomura}.

\subsection{Dirac materials}

It is interesting to note that the first 3D TI material
Bi$_{1-x}$Sb$_x$, whose topological surface state consists of 2D
massless Dirac fermions, has long been known to possess peculiar band
structure to give rise to 3D {\it massive} Dirac fermions in the bulk
\cite{Wolff}. 
This situation is similar to the Kane-Mele model where 1D massless Dirac
fermions emerge out of 2D massive Dirac fermions. Since Dirac fermions
play important roles in TIs, it would be useful to mention the history
of Dirac physics in condensed matter. 

The semimetal Bi has been an important testing ground of quantum physics
ever since the Shubnikov-de Haas (SdH) and de Haas-van Alphen (dHvA)
oscillations were discovered in Bi in 1930 \cite{Shoenberg}. This is
essentially because the extremely low carrier density ($\sim$10$^{-5}$
per atom) and the very long mean free path (reaching $\sim$1 mm) easily
put the system in the ``quantum limit" at relatively low magnetic fields
\cite{Edel'man}. In the mid-20th century, one of the long-standing
puzzles in Bi was its large diamagnetism, which defies the common wisdom
for magnetism in metals involving Pauli paramagnetism and Landau
diamagnetism \cite{Fukuyama}. Intriguingly, in Bi$_{1-x}$Sb$_x$ at low
Sb concentration, the carrier density becomes even lower than in Bi,
while the diamagnetic susceptibility increases, which is also opposite
to the expectation from Landau diamagnetism \cite{Fukuyama}. To address
the unusual electronic properties of Bi, an effective two-band model was
constructed by Cohen and Blount in 1960 \cite{CohenBlount}. In 1964,
Wolff recognized that this two-band model can be transformed into the
four-component massive Dirac Hamiltonian, and he presented an elegant
analysis of the selection rules using the Dirac theory \cite{Wolff}. This
was the beginning of the notion of Dirac fermions in solid states,
although some of the peculiar physics of massless Dirac fermions were
recognized in as early as 1956 by McClure in the context of graphite
\cite{McClure}.

Speaking of graphite, the mapping of the $\mathbf{k}\cdot\mathbf{p}$
Hamiltonian of its 2D sheet (i.e. graphene) \cite{Slonczewski-Weiss} to
the massless Dirac Hamiltonian was first used by Semenoff in 1984
\cite{Semenoff}. With the experimental realization of graphene
\cite{graphene}, this system has become a prototypical Dirac material.
One of the distinguishing properties of massless Dirac fermions is the
Berry phase of $\pi$; an important consequence of the $\pi$ Berry phase
in the condensed matter setting is the absence of backscattering, which
was pointed out first by Ando, Nakanishi, and Saito in 1998
\cite{T.Ando}.

An important aspect of the Dirac physics is that magnetic fields
necessarily cause interactions between upper and lower Dirac cones. In
fact, the Dirac formalism allows one to naturally include such
``interband effects" of magnetic fields into calculations. In 1970
Fukuyama and Kubo showed, by explicitly performing the calculations, that the
large diamagnetism in Bi and Bi$_{1-x}$Sb$_x$ is in fact a consequence
of their Dirac nature \cite{Fukuyama}. Intriguingly, due to such
interband effects of magnetic fields, the Hall coefficient does {\it not}
diverge but becomes zero when the carrier density vanishes at the Dirac
point \cite{Fukuyama2007}, which is indeed observed in a topological
insulator \cite{Fuhrer2012}.

\subsection{Old proposals for interface Dirac fermions}

It is prudent to mention that already in 1985, the appearance of 2D
massless Dirac fermions at a certain interface was anticipated
\cite{Volkov,Pankratov}. Specifically, Volkov and Pankratov discussed
that the relative band inversion occurring in SnTe/PbTe and in CdTe/HgTe
would lead to a gapless interface state \cite{Volkov}. 
Also, Fradkin, Dagotto, and
Boyanovsky discussed in 1986 that an antiphase domain wall in PbTe
(which creates the atomic arrangements of Pb-Te-{\bf Pb-Pb}-Te-Pb or
Te-Pb-{\bf Te-Te}-Pb-Te) can be viewed as a parity anomaly in 3+1 dimension
and would lead to a gapless interface state \cite{Fradkin}.

What distinguishes the recent discussions of the topological surface
states from those old works is the notion of {\it topological
protection}, which guarantees the stability of the gapless state. For
example, SnTe is a band-inverted semiconductor, but its band inversion
takes place at an {\it even number} of points in the Brillouin zone, which
makes 3D SnTe topologically-trivial in the $Z_2$ topological 
classification \cite{FuKane3D}. 
This means that, even if the interface state is expected to appear at the
SnTe/PbTe interface as a result of band inversion, its gapless nature is
not protected by the $Z_2$ topology. (Nevertheless, SnTe was recently
found to be a {\it topological crystalline insulator} \cite{HsiehFu, TanakaTCI}, 
and its gapless surface
state is actually topologically protected by mirror symmetry, as discussed 
later.) In the
case of the antiphase domain wall in PbTe, detailed symmetry arguments
highlighted the difficulty in realizing the parity anomaly.
\cite{Tchernyshyov} 

\section{Basics of the Topology in Topological Insulators}

\subsection{Berry phase}

The notion of Berry phase \cite{Berry} 
is important in the discussions of topological phases. 
Here, I briefly discuss its definition and meaning. 

Let $\mathbf{R}(t)$ be a set of time-dependent parameters and consider it 
as a vector in the parameter space. We write the 
Hamiltonian of a system specified by the parameter $\mathbf{R}(t)$ as 
$H[\mathbf{R}(t)]$ and its $n$th eigenstate as 
$|n, \mathbf{R}(t)\rangle$. 
The Schr\"odinger equation for this system is
\begin{equation}
H[\mathbf{R}(t)]\, |n, \mathbf{R}(t)\rangle = 
E_n[\mathbf{R}(t)]\, |n, \mathbf{R}(t)\rangle.
\end{equation}
Suppose that $\mathbf{R}$ changes adiabatically from the $t$ = 0
value $\mathbf{R}_0$. The time evolution of the state follows 
\begin{equation}
H[\mathbf{R}(t)]\, |n, t\rangle = 
i\hbar \frac{\partial}{\partial t} |n, t\rangle
\label{eq:t-dep1}
\end{equation}
and the state at time $t$ is given by
\begin{equation}
|n, t\rangle = \exp\left(\frac{i}{\hbar} \int_0^t dt' L_n[\mathbf{R}(t')] \right)
|n, \mathbf{R}(t)\rangle,
\label{eq:t-dep2}
\end{equation}
where
\begin{equation}
L_n[\mathbf{R}(t)] = i \hbar \dot{\mathbf{R}}(t) \cdot 
\langle n, \mathbf{R}(t)| \nabla_R | n, \mathbf{R}(t) \rangle
-E_n[\mathbf{R}(t)].
\label{eq:Ln}
\end{equation}
This can be easily confirmed by putting Eq. (\ref{eq:t-dep2}) into
the r.h.s. of Eq. (\ref{eq:t-dep1}).
Using $L_n[\mathbf{R}(t)]$ given in Eq. (\ref{eq:Ln}), one may write
the time-dependent state as
\begin{eqnarray}
|n, t\rangle = \exp\left(- \int_0^t dt'  \dot{\mathbf{R}}(t') \cdot 
\langle n, \mathbf{R}(t')| \nabla_R | n, \mathbf{R}(t') \rangle \right)
 | n, \mathbf{R}(t) \rangle \nonumber \\
\times \exp\left(\frac{i}{\hbar} \int_0^t dt' E_n[\mathbf{R}(t')] \right).
\label{eq:time.evol}
\end{eqnarray}
In this expression of $|n, t\rangle$, the first exponential term represents
the nontrivial effect of the quantum-mechanical phase accumulated 
during the time evolution, and the last exponential term is a trivial 
one called dynamical term.

When $\mathbf{R}$ moves on a closed loop $C$ from $t$ = 0 and returns
to the original position at $t = T$, i.e. $\mathbf{R}(T) = \mathbf{R}_0$,
the Berry phase $\gamma_n[C]$ for this loop $C$ is defined as
\begin{eqnarray}
\gamma_n[C] &\equiv& \int_0^T dt  \dot{\mathbf{R}}(t) \cdot 
   i \langle n, \mathbf{R}(t)| \nabla_R | n, \mathbf{R}(t) \rangle 
\label{eq:Berry} \\
&=& \oint_C d\mathbf{R} \cdot 
   i \langle n, \mathbf{R}| \nabla_R | n, \mathbf{R} \rangle \\
&\equiv& - \oint_C d\mathbf{R} \cdot \mathbf{A}_n(\mathbf{R}) \\
&=& - \int_S d\mathbf{S} \cdot \mathbf{B}_n(\mathbf{R})
\end{eqnarray}
The last equality comes from the Stokes' theorem. 
Here, we define the Berry connection
\begin{equation}
\mathbf{A}_n(\mathbf{R}) 
= - i \langle n, \mathbf{R}| \nabla_R | n, \mathbf{R} \rangle,
\label{eq:BC}
\end{equation}
and its rotation is the Berry curvature
\begin{equation}
\mathbf{B}_n(\mathbf{R})
= \nabla_R \times \mathbf{A}_n(\mathbf{R}).
\end{equation}
From Eqs. (\ref{eq:time.evol}) and (\ref{eq:Berry}), 
one can see that the Berry phase 
means the accumulated phase factor of a quantum-mechanical 
system after it completes a closed path in the parameter space.  
The Berry connection corresponds to the gauge field defined on 
that parameter space, similar to the vector potential for 
electromagnetic fields in real space.

\subsection{TKNN invariant}

The topological invariant defined for the integer quantum Hall 
system, the TKNN invariant \cite{TKNN}, is closely related to the Berry phase.
To see this, we derive the TKNN invariant by calculating the 
Hall conductivity of a 2D electron system of size $L \times L$ 
in perpendicular magnetic
fields, where the electric field $E$ and the magnetic field 
$B$ are applied along the 
$y$ and $z$ axes, respectively. By treating the effect of the electric
field $E$ as a perturbation potential $V = -eEy$, one may use the
perturbation theory \cite{NomuraNote} to approximate the perturbed eigenstate 
$|n\rangle_E$ as
\begin{equation}
|n \rangle_E = |n\rangle + \sum_{m (\neq n)} 
\frac{\langle m|\, (-eEy)\, |n \rangle}{E_n - E_m} |m \rangle + \cdots.
\end{equation}
Using this perturbed eigenstate, one may obtain the expectation value of the
current density along the $x$ axis, $j_x$, in the presence of the $E$ field 
along the $y$ axis as
\begin{eqnarray}
\lefteqn{\langle j_x \rangle_E
= \sum_n f(E_n) \langle n |_E \left( \frac{ev_x}{L^2} \right) \, |n \rangle_E } \nonumber \\
&=& \langle j_x \rangle_{E=0}  + \frac{1}{L^2}\sum_n f(E_n) 
\sum_{m (\neq n)} \left( \frac{\langle n | \, ev_x \, | m \rangle 
\langle m | \, (-eEy) \, |n \rangle}{E_n-E_m} \right. \nonumber \\
& & + \left. \frac{\langle n | \, (-eEy) \, | m \rangle \langle m | \, ev_x \, 
|n \rangle}{E_n-E_m}  \right),
\end{eqnarray}
where $v_x$ is the electron velocity along the $x$ direction and 
$f(E_n)$ is the Fermi distribution function.
The Heisenberg equation of motion $\frac{d}{dt}y = v_y = 
\frac{1}{i\hbar}[y, H]$ leads to 
\begin{equation}
\langle m |\, v_y\, |n\rangle
= \frac{1}{i\hbar}(E_n - E_m) \langle m |\, y\, |n\rangle,
\end{equation}
which allows one to evaluate
\begin{eqnarray}
\sigma_{xy} &=& \frac{\langle j_x \rangle_E}{E} \nonumber \\
&=& -\frac{i\hbar e^2}{L^2}\sum_{n \neq m} f(E_n) \nonumber \\
& & \times \frac{\langle n | \, v_x \, | m \rangle \langle m | \, v_y \, |n \rangle
- \langle n | \, v_y \, | m \rangle \langle m | \, v_x \, |n \rangle}
{(E_n-E_m)^2} .
\label{eq:sigma1}
\end{eqnarray}
When we consider a system in a periodic potential and its Bloch states 
$|u_{n\mathbf{k}}\rangle$ as the eigenstates, the identity
\begin{equation}
\langle u_{m\mathbf{k'}} |\, v_{\mu}\, |u_{n\mathbf{k}}\rangle 
= \frac{1}{\hbar}(E_{n\mathbf{k}} - E_{m\mathbf{k'}}) 
 \langle u_{m\mathbf{k'}}| \frac{\partial}{\partial k_{\mu}}
 |u_{n\mathbf{k}}\rangle
\end{equation}
allows one to rewrite Eq. (\ref{eq:sigma1}) into the form
\begin{eqnarray}
\sigma_{xy} 
&=& -\frac{i e^2}{\hbar L^2} \sum_{\mathbf{k}} \sum_{n \neq m} 
f(E_{n\mathbf{k}}) \nonumber \\
& & \times \left( \frac{\partial}{\partial k_x}\langle u_{n\mathbf{k}}|
 \frac{\partial}{\partial k_y}u_{n\mathbf{k}} \rangle
 - \frac{\partial}{\partial k_y}\langle u_{n\mathbf{k}}|
 \frac{\partial}{\partial k_x}u_{n\mathbf{k}} \rangle \right) .
\label{eq:sigma2}
\end{eqnarray}
Since the Berry connection defined in Eq. (\ref{eq:BC}) is 
written for Bloch states as
\begin{equation}
\mathbf{a}_n(\mathbf{k}) = 
-i \langle u_{n\mathbf{k}}|\, 
 \nabla_k \, |u_{n\mathbf{k}} \rangle 
= -i \langle u_{n\mathbf{k}}|\, 
 \frac{\partial}{\partial \mathbf{k}} \, |u_{n\mathbf{k}} \rangle ,
\label{eq:BC2}
\end{equation}
the Hall conductivity reduces to 
\begin{equation}
\sigma_{xy} = \nu \frac{e^2}{h}
\end{equation}
with
\begin{equation}
\nu = \sum_n \int_{\rm BZ} \frac{d^2\mathbf{k}}{2\pi}
\left( \frac{\partial a_{n,y}}{\partial k_x} 
 - \frac{\partial a_{n,x}}{\partial k_y} \right).
\end{equation}
This $\nu$ can be expressed as $\nu = \sum_n \nu_n$
with $\nu_n$ the contribution from the $n$th band, and 
one can easily see that $\nu_n$ is related to the Berry phase, namely
\begin{eqnarray}
\nu_n &=& \int_{\rm BZ} \frac{d^2\mathbf{k}}{2\pi}
\left( \frac{\partial a_{n,y}}{\partial k_x} 
 - \frac{\partial a_{n,x}}{\partial k_y} \right) \nonumber \\
&=& \frac{1}{2\pi}\oint_{\partial {\rm BZ}} d\mathbf{k} 
\cdot \mathbf{a}_n(\mathbf{k}) \nonumber \\
&=& - \frac{1}{2\pi} \gamma_n[\partial {\rm BZ}] .
\label{eq:nu_n}
\end{eqnarray}
Because of the single-valued nature of the wave function, its change 
in the phase factor after
encircling the Brillouin zone boundary ($\partial {\rm BZ}$) can only be an
integer multiple of $2\pi$, which means 
\begin{equation}
\gamma_n[\partial {\rm BZ}] = 2\pi m \ \ \ (m \in Z).
\end{equation}
Therefore, $\nu_n$ can only
take an integer value, and hence $\sigma_{xy}$ is quantized
to integer multiples of $e^2/h$.
The integer $\nu$ is called TKNN invariant, 
and it plays the role of the topological invariant
of the quantum Hall system, which is a TRS-breaking TI.
It is clear from Eq. (\ref{eq:nu_n}) that the TKNN invariant becomes
nonzero (i.e. the system becomes topological) when the Berry
connection
$\mathbf{a}_n(\mathbf{k})$ is not a single-valued function.

\subsection{Time-reversal operator}

We now consider insulators preserving TRS. 
The TR operator $\Theta$ for a spin 1/2 particle takes the simple
form $\Theta = -i s_y K$, where $K$ is the complex conjugate operator.
Hereafter, $s_{\mu}$ ($\mu = x, y, z$) denotes the spin operator
given by Pauli matrices. An important property of the TR operator is that 
\begin{equation}
\Theta^2 = -1 .
\end{equation}
Taking the eigenstates of $s_z$ as the basis set $\{\, |\sigma\rangle \, \}$,
it follows
\begin{eqnarray}
\langle \psi | \Theta | \phi \rangle &=& \sum_{\sigma,\sigma'}
\langle \psi | \sigma \rangle \langle \sigma | (-is_y) | \sigma' \rangle 
\langle \sigma' | \phi^* \rangle \nonumber \\
&=& -\sum \langle \psi | \sigma^* \rangle \langle \sigma^* | is_y | \sigma' \rangle 
\langle \phi | (\sigma')^* \rangle \nonumber \\
&=& -\sum \langle \sigma | \psi^* \rangle \langle (\sigma')^* |(is_y)^{\dagger}|
\sigma \rangle \langle \phi | (\sigma')^* \rangle \nonumber \\
&=& -\sum \langle \phi | (\sigma')^* \rangle \langle (\sigma')^* |(-is_y)|
\sigma \rangle \langle \sigma | \psi^* \rangle \nonumber \\
&=& - \langle \phi | \Theta | \psi \rangle.
\label{eq:rel1}
\end{eqnarray}
Similarly, one obtains
\begin{eqnarray}
\langle \Theta \psi | \Theta \phi \rangle 
&=& \left( \sum_{\sigma} \langle \sigma^*|\psi \rangle \langle \sigma | (+is_y) \right)
\left( \sum_{\sigma'} (-is_y) | \sigma' \rangle \langle \sigma' | \phi^* \rangle
\right) \nonumber \\
&=& \sum_{\sigma,\sigma'} \langle \phi|(\sigma')^* \rangle 
\langle \sigma | \sigma' \rangle \langle \sigma^* | \psi \rangle \nonumber \\
&=& \sum \langle \phi | (\sigma')^* \rangle \langle (\sigma')^* | \sigma^* \rangle
\langle \sigma^* | \psi \rangle \nonumber \\
&=& \langle \phi | \psi \rangle.
\label{eq:rel2}
\end{eqnarray}
Also, for arbitrary linear operator $A$
\begin{equation}
\langle \Theta \psi |\Theta A \Theta^{-1} | \Theta \phi \rangle 
= \langle \phi | A^{\dagger} | \psi \rangle .
\end{equation}

\subsection{TRS and Bloch Hamiltonian}

\begin{figure}[t]
\begin{center}
\includegraphics[width=4cm]{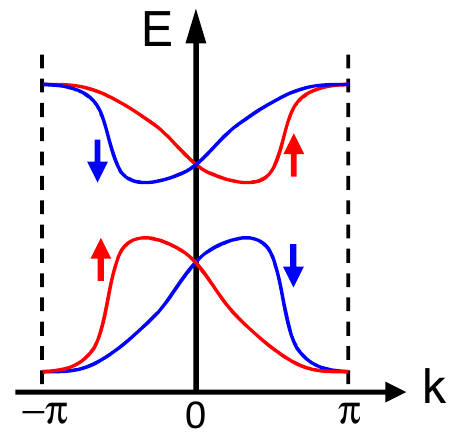}
\caption{(Color online) 
Kramers pairs of bands. Note that each pair of bands
are degenerate at the TR-invariant momentum (TRIM)
where $+\mathbf{k}$ becomes equivalent to $-\mathbf{k}$ due to 
the periodicity of the Brillouin zone. In this figure, there are two
TRIMs, $k$ = 0 and $k = \pi$ (which is equivalent to $k = -\pi$).
The lifting of degeneracy at $k$ values other than 0 and $\pi$ comes 
from spin-orbit coupling.}
\label{fig:Kramers}
\end{center}
\end{figure}

Let $\mathcal{H}$ be the total Hamiltonian of a periodic system:
\begin{equation}
\mathcal{H} | \psi_{n\mathbf{k}} \rangle 
= E_{n\mathbf{k}} | \psi_{n\mathbf{k}} \rangle.
\end{equation}
Bloch's theorem leads to the separation of $\psi_{n\mathbf{k}}$ into
\begin{equation}
| \psi_{n\mathbf{k}} \rangle = e^{i\mathbf{k \cdot r}} | u_{n\mathbf{k}} \rangle,
\end{equation}
where $|u_{n\mathbf{k}}\rangle$ is the cell-periodic eigenstate of the 
Bloch Hamiltonian
\begin{equation}
H(\mathbf{k}) = e^{-i\mathbf{k \cdot r}} \, \mathcal{H} \, e^{i\mathbf{k \cdot r}},
\end{equation}
and $|u_{n\mathbf{k}}\rangle$ satisfies the reduced Schr\"odinger equation
\begin{equation}
H(\mathbf{k}) | u_{n\mathbf{k}} \rangle = E_{n\mathbf{k}} | u_{n\mathbf{k}} \rangle.
\end{equation}

It is important to note that when $\mathcal{H}$ preserves TRS, i.e. 
$[\mathcal{H},\Theta] = 0$, then $H(\mathbf{k})$ satisfies
\begin{equation}
H(-\mathbf{k}) = \Theta H(\mathbf{k}) \Theta^{-1}.
\end{equation}

\begin{figure}[t]
\begin{center}
\includegraphics[width=7.5cm]{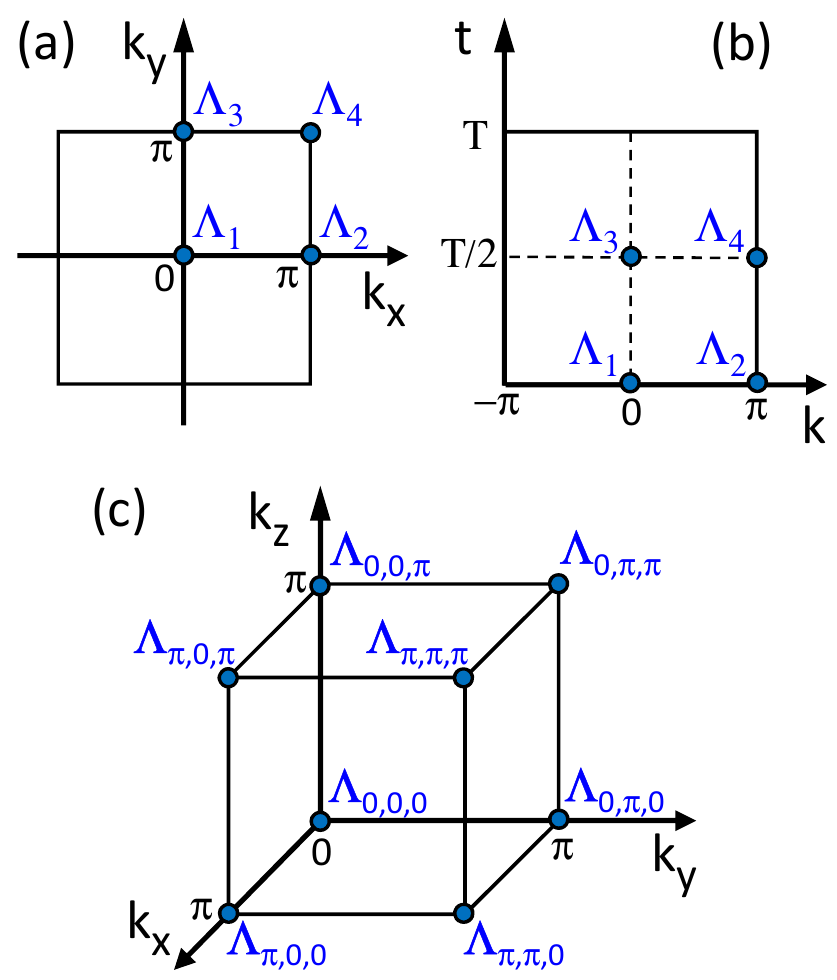}
\caption{(Color online) 
Time-reversal-invariant momenta (TRIMs).
(a) There are 4 TRIMs for a 2D BZ; this figure shows
the case of a square BZ.
(b) The TRIMs for a TR-symmetric 1D system considered in 
Sec. 3.5. One can see that identifying $t \rightarrow k_y$ maps
the periodic $(k,t)$ space to the ordinary 2D BZ shown in (a).
(c) There are 8 TRIMs for a 3D BZ; this figure exemplifies
the case of a cubic BZ.}
\label{fig:TRIM}
\end{center}
\end{figure}

This identity means that the energy bands of a TR symmetric 
system come in pairs (i.e. 
$+\mathbf{k}$ state and $-\mathbf{k}$ state are at the same energy),
and they are called {\it Kramers pairs} (Fig. \ref{fig:Kramers}).
Naturally, Kramers pairs are degenerate at the TR-invariant 
momentum (TRIM)
where $+\mathbf{k}$ becomes equivalent to $-\mathbf{k}$ due to 
the periodicity of the BZ (Fig. \ref{fig:TRIM}).

Now we discuss a matrix representation of the TR operator in the Bloch wave function 
basis. A convenient matrix is
\begin{equation}
w_{\alpha \beta}(\mathbf{k}) = \langle u_{\alpha, -\mathbf{k}} | 
\Theta | u_{\beta, \mathbf{k}} \rangle,
\end{equation} 
where $\alpha$ and $\beta$ are band indices.
This matrix relates the two Bloch states $| u_{\alpha, -\mathbf{k}} \rangle$ and
$|u_{\beta, \mathbf{k}} \rangle$ via
\begin{equation}
| u_{\alpha, -\mathbf{k}} \rangle = \sum_{\beta} 
w_{\alpha \beta}^*(\mathbf{k}) \, \Theta \, | u_{\beta, \mathbf{k}} \rangle.
\label{eq:w-rel1}
\end{equation} 
One can easily confirm that $w_{\alpha \beta}(\mathbf{k})$ is a unitary matrix, i.e.
\begin{equation}
\sum_{\alpha} w_{\gamma \alpha}^{\dagger}(\mathbf{k}) 
w_{\alpha \beta}(\mathbf{k}) =  
\langle u_{\beta, \mathbf{k}} | u_{\gamma, \mathbf{k}} \rangle 
= \delta_{\beta \gamma},
\end{equation} 
and it has the following property:
\begin{equation}
w_{\beta \alpha}(- \mathbf{k}) = - w_{\alpha \beta}(\mathbf{k}) .
\label{eq:w-rel2}
\end{equation} 
This equation means that at a TRIM $\mathbf{\Lambda}$, 
the $w$ matrix becomes antisymmetric, i.e. 
\begin{equation}
w_{\beta \alpha}(\mathbf{\Lambda}) = - w_{\alpha \beta}(\mathbf{\Lambda}).
\end{equation}
For example, if there are only two occupied bands, $w_{\alpha \beta}$ 
becomes a 2 $\times$ 2 matrix and at $\mathbf{\Lambda}$ it can be
explicitly written as
\begin{equation}
w(\mathbf{\Lambda}) 
= \left( \begin{array}{cc}
0 & w_{12}(\mathbf{\Lambda}) \\
-w_{12}(\mathbf{\Lambda}) & 0 \end{array} \right)
= w_{12}(\mathbf{\Lambda}) \left( \begin{array}{cc} 0 & 1 \\
-1 & 0 \end{array} \right).
\end{equation} 

Another convenient matrix to consider for a TR symmetric 
system is the $U(2)$ Berry connection matrix
\begin{equation}
\mathbf{a}_{\alpha \beta}(\mathbf{k}) = -i \,\langle u_{\alpha,\mathbf{k}} | 
\nabla_{\mathbf{k}} | u_{\beta,\mathbf{k}} \rangle.
\label{eq:U(2)}
\end{equation}
Notice that $\mathbf{a}$ is actually a set of three matrices.
By using Eqs. (\ref{eq:rel2}) and (\ref{eq:w-rel1}), it can be shown that 
$\mathbf{a}_{\alpha \beta}(\mathbf{k})$ and 
$\mathbf{a}_{\alpha \beta}(\mathbf{-k})$ are related by
\begin{equation}
\mathbf{a}(\mathbf{-k}) 
= w(\mathbf{k}) \mathbf{a}^*(\mathbf{k})w^{\dagger}(\mathbf{k}) 
+ i w(\mathbf{k}) \nabla _{\mathbf{k}} w^{\dagger}(\mathbf{k}) ,
\end{equation}
and taking the trace gives
\begin{equation}
\tr[\mathbf{a}(\mathbf{-k})] 
= \tr[\mathbf{a^*}(\mathbf{k})] 
+ i \tr[w(\mathbf{k}) \nabla _{\mathbf{k}} w^{\dagger}(\mathbf{k})] .
\end{equation}
Noticing that $\tr[\mathbf{a}]=\tr[\mathbf{a^*}]$ (because 
$\mathbf{a}_{\beta \alpha} = \mathbf{a^*}_{\alpha \beta}$) and 
that $w \nabla w^{\dagger} = -(\nabla w) w^{\dagger}$ (because
$w w^{\dagger} =1$), 
the replacement $\mathbf{-k} \rightarrow \mathbf{k}$ leads to
\begin{equation}
\tr[\mathbf{a}(\mathbf{k})] 
= \tr[\mathbf{a}(\mathbf{-k})] 
+ i \tr[w^{\dagger}(\mathbf{k}) \nabla _{\mathbf{k}} w(\mathbf{k})] .
\label{eq:tr-a-w}
\end{equation}
This relation becomes important in the calculation of the $Z_2$ topological invariant.

\subsection{$Z_2$ time-reversal polarization}

In order to pedagogically derive the topological invariant to characterize 2D 
electron systems preserving TRS, we follow Fu and Kane \cite{FuTRP} and 
consider a 1D system with length $L$ and lattice constant $a$ =1.
We only consider two bands that form a Kramers pair,
and denote their Bloch wave functions as $| u_1(k)\rangle$ and $| u_2(k)\rangle$.
Assume that the band parameters change with time and return to the original values
at $t = T$. Furthermore, we consider the situation when the 1D Hamiltonian satisfies
the following conditions:
\begin{eqnarray}
& & H[t + T] = H[t] 
\label{eq:1D-1} \\
& & H[-t] = \Theta H[t] \Theta^{-1}.
\label{eq:1D-2}
\end{eqnarray}

From the modern theory of ferroelectricity \cite{Resta}, 
it is known that charge polarization $P_{\rho}$ can be
calculated by integrating the Berry connection of the occupied states 
over the BZ. In the present case
of the 1D two-band system, $P_{\rho}$ may be written as
\begin{equation}
P_{\rho} = \int_{-\pi}^{\pi} \frac{dk}{2\pi}A(k),
\end{equation}
where 
\begin{eqnarray}
A(k) 
&=& -i\langle u_1(k)| \nabla_k| u_1(k) \rangle -i\langle u_2(k)| \nabla_k| u_2(k) \rangle \\
&=& a_{11}(k) + a_{22}(k) \nonumber \\
&=& \tr[a].
\end{eqnarray}
The contribution from each band may be called {\it partial polarization} defined by
\begin{equation}
P_i = \int_{-\pi}^{\pi} \frac{dk}{2\pi}a_{ii}(k) \ \ \ \ \ (i = 1,2),
\end{equation}
which gives the relation $P_{\rho} = P_1 + P_2$.
The {\it time-reversal polarization} is defined by
\begin{equation}
P_{\theta} = P_1 - P_2 = 2P_1 - P_{\rho}.
\end{equation}
Intuitively, $P_{\theta}$ gives the difference in charge polarization between spin-up
and spin-down bands, since $| u_1(k)\rangle$ and $| u_2(k)\rangle$ form a 
Kramers pair.

From Eqs. (\ref{eq:1D-1}) and (\ref{eq:1D-2}), one can see that the system
is TR symmetric at $t$ = 0 and $t = T/2$. At these times, the Kramers 
degeneracy must be observed at any $k$, which 
dictates that the time-reversed version of $| u_2(k)\rangle$, 
i.e. $\Theta| u_2(k)\rangle$, is equal to $| u_1(-k)\rangle$ except for 
a phase factor. Hence, at $t$ = 0 and $t = T/2$,
\begin{eqnarray}
& & \Theta| u_2(k)\rangle = e^{-i\chi(k)}| u_1(-k)\rangle \\
& & \Theta| u_1(k)\rangle = - e^{-i\chi(-k)}| u_2(-k)\rangle.
\label{eq:u1-u2}
\end{eqnarray}
Using these relations, the $w$ matrix can be shown to become
\begin{equation}
w(k) = \left( \begin{array}{cc}
0 &  e^{-i\chi(k)} \\
-e^{-i\chi(-k)} & 0 \end{array} \right) .
\label{eq:w2x2}
\end{equation} 

Now we calculate $P_1$ at the TR symmetric times. 
First, using Eq. (\ref{eq:u1-u2}) one may obtain
\begin{equation}
a_{11}(-k) = a_{22}(k) - \frac{\partial}{\partial k}\chi(k),
\end{equation}
which leads to 
\begin{eqnarray}
P_1 &=& \frac{1}{2\pi} \left( \int_0^{\pi}dk\,a_{11}(k) 
 + \int_{-\pi}^0 dk\,a_{11}(k) \right) \nonumber \\
&=& \frac{1}{2\pi} \int_0^{\pi}dk \left( a_{11}(k) + a_{22}(k)  
 - \frac{\partial}{\partial k}\chi(k) \right) \nonumber \\
&=& \int_0^{\pi}\frac{dk}{2\pi}A(k) - \frac{1}{2\pi} [\,\chi(\pi) - \chi(0)] .
\label{eq:P1}
\end{eqnarray}
Since $w_{12}(k) = e^{-i\chi(k)}$ from Eq. (\ref{eq:w2x2}), $\chi(k)$ can be
written as
\begin{equation}
\chi(k) = i \log w_{12}(k)
\end{equation}
and Eq. (\ref{eq:P1}) reduces to
\begin{equation}
P_1 = \int_0^{\pi}\frac{dk}{2\pi}A(k) 
- \frac{i}{2\pi} \log \frac{w_{12}(\pi)}{w_{12}(0)} .
\end{equation}
This expression leads to 
\begin{equation}
P_{\theta} = 2P_1 - P_{\rho} = 
\int_0^{\pi}\frac{dk}{2\pi}[A(k) - A(-k)] 
- \frac{i}{\pi} \log \frac{w_{12}(\pi)}{w_{12}(0)} .
\end{equation}
Noticing $A(k) = \tr[a(k)]$ and using Eq. (\ref{eq:tr-a-w}), one obtains
\begin{eqnarray}
P_{\theta} &=& \int_0^{\pi}\frac{dk}{2\pi} 
i \tr[w^{\dagger}(k) \frac{\partial}{\partial k} w(k)] 
- \frac{i}{\pi} \log \frac{w_{12}(\pi)}{w_{12}(0)} \nonumber \\
&=& i \int_0^{\pi}\frac{dk}{2\pi}  
\frac{\partial}{\partial k} \log(\det[w(k)])
- \frac{i}{\pi} \log \frac{w_{12}(\pi)}{w_{12}(0)}  \nonumber \\
&=& \frac{i}{\pi} \cdot \frac{1}{2} \log \frac{\det[w(\pi)]}{\det[w(0)]}
- \frac{i}{\pi} \log \frac{w_{12}(\pi)}{w_{12}(0)} .
\label{eq:P_theta1}
\end{eqnarray}
Since $\det[w] = w_{12}^2$ in the present case, Eq. (\ref{eq:P_theta1})
reduces to
\begin{equation}
P_{\theta} = \frac{1}{i\pi} \log \left(\frac{\sqrt{w_{12}(0)^2}}{w_{12}(0)}
\cdot \frac{w_{12}(\pi)}{\sqrt{w_{12}(\pi)^2}} \right).
\label{eq:P-theta2}
\end{equation}
The argument of this log is $+1$ or $-1$. Since $\log(-1) = i \pi$ if 
the angle in the complex plane is restricted from 0 to $2\pi$, one can see that 
$P_{\theta}$ is 0 or 1 (mod 2).
Physically, the two values of $P_{\theta}$ corresponds to two different
polarization states which the present 1D system can take at $t = 0$
and $t = T/2$. 

We now consider the change of $P_{\theta}$ from $t = 0$ to $T/2$. 
The wave function $| u_n(k, t)\rangle$ can be viewed as a map from 
the 2D phase space $(k,t)$, which forms a torus due to the periodic boundary
conditions, to the Hilbert space. This Hilbert space can be classified into
two groups depending on the difference in $P_{\theta}$ between 
$t = 0$ and $t = T/2$, 
\begin{equation}
\Delta = P_{\theta}(T/2) - P_{\theta}(0) .
\end{equation}
This $\Delta$ is specified only in mod 2, so it gives a $Z_2$ topological invariant
to characterize the Hilbert space. Intuitively, when $P_{\theta}$ changes
between $t = 0$ to $T/2$, the Hilbert space is ``twisted" and $\Delta =1$, while
the Hilbert space is trivial ($\Delta =0$) when there is no change in $P_{\theta}$.
Using Eq. (\ref{eq:P-theta2}), $\Delta$ can be given in terms of
\begin{equation}
(-1)^{\Delta} = \prod_{i=1}^{4} 
 \frac{w_{12}(\Lambda_i)}{\sqrt{w_{12}(\Lambda_i)^2}},
\end{equation}
where $\Lambda_1$ = $(k,t)$ = $(0,0)$, $\Lambda_2$ = $(\pi,0)$,
$\Lambda_3$ = $(0,T/2)$, and $\Lambda_4$ = $(\pi,T/2)$
[see Fig. \ref{fig:TRIM}(b)]. The physical consequence of a cycle with 
$\Delta = 1$ is spin pumping from one end of the 1D system to the other \cite{FuTRP}.

\subsection{General formula of the $Z_2$ invariant}

Extension of the above argument to a multiband system is not very difficult.
The Hamiltonian still satisfies Eqs. (\ref{eq:1D-1}) and (\ref{eq:1D-2}), and 
in the following we take $T = 2\pi$ for simplicity.
Consider that $2N$ bands are occupied and forming $N$ Kramers pairs.
For each Kramers pair $n$, at the TR symmetric times $t = 0$ and $\pi$ 
the wave functions are related by
\begin{eqnarray}
\Theta| u_2^n(k)\rangle &=& e^{-i\chi_n(k)}| u_1^n(-k)\rangle \\
\Theta| u_1^n(k)\rangle &=& - e^{-i\chi_n(-k)}| u_2^n(-k)\rangle ,
\end{eqnarray}
and the $w$ matrix is given by
\begin{equation}
w(k) = \left( \begin{array}{ccccc}
0 &  e^{-i\chi_1(k)} & 0 & 0 & \ldots  \\
-e^{-i\chi_1(-k)} & 0 & 0 & 0 & \ldots  \\
0 & 0 & 0 & e^{-i\chi_2(k)} & \ldots  \\
0 & 0 & -e^{-i\chi_2(-k)} & 0 & \ldots \\
\vdots & \vdots & \vdots & \vdots & \ddots
\end{array} \right).
\label{eq:wnxn}
\end{equation} 
Hence, at $t$ = 0 and $\pi$, $w(0)$ and $w(\pi)$ become antisymmetric 
and we have
\begin{eqnarray}
w_{12}(\Lambda_i)w_{34}(\Lambda_i) \ldots w_{2N-1,2N}(\Lambda_i)
&=& e^{-i \sum_{n=1}^{N} \chi_n(\Lambda_i)} \nonumber \\
&=& {\rm Pf}[w(\Lambda_i)] .
\end{eqnarray}
Note that in the above equation, $w$ is viewed as a function of $k$ and $t$, 
and we used the formula for the Pfaffian of a $2N \times 2N$ skew-symmetric
matrix with 2 $\times$ 2 blocks on the diagonal. In general, Pfaffian is defined 
for an antisymmetric matrix and is related to the determinant by
\begin{equation}
{\rm Pf}[A]^2 = \det[A].
\end{equation}

It is straightforward to extend the previous calculations for the TR symmetric times 
$t = 0$ and $\pi$ to obtain
\begin{eqnarray}
P_1 &=& \int_0^{\pi}\frac{dk}{2\pi}A(k) 
- \frac{1}{2\pi} \sum_{n=1}^{N}[\,\chi_n(\pi) - \chi_n(0)] \nonumber \\
&=& \int_0^{\pi}\frac{dk}{2\pi}A(k) 
- \frac{i}{2\pi} \log \left( \frac{{\rm Pf}[w(\pi)]}{{\rm Pf}[w(0)]} \right) ,
\end{eqnarray}
which in turn gives the TR polarization
\begin{equation}
P_{\theta} = \frac{1}{i\pi} \log \left(
\frac{\sqrt{\det[w(0)]}}{{\rm Pf}[w(0)]}
\cdot \frac{{\rm Pf}[w(\pi)]}{\sqrt{\det[w(\pi)]}} \right).
\end{equation}
Therefore, the $Z_2$ topological invariant $\nu$ is given by
\begin{equation}
(-1)^{\nu} = \prod_{i=1}^{4} 
 \frac{{\rm Pf}[w(\Lambda_i)]}{\sqrt{\det[w(\Lambda_i)]}},
\label{eq:Z2nu}
\end{equation}
which classifies the Hilbert space into ``twisted" ($\nu$ = 1) and
trivial ($\nu$ = 0) ones.

By reinterpreting the periodic 2D phase space $(k,t)$, which forms a torus, 
as the 2D Brillouin zone $(k_x,k_y)$, the above theory provides a 
$Z_2$ topological classification 
of 2D TR-invariant insulators with $2N$ occupied bands.

\subsection{Extension to 3D systems}

Based on a general homotopy argument, Moore and Balents showed
\cite{3DTI1} that
there are four $Z_2$ invariants for 3D systems. While the construction
of the homotopy was involved, the physical origin of the four invariants
can be easily understood. 
For simplicity, consider a cubic system and 
take the lattice constant $a$ = 1. In the 3D BZ of this system,
there are eight TRIMs denoted as $\Lambda_{0,0,0}$, $\Lambda_{\pi,0,0}$,
$\Lambda_{0,\pi,0}$, $\Lambda_{0,0,\pi}$, $\Lambda_{\pi,0,\pi}$,
$\Lambda_{0,\pi,\pi}$, $\Lambda_{\pi,\pi,0}$, and 
$\Lambda_{\pi,\pi,\pi}$ [see Fig. \ref{fig:TRIM}(c)]. 
At these points in the BZ, the Bloch Hamiltonian becomes
TR symmetric, i.e. $\Theta H(\Lambda_i) \Theta^{-1} = H(\Lambda_i)$,
and each Kramers pair of bands become degenerate.

Notice that the six planes in the 3D BZ, 
$x$ = 0, $x$ = $\pm \pi$, $y$ = 0,  $y$ = $\pm \pi$,
$z$ = 0, and $z$ = $\pm \pi$ possess the symmetries of the 2D BZ, and therefore
they each have a $Z_2$ invariant. The six invariants may be denoted as $x_0$, 
$x_1$, $y_0$, $y_1$, $z_0$, and $z_1$, but those six invariants are
not all independent \cite{3DTI1}. This is because the products $x_0 x_1$, $y_0 y_1$, 
and $z_0 z_1$ are redundant, which stems from the fact that 
those three products involve ${\rm Pf}[w(\Lambda_i)]/\sqrt{\det[w(\Lambda_i)]}$
from all eight TRIMs and hence are the same. This means that there are two
constraining relations $x_0 x_1$ = $y_0 y_1$ = $z_0 z_1$, 
which dictates that only four invariants can be independently 
determined in a 3D system.

The concrete construction of the four $Z_2$ invariants were given by 
Fu and Kane \cite{3DTI2}. For each TRIM $\Lambda_i$, we define 
\begin{equation}
\delta(\Lambda_i) \equiv \frac{{\rm Pf}[w(\Lambda_i)]}{\sqrt{\det[w(\Lambda_i)]}}.
\end{equation}
Using this $\delta(\Lambda_i)$, the four $Z_2$ invariants 
$\nu_0$, $\nu_1$, $\nu_2$, $\nu_3$ are defined as
\begin{eqnarray}
(-1)^{\nu_0} &=& \prod_{n_j = 0,\, \pi} \delta(\Lambda_{n_1,n_2,n_3}) \\
(-1)^{\nu_i} &=& \prod_{n_{j\neq i} = 0,\, \pi \,;\, n_i = \pi} 
 \delta(\Lambda_{n_1,n_2,n_3}) \ \ \ (i = 1, 2, 3) .
\end{eqnarray}
The invariant $\nu_0$ is given by 
a product of all eight $\delta(\Lambda_i)$'s, so it is
unique to a 3D system. On the other hand, other $\nu_i$'s are a product of 
four $\delta(\Lambda_i)$'s and is similar to the $Z_2$ invariant in the 2D case.
For example, 
\begin{equation}
(-1)^{\nu_3} = \delta(\Lambda_{0,0,\pi})\, \delta(\Lambda_{\pi,0,\pi})\, 
\delta(\Lambda_{0,\pi,\pi})\, \delta(\Lambda_{\pi,\pi,\pi})
\end{equation}
corresponds to the $Z_2$ invariant on the $z$ = $\pi$ plane,  
which can be seen by considering the TR polarization defined on this plane,
$P_{\theta}(y=0)_{z=\pi}$ = $\frac{1}{i\pi}\log[\delta(\Lambda_{0,0,\pi})\, 
\delta(\Lambda_{\pi,0,\pi})]$ and $P_{\theta}(y=\pi)_{z=\pi}$ 
= $\frac{1}{i\pi}\log[\delta(\Lambda_{0,\pi,\pi})\, \delta(\Lambda_{\pi,\pi,\pi})]$.
When the TR polarization changes between $y$ = 0 and $y$ = $\pi$
[i.e. $P_{\theta}(y=0)_{z=\pi}$ and $P_{\theta}(y=\pi)_{z=\pi}$ 
are different], then
the $Z_2$ topology is nontrivial and $\nu_3$ becomes 1.

\begin{figure}[t]
\begin{center}
\includegraphics[width=8.2cm]{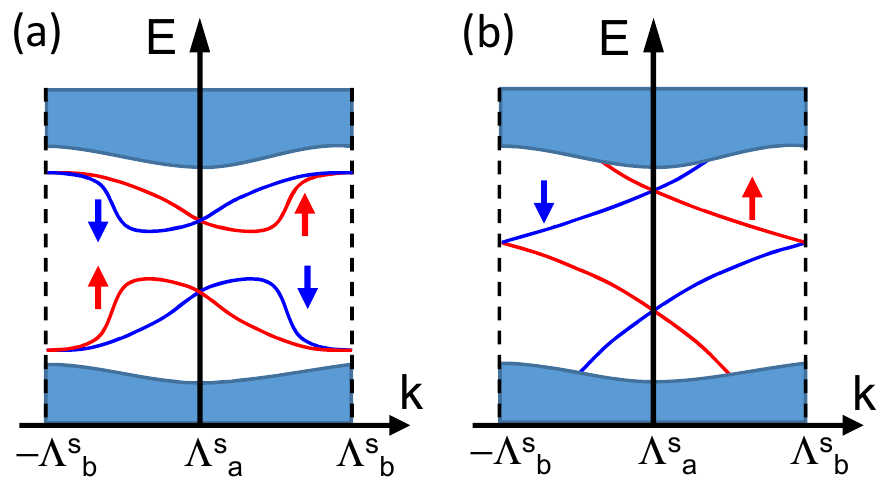}
\caption{(Color online) 
Schematic pictures of the surface states between two surface TRIMs, 
$\Lambda_a^s$ and $\Lambda_b^s$, for (a) topologically trivial
and (b) topologically nontrivial cases. In the latter, the ``switch-partner"
characteristic is a reflection of the change in the TR polarization 
(see text).
The shaded regions represent the bulk continuum states.
To enhance the visibility of the Kramers degeneracy, 
surface state dispersions are shown from $-\Lambda_b^s$ to 
$\Lambda_b^s$ through $\Lambda_a^s$.}
\label{fig:TSS}
\end{center}
\end{figure}

The physical consequence of a nontrivial $Z_2$ invariant is the appearance
of topologically-protected surface states. This is graphically shown 
in Fig. \ref{fig:TSS}, in which topologically trivial and nontrivial surface states 
are compared. In the nontrivial case, the Kramers pairs in the surface state
``switch partners", and as a result, the surface state is guaranteed to 
cross any Fermi energy inside the bulk gap.
This switch-partner characteristic is a reflection of the change in the 
TR polarization discussed above.

It is customary to write the combination of the four invariants in the
form $(\nu_0; \nu_1 \, \nu_2 \, \nu_3)$, because $(\nu_1 \, \nu_2 \, \nu_3)$
can be interpreted as Miller indices to specify the direction of vector 
$\Lambda_i$ in the reciprocal space. A 3D TI is called ``strong" when $\nu_0$ = 1, 
while it is called ``weak" when $\nu_0$ = 0 and $\nu_i$ = 1 for some
$i$ = 1, 2, 3.

\subsection{Topological insulators with inversion symmetry}

Define the space inversion operator $\Pi$ as
\begin{equation}
\Pi | \mathbf{x}, \sigma \rangle = | \mathbf{-x}, \sigma \rangle,
\end{equation}
where $\mathbf{x}$ is the spatial coordinate and $\sigma$ signifies the spin.
One can easily see that 
\begin{equation}
\Pi^2 =1,
\end{equation}
which means that the eigenvalue of the $\Pi$ operator is $\pm 1$.
It is easy to see that $\Pi$ operates on the wave function in momentum
representation as $\Pi | \mathbf{k}, \sigma \rangle = | \mathbf{-k}, 
\sigma \rangle$, and hence Hamiltonian $H = \sum_{\mathbf{k},\sigma,\sigma'}
| \mathbf{k},\sigma \rangle H_{\sigma,\sigma'}(\mathbf{k}) \langle
\mathbf{k}, \sigma'|$ transforms as
\begin{eqnarray}
\Pi \, H\, \Pi^{-1} &=& \sum_{\mathbf{k},\sigma,\sigma'}
| \mathbf{-k},\sigma \rangle H_{\sigma,\sigma'}(\mathbf{k}) \langle
\mathbf{-k}, \sigma'| \nonumber \\
&=& \sum_{\mathbf{k},\sigma,\sigma'}
| \mathbf{k},\sigma \rangle H_{\sigma,\sigma'}(\mathbf{-k}) \langle
\mathbf{k}, \sigma'| .
\end{eqnarray}
When $H$ satisfies $H(\mathbf{-k}) = H(\mathbf{k})$, it also satisfies
\begin{equation}
\Pi \, H\, \Pi^{-1} = H
\end{equation}
and hence the system preserves inversion symmetry (IS).
Our aim is to express the $Z_2$ invariant using the eigenvalues of 
$\Pi$, namely, the parity.

We have previously defined the $U(2)$ Berry connection matrix 
$\mathbf{a}_{\alpha \beta}(\mathbf{k})$ in Eq. (\ref{eq:U(2)})
and calculated its trace, $\tr[\mathbf{a}(\mathbf{k})]$, in 
Eq. (\ref{eq:tr-a-w}). 
Now we denote this trace as 
$\mathbf{a}^c(\mathbf{k})$ and define
\begin{equation}
F(\mathbf{k}) = \nabla_k \times \mathbf{a}^c(\mathbf{k}) , 
\end{equation}
which is the Berry curvature of $\mathbf{a}^c$.
The two symmetries dictate that this $F(\mathbf{k})$ has the following properties:
\cite{Haldane2004}
\begin{eqnarray}
{\rm TRS} &\rightarrow& F(\mathbf{-k}) = - F (\mathbf{k}) \\
{\rm IS} &\rightarrow& F(\mathbf{-k}) = + F(\mathbf{k}) .
\end{eqnarray}
Therefore, the Berry curvature $F(\mathbf{k})$ = 0 for any $\mathbf{k}$,
and hence one can always chose a gauge which makes 
$\mathbf{a}^c(\mathbf{k})$ = 0.

To actually obtain a gauge to set $\mathbf{a}^c(\mathbf{k})$ = 0 for any 
$\mathbf{k}$, we consider the matrix
\begin{equation}
v_{\alpha \beta}(\mathbf{k}) \equiv \langle u_{\alpha \mathbf{k}} |
\, \Pi \, \Theta \, | u_{\beta \mathbf{k}} \rangle .
\end{equation}
One can see that this $v$-matrix is antisymmetric ($v_{\alpha \beta}
= -v_{\beta \alpha}$) and unitary ($\sum_{\beta} v_{\alpha \beta}
v^*_{\gamma \beta} = \delta_{\alpha \gamma}$).
Furthermore, $v$ is related to $\mathbf{a}^c(\mathbf{k})$ via
\begin{equation}
\frac{i}{2} \tr[v^{\dagger} \nabla_k v] = \mathbf{a}^c(\mathbf{k}),
\end{equation}
which can be verified by calculations similar to those employed for deriving
Eq. (\ref{eq:tr-a-w}).
Since $\tr[v^{\dagger} \nabla_k v]$ = $\nabla_k \tr[\log v]$
= $\nabla_k \log(\det[v])$ \cite{FuKane3D}, one can see
\begin{equation}
\mathbf{a}^c(\mathbf{k}) = \frac{i \nabla_k}{2} \log (\det[v(\mathbf{k})])
= i \nabla_k \log {\rm Pf}[v(\mathbf{k})].
\end{equation}
To set $\mathbf{a}^c(\mathbf{k})$ = 0, all we have to do is to 
adjust the phase of $|u_{\alpha \mathbf{k}} \rangle$ 
to make ${\rm Pf}[v(\mathbf{k})]$ = 1.

Now we calculate the $w$-matrix of an inversion symmetric system.
Let $\xi_{\alpha}(\Lambda_i) = \pm 1$ be the eigenvalue of $\Pi$ 
for band $\alpha$ at TRIM $\Lambda_i$, namely, 
$\Pi |u_{\alpha}(\Lambda_i) \rangle = \xi_{\alpha}(\Lambda_i)
|u_{\alpha}(\Lambda_i) \rangle$.
Remembering $\Pi^2$ =1, one may calculate
\begin{eqnarray}
w_{\alpha \beta}(\Lambda_i) &=& \langle u_{\alpha}(-\Lambda_i)| \,
 \Pi \, \Pi \, \Theta \, |u_{\beta}(\Lambda_i)\rangle \nonumber \\
&=& \xi_{\alpha}(\Lambda_i) \langle u_{\alpha}(\Lambda_i)
 |\, \Pi \, \Theta \, |u_{\beta}(\Lambda_i)\rangle \nonumber \\
&=& \xi_{\alpha}(\Lambda_i) v_{\alpha \beta}(\Lambda_i).
\end{eqnarray}
Since $w_{\alpha \beta}$ and $v_{\alpha \beta}$ are antisymmetric,
$\xi_{\alpha} = \xi_{\beta}$ when $w_{\alpha \beta}$ = 
$-w_{\beta \alpha} \neq 0$. Such a non-zero $w_{\alpha \beta}$
is obtained only when the bands $\alpha$ and $\beta$ form a Kramers
pair. Hence, if the bands $\alpha$ and $\beta$ are 
the $n$th Kramers pair in the total of $2N$ bands,
we may write $\xi_{\alpha} = \xi_{\beta} \equiv \xi_{2n}$.
Remembering that $w_{\alpha \beta}$ has the peculiar skew-symmetric 
from of Eq. (\ref{eq:wnxn}), one can see
\begin{equation}
{\rm Pf}[w(\Lambda_i)] = {\rm Pf}[v(\Lambda_i)] 
\prod_{n=1}^{N} \xi_{2n}(\Lambda_i) .
\end{equation}
Since we have chosen the gauge to give ${\rm Pf}[v(\mathbf{k})]$ = 1,
we finally obtain
\begin{equation}
\delta(\Lambda_i) = \frac{{\rm Pf}[w(\Lambda_i)]}{\sqrt{\det[w(\Lambda_i)]}} 
= \prod_{n=1}^{N} \xi_{2n}(\Lambda_i) .
\label{eq:Z2_inv}
\end{equation}
This means that the $Z_2$ invariant can be calculated simply by the parity
eigenvalues $\xi_{2n}$ at TRIMs $\Lambda_i$.
For example, for a 2D system with $2N$ bands preserving IS and TRS, 
the $Z_2$ invariant $\nu$ is given by
\begin{equation}
(-1)^{\nu} 
= \prod_{i=1}^{4} \delta(\Lambda_i)
= \prod_{i=1}^{4} \prod_{n=1}^{N} \xi_{2n}(\Lambda_i) .
\label{eq:Z2_inv2}
\end{equation}
This simplification of the calculation of the $Z_2$ index, accomplished by 
Fu and Kane \cite{FuKane3D}, greatly facilitated the theoretical predictions 
of candidate TI materials based on {\it ab initio} band calculations.

\subsection{BHZ model}

Now that we have learned how the topology of a TR-invariant insulator 
can be specified, let us consider a concrete example that gives rise to a 
topologically nontrivial state. As mentioned in Sec. 2.2, the first
theoretical model proposed in this context was the Kane-Mele model 
based on graphene \cite{KM_QSH}, but it was the BHZ model 
\cite{BHZ} that actually led to the experimental discovery
of the first 2D TI material. In this section, we discuss the BHZ model
within the framework developed by Fu and Kane \cite{FuKane3D} 
to make its $Z_2$ topology transparent.

The BHZ model is motivated by CdTe/HgTe/CdTe quantum well, where
the $s$- and $p$-orbital bands invert at $\mathbf{k}$ = 0 as a function
of the well width. In this quantum-well structure, two 
$s$-orbital states and two $p_{m=3/2}$-orbital states
appear near the Fermi energy. Hence, we consider the following 4 bands:
\begin{equation}
| s, \uparrow \rangle, \ \  |s, \downarrow \rangle, \ \  
|p_x + i p_y, \uparrow \rangle, \ \ |p_x - i p_y, \downarrow. \rangle \nonumber
\end{equation}
Since bulk HgTe crystalizes in the cubic 
zinc-blende structure, we consider 2D square lattice
with two orthogonal lattice vectors $\mathbf{a_1}$ and $\mathbf{a_2}$.
The Hamiltonian is written as
\begin{eqnarray}
H_{BHZ} &=& \sum_i \sum_{\alpha = s,p} \, \sum_{s_z=\pm} 
\epsilon_{\alpha} C_{i,\alpha,s_z}^{\dagger}C_{i,\alpha,s_z} \nonumber \\
& & -\sum_i \sum_{\alpha,\beta} \sum_{\mu=\pm x,\pm y} \, \sum_{s_z=\pm} 
t_{\mu s_z}^{\alpha \beta} C_{i+\mu,\alpha,s_z}^{\dagger}C_{i,\beta,s_z},
\label{eq:BHZ}
\end{eqnarray}
where $\alpha,\beta$ are $s$ or $p$ bands, $s_z$ is the $z$ component
of the spin, and $\mu$ specifies the four nearest-neighbor bonds.  
The hopping term is a matrix in the band basis:
\begin{equation}
t_{\mu s_z} = \left( \begin{array}{cc} t_{ss} & t_{sp}e^{i s_z \theta_{\mu}} \\
t_{sp}e^{-i s_z \theta_{\mu}} & -t_{pp} \end{array} \right) ,
\end{equation}
where $\theta_{\mu}$ is the angle between the $x$ axis and the direction of the 
bond $\mu$, which can be 0, $\frac{\pi}{2}$, $\pi$, and $\frac{3\pi}{2}$.

Fourier transformation of the first term in Eq. (\ref{eq:BHZ}) gives
\begin{equation}
\sum_{i,\alpha,s_z} 
\epsilon_{\alpha} C_{i,\alpha,s_z}^{\dagger}C_{i,\alpha,s_z} = 
\sum_{\mathbf{k}}C_{\mathbf{k}}^{\dagger} \left(
\frac{\epsilon_s + \epsilon_p}{2} I \otimes I + \frac{\epsilon_s - \epsilon_p}{2}
\sigma_z \otimes I \right) C_{\mathbf{k}} .
\end{equation}
Similarly, the hopping term reduces in the $\mathbf{k}$ representation to
\begin{equation*}
-\sum_{i,\alpha,\beta,\mu,s_z} t_{\mu s_z}^{\alpha \beta} 
C_{i+\mu,\alpha,s_z}^{\dagger}C_{i,\beta,s_z}  
\end{equation*}
\begin{eqnarray}
= -\sum_{\mathbf{k}}C_{\mathbf{k}}^{\dagger} \biggl[ 
(t_{ss}-t_{pp})\sum_{\mu} (\cos \mathbf{k} \cdot \mathbf{a_{\mu}}) \,
I \otimes I \nonumber \\
+\, (t_{ss}+t_{pp})\sum_{\mu} (\cos \mathbf{k} \cdot \mathbf{a_{\mu}})
\, \sigma_z \otimes I & \nonumber \\
+\, (2t_{sp} \sin \mathbf{k} \cdot \mathbf{a_1})\, \sigma_y \otimes I
& \nonumber \\
+\, (2t_{sp} \sin \mathbf{k} \cdot \mathbf{a_2})\, \sigma_x \otimes s_z \,
\biggr] C_{\mathbf{k}} & . & 
\end{eqnarray}

Therefore, by defining the following Dirac $\Gamma$ matrices
\begin{eqnarray}
\Gamma^1 = \sigma_x \otimes s_x = \left( \begin{array}{cc} 0 & s_x \\
s_x & 0 \end{array} \right) , \\
\Gamma^2 = \sigma_x \otimes s_y = \left( \begin{array}{cc} 0 & s_y \\
s_y & 0 \end{array} \right) , \\
\Gamma^3 = \sigma_x \otimes s_z = \left( \begin{array}{cc} 0 & s_z \\
s_z & 0 \end{array} \right) , \\
\Gamma^4 = \sigma_y \otimes I = \left( \begin{array}{cc} 0 & -i I \\
i I & 0 \end{array} \right) , \\
\Gamma^5 = \sigma_z \otimes I = \left( \begin{array}{cc} I & 0 \\
0 & -I \end{array} \right) ,
\end{eqnarray}
the total Hamiltonian Eq. (\ref{eq:BHZ}) is written in the form
\begin{equation}
H(\mathbf{k}) = d_0(\mathbf{k}) I + \sum_{a=1}^{5} d_a(\mathbf{k})\Gamma^a
\end{equation}
with the coefficients
\begin{eqnarray}
d_0(\mathbf{k}) &=& (\epsilon_s + \epsilon_p)/2 \nonumber \\
& & - (t_{ss}-t_{sp})
(\cos \mathbf{k} \cdot \mathbf{a_1} + \cos \mathbf{k} \cdot \mathbf{a_2}) , \\
d_1(\mathbf{k}) &=& 0 ,\\
d_2(\mathbf{k}) &=& 0 ,\\
d_3(\mathbf{k}) &=& 2t_{sp} \sin \mathbf{k} \cdot \mathbf{a_2},\\
d_4(\mathbf{k}) &=& 2t_{sp} \sin \mathbf{k} \cdot \mathbf{a_1},\\
d_5(\mathbf{k}) &=& (\epsilon_s - \epsilon_p)/2 \nonumber \\
& & - (t_{ss}+t_{sp})
(\cos \mathbf{k} \cdot \mathbf{a_1} + \cos \mathbf{k} \cdot \mathbf{a_2}) .
\label{eq:d5}
\end{eqnarray}
The energy eigenvalues are given by
\begin{equation}
E(\mathbf{k}) = d_0(\mathbf{k}) \pm \sqrt{\sum_a d_a(\mathbf{k})^2} \,.
\end{equation}

Note that, since $s$-orbital is parity even and $p$-orbital is parity odd, the
inversion operator in this band basis is constructed as
\begin{equation}
\Pi = \sigma_z \otimes I = \Gamma^5. 
\label{eq:gamma5}
\end{equation}

One may notice that the $\Gamma$ matrices defined above 
have the following symmetry:
\begin{eqnarray}
\Theta \, \Gamma^a \, \Theta^{-1} &=& \left\{ \begin{array}{c} 
-\Gamma^a \ \ \ (a = 1,2,3,4) \\
+\Gamma^a \ \ \ (a = 5) \ \ \ \ \ \ \ \ \ \ \ \ \end{array} \right. \\
\Pi \, \Gamma^a \, \Pi^{-1} &=& \left\{ \begin{array}{c} 
-\Gamma^a \ \ \ (a = 1,2,3,4) \\
+\Gamma^a \ \ \ (a = 5) \ \ \ \ \ \ \ \ \ \ \ \ \end{array}. \right.
\end{eqnarray}
Since only $\Gamma^5$ is even under TR and inversion, at a TRIM $\Lambda_i$
where the system preserves both TRS and IS, the Hamiltonian must have the form
\begin{equation}
H(\mathbf{k}=\Lambda_i) = d_0(\Lambda_i) I + d_5(\Lambda_i)\Gamma^5 .
\label{eq:BHZ2}
\end{equation}
Also, Kramers theorem dictates that $|s,\uparrow\rangle$
and $|s,\downarrow\rangle$ are degenerate at $\Lambda_i$, and so are 
$|p_x+ip_y,\uparrow\rangle$ and $|p_x-ip_y,\downarrow\rangle$.
Remembering that the $s$- and $p$-orbital bands have even $(+)$ and odd $(-)$ 
parities, respectively, we denote those degenerate states at $\Lambda_i$ as
$|+\rangle$ and $|-\rangle$. One can easily see that the eigenvalues of 
$H(\Lambda_i)$ given by Eq. (\ref{eq:BHZ2}) for those states are 
\begin{eqnarray}
\langle +| H(\Lambda_i) |+ \rangle &=& d_0(\Lambda_i) + d_5(\Lambda_i) 
\equiv E_+ \nonumber \\
\langle -| H(\Lambda_i) |- \rangle &=& d_0(\Lambda_i) - d_5(\Lambda_i) 
\equiv E_- ,
\end{eqnarray}
because $\langle +|\Gamma^5 |+ \rangle = 1$ and 
$\langle -|\Gamma^5 |- \rangle = -1$ due to Eq. (\ref{eq:gamma5}).

When $d_5(\Lambda_i) >0$ at a TRIM $\Lambda_i$, 
$E_+ > E_-$ and the $|-\rangle$ state is occupied at that $\Lambda_i$
if the system is half-filled; in this case,
the parity of the occupied state at $\Lambda_i$, $\delta(\Lambda_i)$ in Eq. 
(\ref{eq:Z2_inv}), is $-1$. On the other hand, when $d_5(\Lambda_i) <0$, 
$E_+$ becomes
smaller than $E_-$ and hence the $|+\rangle$ state is occupied, making 
$\delta(\Lambda_i) = +1$. Therefore, in the present 4-band model
one may conclude
\begin{equation}
\delta(\Lambda_i) = -{\rm sgn}[d_5(\Lambda_i)] .
\label{eq:delta-BHZ}
\end{equation}
If we set $|\mathbf{a_1}|=|\mathbf{a_2}|=\pi$, the four TRIMs are $(0,0)$,
$(0,1)$, $(1,0)$, and $(1,1)$. Putting Eq. (\ref{eq:d5}) into Eq. (\ref{eq:delta-BHZ}),
one obtains
\begin{equation}
\delta(\Lambda(n_1,n_2))
= -{\rm sgn}\left[\frac{\epsilon_s - \epsilon_p}{2} 
- (t_{ss}+t_{sp})\Bigl\{ (-1)^{n_1} + (-1)^{n_2} \Bigl\} \right] .
\end{equation}
This leads to the following conclusion regarding the topology of this 4-band system:
\begin{itemize}
\item If $\epsilon_s - \epsilon_p > 4(t_{ss}+t_{sp})$, $\delta <0$ for all $\Lambda_i$, 
which leads to the $Z_2$ invariant $\nu$ given by Eq. (\ref{eq:Z2_inv2}) to become 0. 
This means that the system is a trivial insulator.
\item If $\epsilon_s - \epsilon_p < 4(t_{ss}+t_{sp})$, $\delta >0$ at $\mathbf{k}=0$
but $\delta <0$ for other three TRIMs, which leads to $\nu$ = 1. 
This means that the system is a TI.
\end{itemize}

It is instructive to note that in the trivial state the band order is
$E_- < E_+$, which means $p$-orbital bands lie below the $s$-orbital
bands, which is normal for a band insulator. When the system becomes a
TI, the band order flips only at the $\Gamma$ point in the 2D BZ, and
such a band inversion occurring at an odd number of TRIMs in the BZ is
the key to realizing the nontrivial $Z_2$ topology. In the case of HgTe,
this band inversion occurs due to strong SOC.
Therefore, this example tells us that a viable strategy for discovering
a TI is to look for materials having strong SOC that causes band
inversion at an odd number of TRIMs in the BZ.

\section{TI Materials Discovered to Date}

Table I gives a summary of the TI materials discussed in this section.
This table lists only those materials that have
been experimentally addressed as of May 2013, since there are too many materials that
have been predicted to be TIs but not tested experimentally.

\begin{table*}[t]
\caption{Summary of topological insulator materials that have bee
experimentally addressed. The definition of (1;111) {\it etc.} is introduced 
in Sec. 3.7. (In this table, S.S., P.T., and SM stand for surface state,
phase transition, and semimetal, respectively.) } 
\centering 


\begin{tabular}{c c c c c c} 
\hline 
Type  &  Material  &  Band gap  &  Bulk transport  &  Remark  &  Reference   \\ 
[0.5ex] 
\hline 
2D, $\nu$ = 1 & CdTe/HgTe/CdTe  & $<10$ meV & insulating & high mobility & \cite{Konig2007} \\
2D, $\nu$ = 1 & AlSb/InAs/GaSb/AlSb & $\sim$4 meV & weakly insulating & gap is too small & \cite{GaSb1} \\
3D (1;111) & Bi$_{1-x}$Sb$_x$ & $<30$ meV & weakly insulating & complex S.S. & \cite{Hsieh2008,Nishide} \\
3D (1;111) & Sb & semimetal & metallic & complex S.S. & \cite{Hsieh2009} \\
3D (1;000) & Bi$_2$Se$_3$ & 0.3 eV & metallic & simple S.S. & \cite{Hasan_Bi2Se3} \\
3D (1;000) & Bi$_2$Te$_3$ & 0.17 eV & metallic & distorted S.S. & \cite{Chen_Bi2Te3, Hasan_Bi2Te3} \\
3D (1;000) & Sb$_2$Te$_3$ & 0.3 eV & metallic & heavily $p$-type & \cite{Sb2Te3} \\
3D (1;000) & Bi$_2$Te$_2$Se & $\sim$0.2 eV & reasonably insulating & $\rho_{xx}$ up to 6 $\Omega$cm & \cite{XuBTS, RenBTS, XiongBTS} \\
3D (1;000) & (Bi,Sb)$_2$Te$_3$ & $<0.2$ eV & moderately insulating & mostly thin films & \cite{XueBST} \\
3D (1;000) & Bi$_{2-x}$Sb$_{x}$Te$_{3-y}$Se$_{y}$ & $<0.3$ eV & reasonably insulating & Dirac-cone engineering & \cite{ RenBSTS, TaskinBSTS, Arakane} \\
3D (1;000) & Bi$_2$Te$_{1.6}$S$_{1.4}$ & 0.2 eV & metallic & $n$-type & \cite{JiBTS} \\
3D (1;000) & Bi$_{1.1}$Sb$_{0.9}$Te$_2$S & 0.2 eV & moderately insulating & $\rho_{xx}$ up to 0.1 $\Omega$cm & \cite{JiBTS} \\
3D (1;000) & Sb$_2$Te$_2$Se & ? & metallic & heavily $p$-type & \cite{XuBTS} \\
3D (1;000) & Bi$_2$(Te,Se)$_2$(Se,S) & 0.3 eV & semi-metallic & natural Kawazulite & \cite{Kawazulite} \\
3D (1;000) & TlBiSe$_2$ & $\sim$0.35 eV & metallic & simple S.S., large gap & \cite{SatoTl, KurodaTl, ChenTl} \\
3D (1;000) & TlBiTe$_2$ & $\sim$0.2 eV & metallic & distorted S.S. & \cite{ChenTl} \\
3D (1;000) & TlBi(S,Se)$_2$ & $<0.35$ eV & metallic & topological P.T. & \cite{XuTl, SatoNatPhys} \\
3D (1;000) & PbBi$_2$Te$_4$ & $\sim$0.2 eV & metallic & S.S. nearly parabolic & \cite{SoumaPBT, KurodaPBT} \\
3D (1;000) & PbSb$_2$Te$_4$ & ? & metallic & $p$-type & \cite{SoumaPBT} \\
3D (1;000) & GeBi$_2$Te$_4$ & 0.18 eV & metallic & $n$-type & \cite{XuBTS, Neupane, Okamoto} \\
3D (1;000) & PbBi$_4$Te$_7$ & 0.2 eV & metallic & heavily $n$-type & \cite{Eremeev} \\
3D (1;000) & GeBi$_{4-x}$Sb$_x$Te$_7$ & 0.1--0.2 eV & metallic & $n$ ($p$) type at $x$ = 0 (1) & \cite{Muff_GBT} \\ 
3D (1;000) & (PbSe)$_5$(Bi$_2$Se$_3$)$_{6}$ & 0.5 eV & metallic & natural heterostructure & \cite{Nakayama} \\ 
3D (1;000) & (Bi$_2$)(Bi$_2$Se$_{2.6}$S$_{0.4}$) & semimetal & metallic & (Bi$_2$)$_n$(Bi$_2$Se$_3$)$_m$ series & \cite{Valla2012} \\ 
3D (1;000) & (Bi$_2$)(Bi$_2$Te$_3$)$_2$ & ? & ? & no data published yet & \cite{CavaReview} \\ 
3D TCI & SnTe & 0.3 eV (4.2 K) & metallic & Mirror TCI, $n_{\mathcal{M}} = -2$ & \cite{TanakaTCI} \\
3D TCI & Pb$_{1-x}$Sn$_x$Te & $<0.3$ eV & metallic & Mirror TCI, $n_{\mathcal{M}} = -2$ & \cite{XuTCI2} \\
3D TCI & Pb$_{0.77}$Sn$_{0.23}$Se & invert with $T$ & metallic & Mirror TCI, $n_{\mathcal{M}} = -2$ & \cite{Dziawa} \\
2D, $\nu$ = 1? & Bi bilayer & $\sim$0.1 eV & ? & not stable by itself & \cite{Yang2012,Sabater2013} \\
3D (1;000)? & Ag$_2$Te & ? & metallic & famous for linear MR & \cite{Ag2Te_AB1,Ag2Te_AB2} \\
3D (1;111)? & SmB$_6$ & 20 meV & insulating & possible Kondo TI & \cite{SmB6-1, SmB6-2, SmB6-3, SmB6ARPES} \\
3D (0;001)? & Bi$_{14}$Rh$_3$I$_9$ & 0.27 eV & metallic & possible weak 3D TI & \cite{BiRhI} \\
3D (1;000)? & $R$BiPt ($R$ = Lu, Dy, Gd) & zero gap & metallic & evidence negative & \cite{Kaminski} \\
Weyl SM? & Nd$_2$(Ir$_{1-x}$Rh$_x$)$_2$O$_7$ & zero gap & metallic & too preliminary & \cite{Nd2Ir2O7} \\
[0.5ex] 
\hline 
\end{tabular}


\end{table*}

\subsection{Two-dimensional TIs}

The first material that was experimentally identified as a TR-invariant
TI was CdTe/HgTe/CdTe quantum well \cite{Konig2007}, 
namely, a thin layer of HgTe
sandwiched by CdTe. This is a 2D system where the degree of freedom for
the perpendicular direction is quenched due to the quantum confinement
of the electronic states in the HgTe unit and the resulting subband
formation. Both HgTe and CdTe crystallize in zinc-blende structure, and
their superlattices have been actively studied because of their
application to infrared detectors. Therefore, the necessary technology
to synthesize the required quantum well was already developed before the
prediction of its TI nature, although it involves a very specialized
molecular-beam epitaxy (MBE) technique to deal with mercury \cite{Konig2007}.

As we have seen in the analysis of the BHZ model in the previous
section, the inversion between $p$- and $s$-orbital bands is essential
for the system to obtain the TI nature. Bulk HgTe realizes such a band
inversion, while CdTe does not. Therefore, HgTe is a good starting
material for conceiving a TI phase; however, there is a problem in the
band structure of bulk HgTe, that is, a crystal-symmetry-protected
degeneracy at the $\Gamma$ point makes the system to be intrinsically a
zero-gap semiconductor \cite{BHZ}, 
which means that there is no band gap between
the $p$- and $s$-orbital bands and the system is not qualified as an insulator.
However, by sandwiching HgTe by
CdTe, which has slightly larger lattice constant, the epitaxial strain
exerted on HgTe breaks the cubic lattice symmetry and leads to a gap
opening, and hence the system can become a genuine insulator. It was
predicted by BHZ that above a certain critical
thickness the strained HgTe unit retains the band inversion and should
be a TI \cite{BHZ}. 
This prediction was confirmed by a group at the University of W\"urzburg
led by Molenkamp via transport experiment of
micro-fabricated samples \cite{Konig2007}. 
They found that, when the thickness of the
HgTe unit is above the critical thickness $d_c$ of 6.3 nm, their samples
show a ``negative energy gap" (i.e. band is inverted) and quantization
of the conductance to $2e^2/h$ in zero magnetic field was observed when
the chemical potential is tuned into the gap (Fig. \ref{fig:Koenig}). In contrast, when the
thickness is below $d_c$, the band inversion is lost and they observed
diverging resistivity.

The W\"urzburg group later reported the observation of nonlocal transport
\cite{Roth}, which gave further support to the existence of an edge state. 
They also showed that the 1D edge state
responsible for the quantized transport in the 2D TI phase is likely to
be helically spin polarized, by fabricating an elaborate device structure
which relies on the conventional spin Hall effect that occurs in
metallic (doped) HgTe \cite{Brune2012}. 

The CdTe/HgTe/CdTe quantum well can be made very clean with the carrier
mobility of up to $\sim$10$^5$ cm$^2$/Vs, which makes it possible to
study quantum transport properties. On the other hand, the drawback of
this system is that the bulk band gap that opens due to the epitaxial
strain is very small (up to $\sim$ 10 meV, depending on the thickness
\cite{Konig2012arXiv}), which makes the detection of the TI phase to be
possible only at very low temperature. Also, since the synthesis of
CdTe/HgTe/CdTe quantum wells requires dedicated MBE machines, the sources
of samples for basic physics experiments are currently very limited, which
has made the progress of experimental studies of 2D TIs (QSH insulators) 
relatively slow.

Recently, another 2D TI system, AlSb/InAs/GaSb/AlSb quantum well, was
theoretically predicted \cite{Liu2008} and experimentally confirmed
\cite{GaSb1,GaSb2,GaSb3}. The essential workings of this system are the
following: The valence-band top of GaSb lies above the conduction-band
bottom of InAs. Hence, when InAs and GaSb are in direct contact and they
are both quantum confined (by the outer units of AlSb which has a large
band gap and works as a barrier), the resulting hole subband in GaSb may
lie above the electron subband in InAs, and therefore the band order of
this quantum well is inverted. The band gap in this quantum well arises
from anti-crossing of the two subbands at finite momentum and hence is
very small ($\sim$4 meV), which makes clean observation of the helical
edge state very difficult \cite{KnezRPB2010,GaSb1}. Reasonably
convincing evidence for the TI phase was obtained via observation of the
$2e^2/h$ quantization of the zero-bias Andreev reflection conductance
through Nb point contacts \cite{GaSb2}, but more recently, direct
observation of the conductance quantization to $2e^2/h$ has been achieved
by introducing disorder to the InAs/GaSb interface by Si doping to
localize the unwanted bulk carriers \cite{GaSb3}.

Other possible candidates of 2D TIs include Bi bilayer
\cite{Bi-bilayer}, Na$_2$IrO$_3$ \cite{NaIrO}, and graphene with
artificially enhanced SOC \cite{graSOC,Hu2012}. In particular, for the Bi
bilayer there have been experimental efforts to address its topological
nature \cite{Hirahara2011,Yang2012,Sabater2013}, and suggestive evidence
for the existence of edge states has been reported
\cite{Yang2012,Sabater2013}. It is worth mentioning that silicene, an
analog of graphene consisting of silicon atoms instead of carbon atoms
\cite{Takeda1994}, has also been predicted to be a 2D TI with a high
tunability \cite{silicene}. Since the experimental studies of 2D TIs
have been hindered by the scarcity of samples, new discoveries of 2D TI
materials are strongly call for.

\subsection{Three-dimensional TIs}

As already mentioned in Sec. 2, the first 3D TI material that was
experimentally identified was Bi$_{1-x}$Sb$_x$ \cite{Hsieh2008}, 
following the very
specific prediction by Fu and Kane \cite{FuKane3D}. 
This material is an alloy of Bi
and Sb and it naturally possesses the two essential features, (i) band
inversion at an odd number of TRIMs and (ii) opening of a bulk band gap,
in the Sb concentration range of 0.09 to 0.23 \cite{Lenoir}. 
The 3D $Z_2$ invariant has been identified as (1;111).

Unfortunately, it turned out that this system is not very suitable for
detailed studies of the topological surface state due to the
complicated surface-state structure \cite{Hsieh2008,Nishide},
as can be seen in Fig. \ref{fig:BiSb}. 
This is because its parent
material, Bi, already harbors prominent spin-non-degenerate surface
states due to the strong Rashba effect on the surface of this material
\cite{Hirahara}.
In Bi$_{1-x}$Sb$_x$, such non-topological, Rashba-split surface states
are responsible for 2 or 4 Fermi-level crossings of the surface states
(depending on the chemical potential) \cite{Hsieh2008,Nishide}, 
and the topological one
contributes just one additional Fermi-level crossing. First-principle
calculations of the surface states of Bi$_{1-x}$Sb$_x$ have been
reported \cite{Teo,ZhangBiSb}, 
but the predicted surface-state structure does not really
agree with the experimental results. Such an incomplete understanding of
the nature of the surface state is partly responsible for confusions
occasionally seen in interpretations of experimental data. For example,
in the STS work which addressed the
protection from backscattering in the surface state of Bi$_{1-x}$Sb$_x$
\cite{Roushan}, the analysis considered the spin polarizations of 
only those surface states that are also present in topologically trivial Bi
\cite{Hirahara}, and yet, it was concluded that the result
gives evidence for topological protection.

Nevertheless, Bi$_{1-x}$Sb$_x$ is unique among known 3D TI
materials in that it has an intrinsically high 2D carrier mobility of
$\sim$10$^4$ cm$^2$/Vs (despite the fact that it is an alloy),
which makes it easy to study novel 2D quantum
transport \cite{Taskin2009,Taskin2010}. 
Also, it is relatively easy for this system to reduce the
bulk carrier density to $\sim$10$^{16}$ cm$^{-3}$ in high-quality single
crystals, making it possible, for example, to perform Landau level
spectroscopy of the surface states via magneto-optics \cite{Basov}. The
bulk band gap of Bi$_{1-x}$Sb$_x$ is not very large (up to $\sim$30 meV
depending on $x$) \cite{Lenoir}, 
but it is large enough to detect the 2D transport properties at 4 K.

\begin{figure}[t]
\begin{center}
\includegraphics[width=5cm]{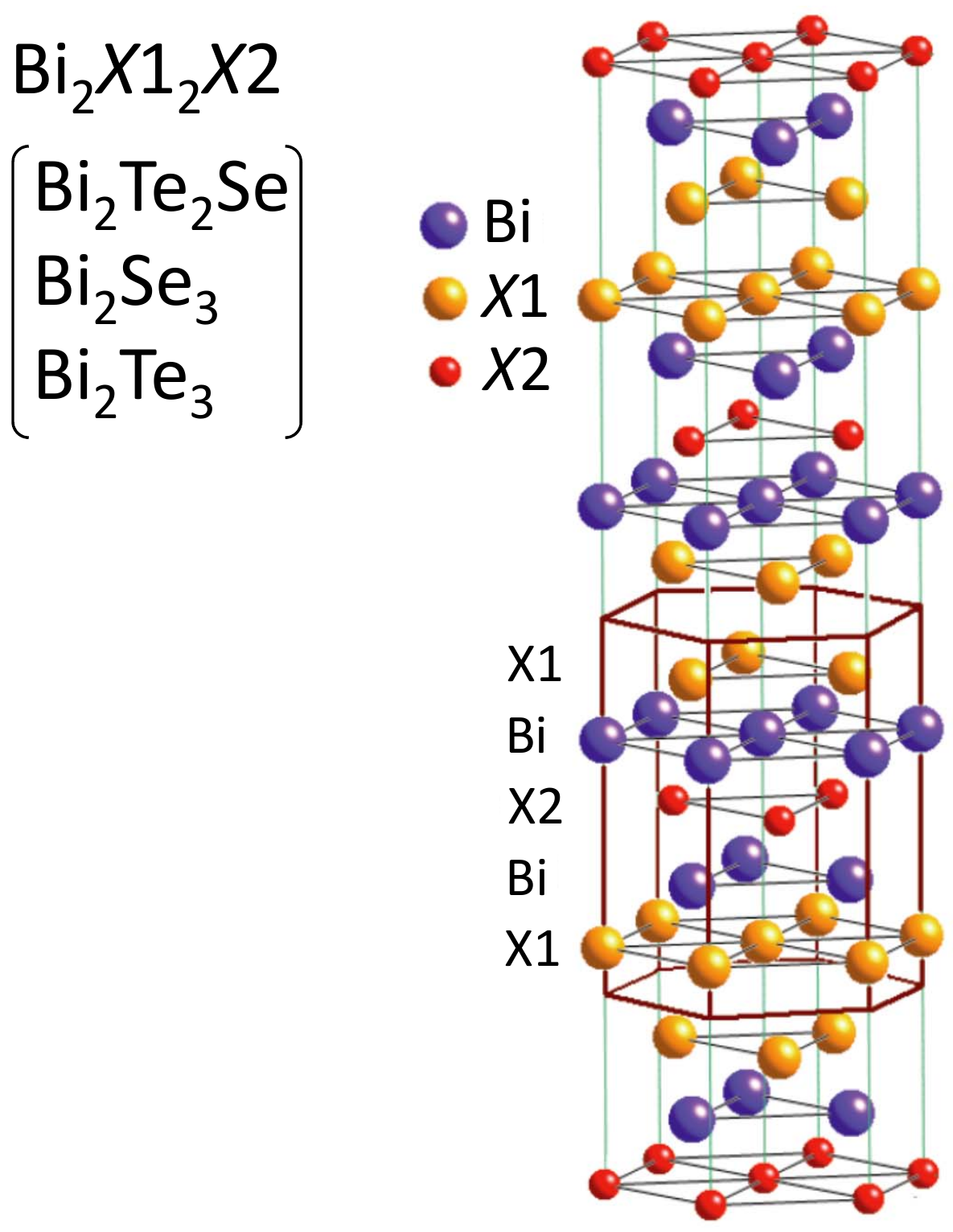}
\caption{(Color online) 
Crystal structure of tetradymite chalcogenides,
Bi$_2$Se$_3$ ($X1$ = $X2$ = Se), Bi$_2$Te$_3$ ($X1$ = $X2$ = Te), 
and Bi$_2$Te$_2$Se ($X1$ = Te; $X2$ = Se).
The quintuple layer (QL) enclosed by brown cage is the 
building block of this type of materials. The stack of the QLs
are in the -A-B-C-A-B-C- manner, and hence the unit cell
consists of a stack of 3 QLs.
The Bi$_{2-x}$Sb$_x$Te$_{3-y}$Se$_y$ ($y \ge 1$) compound also 
takes this structure, in which $X2$ = Se, $X1$ = Se/Te, and
the original Bi site becomes a mixture of Bi/Sb. 
Taken from Ref. \citen{RenBTS}; copyright
American Physical Society (2010).}
\label{fig:tetradymite}
\end{center}
\end{figure}

Along with Bi$_{1-x}$Sb$_x$, Fu and Kane predicted \cite{FuKane3D} 
that HgTe, $\alpha$-phase of Sn (called ``gray tin"), 
and Pb$_{1-x}$Sn$_x$Te would become 3D
TIs under uniaxial strain to break the cubic lattice symmetry. They also
suggested that Bi$_2$Te$_3$ would be a candidate, but they did not
perform band calculations to elucidate the parity eigenvalues. Such
calculations were done by Zhang {\it et al.} \cite{ZhangBi2Se3}, 
who came up with a
concrete prediction that Bi$_2$Se$_3$, Bi$_2$Te$_3$, and Sb$_2$Te$_3$
should be 3D TIs but Sb$_2$Se$_3$ is not; furthermore, Zhang {\it et
al.} proposed a low-energy effective model to describe the bulk band
structure of this class of materials. This model, with some corrections
made later \cite{LiuBi2Se3}, 
has become a popular model for theoretically discussing the
properties of a 3D TI. Experimentally, existence of a single Dirac-cone
surface state was reported in 2009 for Bi$_2$Se$_3$ by Xia {\it et al.}
\cite{Hasan_Bi2Se3} and for Bi$_2$Te$_3$ by Chen {\it et al.}
\cite{Chen_Bi2Te3} and also by Hsieh {\it et al.} \cite{Hasan_Bi2Te3};
Sb$_2$Te$_3$ was measured by Hsieh {\it et al.} \cite{Hasan_Bi2Te3}
along with Bi$_2$Te$_3$, but the existence of the topological
surface state was left unconfirmed due to the heavily $p$-type nature of
the measured samples. The topological nature of Sb$_2$Te$_3$ was
confirmed only recently in thin-film samples using STS \cite{Sb2Te3}.

Bi$_2$Se$_3$, Bi$_2$Te$_3$, and Sb$_2$Te$_3$ all crystallize in
tetradymite structure, which consists of covalently bonded quintuple
layers (e.g. Se-Bi-Se-Bi-Se) that are stacked in -A-B-C-A-B-C- manner
and are weakly interacting with van der Waals force 
(Fig. \ref{fig:tetradymite}); therefore, those
materials cleave easily between quintuple layers (QLs). Since each QL is
about 1 nm thick, the lattice constant along the $c$-axis is about 3 nm.

\begin{figure}[t]
\begin{center}
\includegraphics[width=8cm]{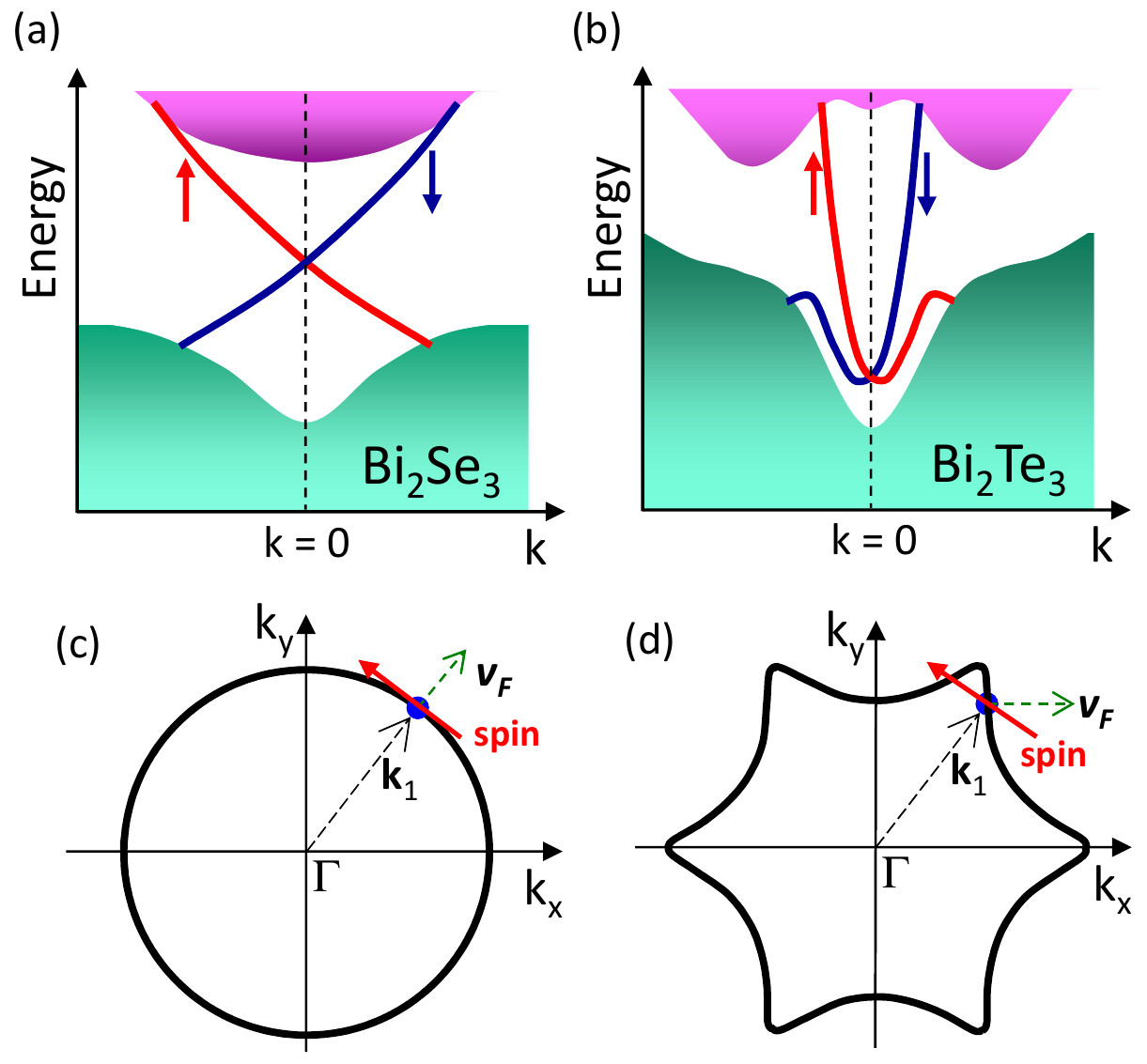}
\caption{(Color online) 
Schematic bulk and surface band structures of
(a) Bi$_2$Se$_3$ and (b) Bi$_2$Te$_3$.
Note that the surface states are spin non-degenerate
and are helically spin polarized. 
Representative constant-energy contours of the Dirac cones 
for (c) Bi$_2$Se$_3$ and (d) Bi$_2$Te$_3$ are also 
schematically shown. Note that the spin vector is always 
perpendicular to the wave vector $\mathbf{k}_1$ in both
Bi$_2$Se$_3$ and Bi$_2$Te$_3$, but the Fermi velocity
vector $\mathbf{v}_F$ can be non-orthogonal to the spin vector in 
Bi$_2$Te$_3$ due to the hexagonal warping, which leads to 
strong quasiparticle interference. }
\label{fig:BiSe-BiTe}
\end{center}
\end{figure}

The 3D $Z_2$ invariant of these tetradymite systems is (1;000), which
means that topological Dirac-cone surface state is centered at the
$\bar{\Gamma}$ point of the surface BZ 
(Fig. \ref{fig:BiSe-BiTe}). This simplicity
of the topological surface state and the absence of non-topological
surface states make those materials well suited for experimentally
addressing the properties of the topological surface state. Also,
single-crystal growth of those materials is relatively simple, which
made it easy for many experimentalists to start working on them.
Furthermore, the bulk band gap of Bi$_2$Se$_3$ is relatively large, 
0.3 eV, and thus one can see technological relevance that topological
properties of this material may potentially be exploited at room
temperature. All those factors helped initiate a surge of research
activities on 3D TIs.

The surface-state structure of Bi$_2$Se$_3$ is relatively simple and
presents an almost idealized Dirac cone with only slight curvature, as
shown schematically in Fig. \ref{fig:BiSe-BiTe}(a). In contrast, the
surface state of Bi$_2$Te$_3$ is a bit more complicated [see Fig.
\ref{fig:BiSe-BiTe}(b)] and the Dirac point is located beneath the top
of the valence band, which makes it difficult to probe the surface
transport properties near the Dirac point without being disturbed by
bulk carriers in Bi$_2$Te$_3$. 

Another difference between the two materials is that the constant-energy
contour of the Dirac cone is almost spherical in Bi$_2$Se$_3$
[Fig. \ref{fig:BiSe-BiTe}(c)], while it
presents significant hexagonal warping in Bi$_2$Te$_3$
[Fig. \ref{fig:BiSe-BiTe}(d)]. This warping is
caused by a $k^3$ term which stems from cubic Dresselhaus spin-orbit
coupling at the surface of rhombohedral crystal systems
\cite{Fu_warping}. Intriguingly, this hexagonal warping gives rise to
peculiar physics such as strong quasiparticle interference
\cite{XueQP,Alpichshev} and appearance of a finite 
out-of-plane spin polarization \cite{SoumaSz}.

While Bi$_2$Se$_3$ and Bi$_2$Te$_3$ are conceptually simple TI materials,
their chemistry is not so simple and they are always degenerately doped due to 
naturally occurring crystalline defects, which cause their transport properties 
to be dominated by bulk carriers. In this respect,
a promising TI material having the tetradymite structure is 
Bi$_2$Te$_2$Se, which has a chalcogen-ordered structure shown in 
Fig. \ref{fig:tetradymite}. 
The topological surface state of this material was first
reported by Xu {\it et al.} in a preprint that has not been published
\cite{XuBTS}, and they reported metallic $n$-type nature for the 
stoichiometric composition of this material.
However, Ren {\it et al.} discovered \cite{RenBTS} that by growing crystals 
from a slightly Se-rich starting
composition of Bi$_2$Te$_{1.95}$Se$_{1.05}$, one can obtain crystals
showing a large bulk resistivity exceeding 1 $\Omega$cm. 
Furthermore, they
demonstrated that in such crystals, the chemical potential is located
within the bulk band gap and one can observe clear SdH
oscillations coming from the topological surface state, which
contributed $\sim$6\% of the total conductance in a 260-$\mu$m-thick
bulk crystal \cite{RenBTS}. 
This value is to be contrasted with the preceding surface
transport study by Qu {\it et al.} on Bi$_2$Te$_3$, which found the
surface contribution of only $\sim$0.3\% in a 100-$\mu$m-thick 
sample \cite{QuBi2Te3}.
In fact, Bi$_2$Te$_2$Se was the first 3D TI material to present a
reasonably bulk-insulating behavior, which opened the door for detailed
transport studies of the topological surface state. This discovery by
Ren {\it et al.} was followed by an independent report by 
Xiong {\it et al.} \cite{XiongBTS}, who reported an even larger resistivity 
of 6 $\Omega$cm. Detailed defect chemistry in Bi$_2$Te$_2$Se was
discussed by Jia {\it et al.} \cite{JiaBTS}.
In passing, a tetradymite compound having a similar chalcogen-ordered
structure, Sb$_2$Te$_2$Se, has also been confirmed to be topological \cite{XuBTS},
but it is heavily $p$-type doped.

Ren {\it et al.} tried to further improve the bulk-insulating property
of Bi$_2$Te$_2$Se, and they reported two possible routes. One is to
expand the phase space of the compositions into
Bi$_{2-x}$Sb$_x$Te$_{3-y}$Se$_y$ \cite{RenBSTS,TaskinBSTS}, 
and they found a series of particular
combinations of $(x,y)$ where the samples show maximally bulk-insulating
behavior \cite{RenBSTS}. 
(This material is discussed in detail in Sec. 7.) The second route to
improve the insulating property of Bi$_2$Te$_2$Se is to employ Sn doping to the Bi
site \cite{RenSnBTS}.
Using both routes, it has been shown to be possible to achieve
surface-dominated transport in bulk single crystals 
\cite{TaskinBSTS, RenSnBTS}.

\begin{figure}[t]
\begin{center}
\includegraphics[width=6.5cm]{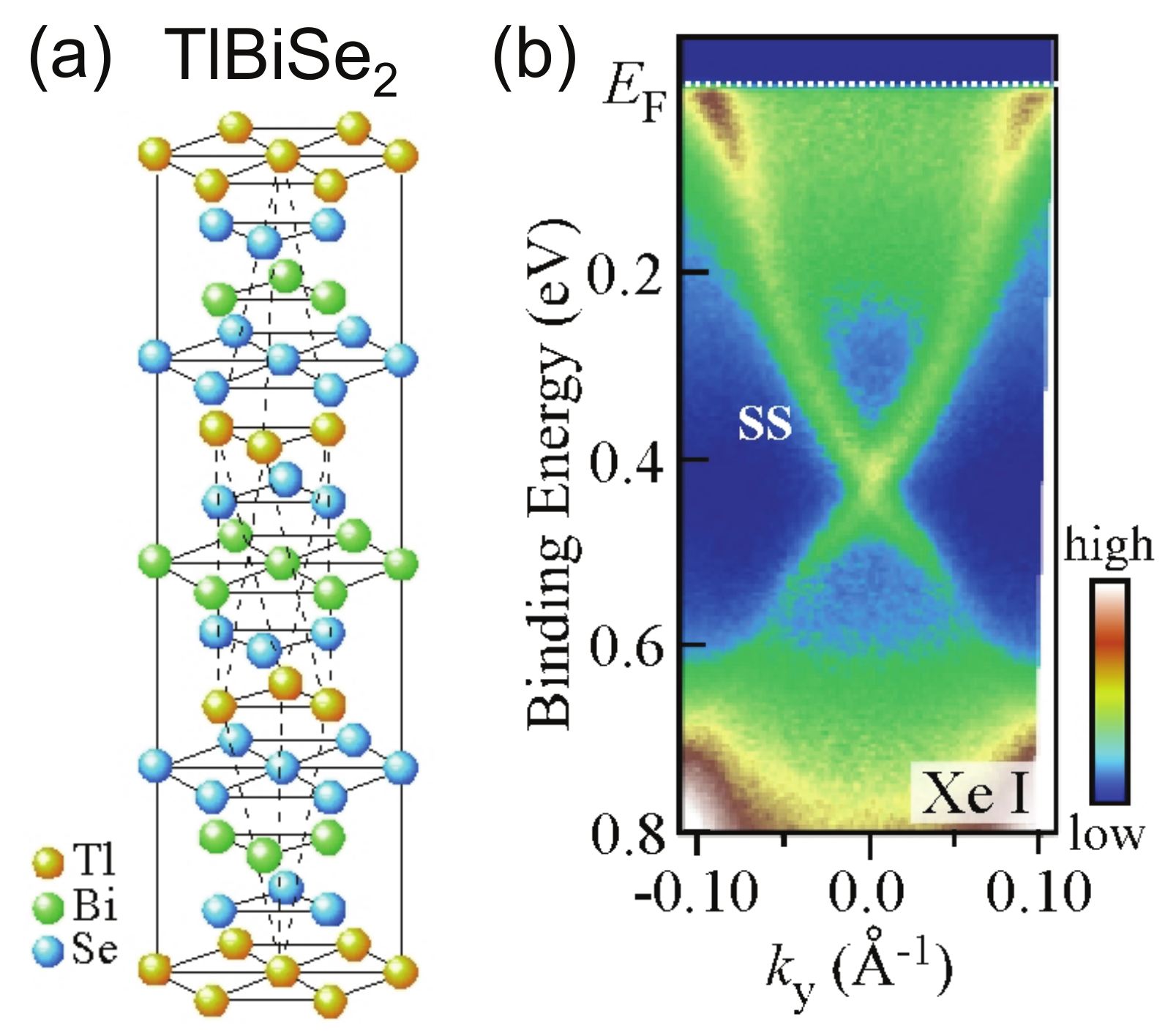}
\caption{(Color online) 
(a) Crystal structure of TlBiSe$_2$.
(b) Topological surface state observed in TlBiSe$_2$ by ARPES experiment
using Xe I line ($h\nu$ = 8.437 eV).
Taken from Ref. \citen{SatoTl}; copyright
American Physical Society (2010).}
\label{fig:TBS}
\end{center}
\end{figure}

The discovery of tetradymite TI materials demonstrated that elucidating
the parity eigenvalues based on {\it ab initio} band calculations is a
practical and powerful strategy for making predictions for new TI
materials. In fact, the next TI material discovered after Bi$_2$Se$_3$
and Bi$_2$Te$_3$ was TlBiSe$_2$,
and its discovery \cite{SatoTl, KurodaTl, ChenTl}, done in 2010, was
also led by theoretical predictions \cite{YanTl, LinTl}. 
By 2010, the competition in the TI
research was already very heated and two theoretical groups
independently made predictions that thallium-based III-V-VI$_2$ ternary
chalcogenides Tl$M'X_2$ [$M'$ = Bi and Sb; $X$ = S, Se, and Te; see Fig.
\ref{fig:TBS}(a)] should
be TIs; experimentally, the TI nature of TlBiSe$_2$ was reported first
by Sato {\it et al.} \cite{SatoTl}, which was soon followed by independent reports by
Kuroda {\it et al.} \cite{KurodaTl} and by Chen {\it et al.}
\cite{ChenTl}. The latter also confirmed TlBiTe$_2$ to be a
TI \cite{ChenTl}. The surface state structure of TlBiSe$_2$ is similar to that in
Bi$_2$Se$_3$ [see Fig. \ref{fig:TBS}(b)], and its simplicity makes it suitable for studying
fundamental properties of the Dirac cone without being bothered by
additional complications. Also, Sato {\it et al.} elucidated that the
bulk band gap of TlBiSe$_2$ is 0.35 eV \cite{SatoTl}, 
which is among the largest so far reported for 3D TI materials.

Intriguingly, TlBiS$_2$, which was initially predicted to be a TI \cite{YanTl, LinTl},
turned out to be non-topological. This indicates that {\it ab initio}
calculations are not always accurate in every details, and one should
interpret the results with due care \cite{Zunger}. The non-topological
nature of TlBiS$_2$ naturally points to a topological phase transition,
accompanied by a closing of the bulk band gap,
occurring in the solid solution TlBi(S$_{1-x}$Se$_x$)$_2$. Such a
transition was studied independently by Xu {\it et al.} \cite{XuTl} 
and by Sato {\it et al.} \cite{SatoNatPhys}
and was found to be located at $x$ = 0.5. Intriguingly, Sato {\it
et al.} found an unexpected gap opening at the Dirac point when the
composition is on the topological side and is close to the transition
\cite{SatoNatPhys}. This is surprising, because such a gap opening
points to a lifting of the Kramers degeneracy, but in this system
TRS is not explicitly broken.
While the origin of this ``Dirac gap" is not clear and its existence has
been a matter of debate, a recent follow-up experiment substantiated its
unique properties \cite{SoumaTl}.

After the discovery of TlBiSe$_2$, various ternary compounds have been
identified to be 3D TIs. Among them, GeBi$_2$Te$_4$ was suggested to be an
intrinsic insulator \cite{XuBTS, Neupane}. 
However, a follow-up work by Okamoto {\it et al.} showed
that GeBi$_2$Te$_4$ is naturally electron doped \cite{Okamoto}. 
The GeBi$_2$Te$_4$
crystals grown in the present author's laboratory are also highly
conducting with $n$-type carriers.

Another relatively new TI compound is Pb(Bi$_{1-x}$Sb$_x$)$_2$Te$_4$,
for which the observation of a topological surface state, as well as a
switching between $p$- and $n$-type surface carriers with $x$, has been
reported by Souma {\it et al.} \cite{SoumaPBT}.
This work was also motivated by
theoretical predictions \cite{XuBTS,PBT1,PBT2}. Kuroda {\it et al.}
independently reported the topological surface state for one of its end
members, PbBi$_2$Te$_4$ \cite{KurodaPBT}. Interestingly, Pb-based
materials form various homologous series of compounds, such as
$n$PbTe-$m$Bi$_2$Te$_3$. In fact, PbBi$_2$Te$_4$ can be viewed as a
member of this homologous series with $n$ = $m$ =1. Observation of a
topological surface state was also reported for its $n$ = 1, $m$ =2
member, PbBi$_4$Te$_7$ \cite{Eremeev}. In passing, Ge-based materials
form similar homologous series as Pb-based ones. GeBi$_2$Te$_4$
\cite{XuBTS, Neupane} and GeBi$_{4-x}$Sb$_x$Te$_7$ \cite{Muff_GBT} are
examples with $n$ = $m$ =1 and $n$ = 1, $m$ =2, respectively.

It is known that Bi bilayer and Bi$_2$Se$_3$ (or Bi$_2$Te$_3$) also form a
homologous series of compounds, (Bi$_2$)$_n$(Bi$_{2}X_3$)$_m$ ($X$ = Se, Te)
\cite{Valla2012}.
Topological surface state has been reported for a partially-S-substituted variant,
(Bi$_2$)(Bi$_2$Se$_{2.6}$S$_{0.4}$) (= Bi$_4$Se$_{2.6}$S$_{0.4}$) 
which is a semimetal \cite{Valla2012}.
In this type of homologous series, (Bi$_2$)(Bi$_2$Te$_3$)$_2$ (= BiTe) 
has also been synthesized and was found to be topological \cite{CavaReview}

\begin{figure}[t]
\begin{center}
\includegraphics[width=8cm]{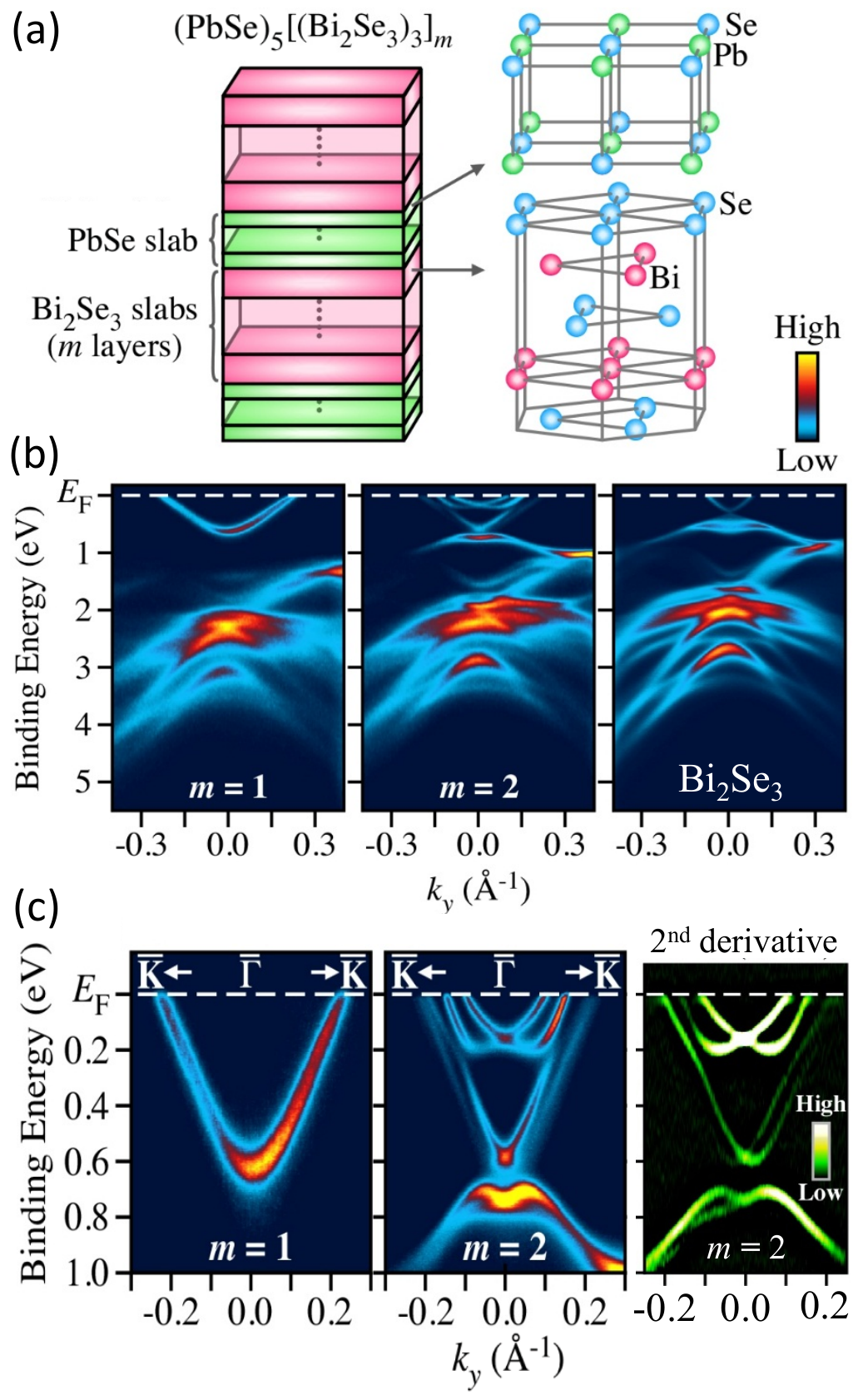}
\caption{(Color online) 
(a) Schematic picture of the crystal structure of 
(PbSe)$_5$(Bi$_2$Se$_3$)$_{3m}$, which forms a natural 
multilayer heterostructure consisting of topological insulator 
(Bi$_2$Se$_3$) and an ordinary insulator (PbSe) units.
(b) ARPES intensities of (PbSe)$_5$(Bi$_2$Se$_3$)$_{3m}$ 
for $m$ = 1 (left) and 2 (center), together with that of Bi$_2$Se$_3$
(right), plotted as a function of 
the binding energy and wave vector measured along the 
$k_y$ axis ($\bar{\Gamma}-\bar{K}$ cut) at $T$ = 30 K 
with $h\nu$ = 60 eV.
(c) Magnified view of the same data for $m$ = 1 (left) and 2 (center) 
at low energy; 
the right panel shows the second derivative of the $m$ = 2 data, 
in which the gap opening at the Dirac point is evident. 
Taken from Ref. \citen{Nakayama}; copyright
American Physical Society (2012).}
\label{fig:PBS}
\end{center}
\end{figure}

As mentioned in conjunction with Bi$_{1-x}$Sb$_x$, one of the initial
predictions of Fu and Kane was that strained HgTe would be a 3D TI with
the 3D $Z_2$ invariant (1;000) \cite{FuKane3D}. 
This was experimentally verified in 2011
using MBE-grown films by Br\"une {\it et al.} \cite{HgTe3D} 
who reported that a 70-nm-thick HgTe film grown on a CdTe substrate 
is uniformly strained (0.3\%)
and obtains a band gap of $\sim$22 meV; a strange pattern of quantum
Hall effect was observed with Hall plateaus appearing at filling factors
$\nu$ = 2, 3, 4, 5, 7, 9. This was interpreted \cite{HgTe3D} 
to be a result of a
combination of two quantum Hall effects occurring in parallel at the top and
bottom surfaces that contain slightly different carrier densities. So
far, strained HgTe is unique in that the mobility of the surface
carriers can be high enough to present quantum Hall effect. 

An interesting material which was experimentally found to be topological
{\it without} any theoretical prediction is the
(PbSe)$_5$(Bi$_2$Se$_3$)$_{3m}$ compound ($m$ = 1, 2) \cite{Nakayama}.
This material forms a natural multilayer heterostructure consisting of
topological insulator (Bi$_2$Se$_3$) and an ordinary insulator (PbSe)
units (Fig. \ref{fig:PBS}). The precise atomic positions within the unit cell has not
been determined yet for this material, and hence no {\it ab initio} band
calculation has been performed. Nakayama {\it et al.} found that in
this material the PbSe unit provides a barrier for the electronic states
in the Bi$_2$Se$_3$ unit to be quantum confined, which, for $m$ = 2,
leads to hybridization of the topological interface states at the top
and bottom of each Bi$_2$Se$_3$ unit, resulting in a gapped Dirac-cone
structure. Also, the bulk band gap of this $m$ = 2 compound was found to be 
as large as 0.5 eV, which is probably due to the quantum confinement
effect on the bulk band.
For $m$ = 1, the quantum confinement apparently lifts the
non-trivial $Z_2$ topology from the Bi$_2$Se$_3$ unit and gives rise to
a degenerate parabolic band (Fig. \ref{fig:PBS}).

\subsection{Candidate 3D TIs}

There are other candidate materials of 3D TIs for which the topological
surface states have not been observed. There have been many predictions
for 3D TIs based on {\it ab initio} band calculations \cite{YanZhang}, 
but in the following I will discuss only those materials that have been
experimentally addressed.

Ag$_2$Te has been known to be a material presenting large, linear-in-$B$
magnetoresistance over an exceedingly wide magnetic-field range
\cite{Ag2Te_MR}, and it has been proposed that this unusual property
may be associated with its 3D-TI nature \cite{Ag2Te_theory}. 
ARPES measurements of this material have not been reported, but 
Aharonov-Bohm (AB) oscillations have been observed in 
nanowires of Ag$_2$Te \cite{Ag2Te_AB1, Ag2Te_AB2}.
Such an observation suggests the existence of some surface conduction layer,
but its topological nature is still left to be confirmed, because similar
AB oscillations are known to originate from trivial surface states in InN
nanowires \cite{InN}. 

A very interesting candidate is SmB$_6$, which may be a {\it topological
Kondo insulator} \cite{Dzero, Takimoto, FengLu}, 
because the band bap of this material stems from Kondo
effect. If this material is indeed a TI, it will be the first material
where electron correlations and nontrivial band topology both play
important roles. So far, strong evidence for surface-dominated transport
at low temperature has been reported \cite{SmB6-1, SmB6-2, SmB6-3}, 
but existing ARPES data do not
resolve topological surface state within the small gap ($\sim$20
meV) \cite{SmB6ARPES}. 
A puzzling feature in the reported surface-transport data
\cite{SmB6-3} is that the observed surface carrier density of
1.1 $\times$ 10$^{15}$ cm$^{-2}$ is about 1000 times larger than that found
in other TI materials and is obviously too large for the expected Dirac
carrier density \cite{FengLu}. One should keep in mind that the
surface-dominated transport can be observed in insulators in ambient
condition due to the formation of a trivial accumulation (or inversion) layer,
as reported for pure Te by von Klitzing and Landwehr \cite{KlitzingTe}, 
and hence such an
observation alone does not give evidence for a TI phase.

Another interesting candidate is Bi$_{14}$Rh$_3$I$_9$, which may be a weak
3D TI with the $Z_2$ index (0;001) \cite{BiRhI}. The crystal structure of this
material can be viewed as a stack of intermetallic
[(RhBi$_4$)$_3$I]$^{2+}$ layers and intervening [Bi$_2$I$_8$]$^{2-}$
zigzag-chain layers; the former consist of graphene-like honeycomb
network of edge-sharing RhBi$_8$ cubes, giving rise to an electronic
structure which is similar to that in graphene but is enriched by strong
SOC due to Bi. The {\it ab initio} band calculations predict that this
is a quasi-2D system which essentially consists of a stacked 2D TI
layers; as a result, the system is predicted to be a weak 3D TI with the
topological surface state expected only on the sides of the stacks. The
bulk band structure seen by ARPES is in reasonable agreement with the calculations,
but no surface state was detected because the ARPES was done on the
cleaved top surface where the surface state is not expected. Obviously,
some transport measurements to detect the topological surface state on
the side surface is necessary to elucidate the TI nature of this
material.

\subsection{Topological semimetal}

The term ``topological semimetal" can be used for three different classes
of materials. The first is an ordinary semimetal possessing separate
electron and hole pockets compensating each other, but the parity
eigenvalues of the valence band gives rise to a nontrivial $Z_2$ topology;
an example is pure Sb, whose topology is characterized by the 3D $Z_2$
invariant (1;111) \cite{FuKane3D,Hsieh2009}. 
The second is a zero-gap semiconductor which possesses
band inversion, such as unstrained HgTe; this class of materials can be
made genuine TIs by lowering the crystal symmetry to open a gap. The
third is called {\it Weyl semimetal} \cite{Savrasov}, which is 
probably the most fundamentally
interesting. In the following, I will discuss the latter two classes.

Interesting examples of the candidate topological zero-gap semimetals 
are Heusler or half-Heusler compounds \cite{YanZhang}. 
They are ternary intermetallic compounds with
$X_2YZ$ (Heusler) or $XYZ$ (half-Heusler) compositions. Their crystal
structures consist of basic $YZ$ sublattice (which takes the zinc-blende
structure similar to HgTe) stuffed with X in the void space. Many
Heusler or half-Heusler compounds were predicted to have band inversion
and be topological based on {\it ab initio} band calculations
\cite{Heusler1, Heusler2, Heusler3, Heusler4, Heusler5}; under
uniaxial pressure which lowers the crystal symmetry, they obtain a gap
and are expected to become TIs. The ARPES experiments done on $R$PtBi
($R$ = Lu, Dy, Gd) found metallic surface states distinct from the bulk
bands, but the surface-band structure does not appear to support a
topological origin \cite{Kaminski}. 
Interestingly, superconductivity has been found in
LaPtBi \cite{LaPtBi} and YPtBi \cite{YPtBi}; 
if those materials are indeed topological, the fate of
the surface states in the superconducting state is an intriguing issue.

The idea of Weyl semimetal was theoretically proposed by Wan {\it et al.} 
in 2011 \cite{Savrasov}. 
Weyl fermions have definite chirality, which protects them from
gapping, and they are described by a massless two-component Dirac
equation. In solids, when TRS is broken in an inversion
symmetric system and non-degenerate valence and conduction bands touch at
an accidental degeneracy point in a 3D BZ, the low-energy physics is
approximated by massless two-component Dirac equation and electrons
obey the Weyl Hamiltonian $H = \pm
\hbar v_F \boldsymbol{\sigma}\cdot\mathbf{k}
= \pm \hbar v_F (k_x\sigma_x + k_y\sigma_y + k_z\sigma_z)$. 
By symmetry, such a degeneracy point
(Weyl point) must come in pairs, and Wan {\it et al.} showed that a pair
of Weyl points give rise to an arc of zero-energy excitation (Fermi arc)
to connect them in the projected surface BZ \cite{Savrasov}. 
The appearance of this
gapless state on the surface of Weyl semimetal can be viewed as a result
of the bulk-boundary correspondence to signify the non-trivial topology
\cite{BurkovWeyl, TurnerWeyl}. It was proposed based on {\it ab initio} band
calculations that pyrochlore iridates such as Y$_2$Ir$_2$O$_7$ in the
antiferromagnetic phase may realize such a Weyl semimetal 
\cite{Savrasov, Y2Ir2O7theory}. In this
context, a recent study of Nd$_2$(Ir$_{1-x}$Rh$_x$)$_2$O$_7$ found a
transition from a correlated metal to a Mott insulator with decreasing
$x$ that is accompanied by a gradual reduction of the spectral weight,
and it is possible that the Weyl semimetal phase exists at the critical
point \cite{Nd2Ir2O7}.

The Weyl semimetal phase is also conceivable in a TR symmetric system
with broken inversion symmetry, and it was proposed that such a phase
may be achieved by breaking the inversion symmetry through
elaborate multilayer structures using HgTe/CdTe \cite{Halasz} or 
the TlBi(S$_{1-x}$Se$_x$)$_2$ system at the topological phase transition point 
\cite{BansilWeyl}.

\subsection{Topological crystalline insulator}

We have already seen that band insulators can be topologically
classified by evaluating the $Z_2$ invariant from valence-band Bloch
wave functions. This classification is based on TRS of the
system. It turned out that this is not the only possible topological
classification of band insulators, but it is also possible to classify
band insulators based on topologies protected by 
point-group symmetries of the crystal lattice \cite{FuTCI}. 
Those insulators that have nontrivial topology protected by 
point-group symmetries are called {\it topological
crystalline insulators} (TCIs) \cite{FuTCI}. 
So far, concrete topological invariants
are elucidated for systems possessing four-fold ($C_4$) or six-fold
($C_6$) rotation symmetry \cite{FuTCI} and also for systems with 
mirror symmetry \cite{HsiehFu}. In
particular, the latter case gained significant attention after the
prediction of a concrete example, SnTe, was made by Hsieh {\it et al.}
\cite{HsiehFu}. 

\begin{figure}[b]
\begin{center}
\includegraphics[width=8.2cm]{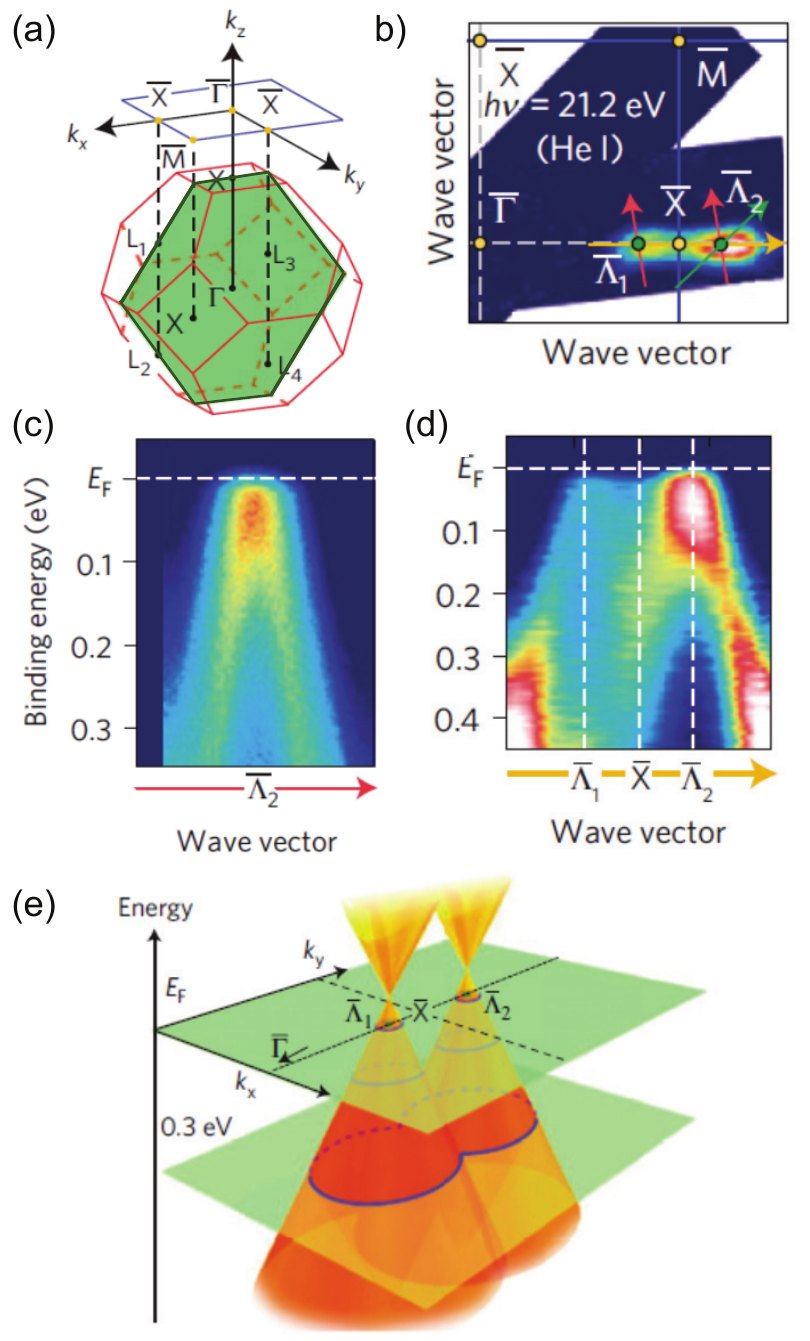}
\caption{(Color online) ARPES data on cleaved (001) surface of SnTe.
(a) The bulk BZ and the corresponding (001) surface BZ of SnTe;
$\Gamma$, $L$, $X$ are the symmetry points in the bulk 3D BZ, 
whereas $\bar{\Gamma}$, $\bar{X}$, $\bar{M}$ are in the surface 2D BZ.
The (110) mirror plane is indicated by the shaded area.
Note that two $L$ points [e.g. $L_1$ and $L_2$ in panel (a)] 
are projected to the same $\bar{X}$ point.
(b) ARPES intensity mapping in the surface BZ at $E_F$ 
measured with $h\nu$ = 21.2 eV at 30 K.
(c) Near-$E_F$ ARPES intensity as a function of the wave vector 
and the binding energy measured along the cut which is nearly
parallel to the $\bar{X}-\bar{M}$ direction and is crossing the 
$\bar{\Lambda}_2$ point [one of the red arrows in panel (b)].
(d) Near-$E_F$ ARPES intensity along the $\bar{\Gamma}-\bar{X}$ cut 
[yellow arrow in panel (b)].
(e) Schematic 2D band dispersions near $E_F$ concluded from the 
data in panels (c) and (d) depicting the 
characteristic double Dirac-cone structure.
Taken from Ref. \citen{TanakaTCI}.}
\label{fig:TCI}
\end{center}
\end{figure}

In TCIs with mirror symmetry, the topology is specified by the
topological invariant $n_{\mathcal{M}}$ called {\it mirror Chern
number}, which evaluates the Chern number in only one of the two Hilbert
subspaces divided according to the mirror eigenvalues \cite{Teo}. 
Based on {\it ab
initio} calculations, Hsieh {\it et al.} showed that SnTe has band
inversions at four TRIMs (four equivalent $L$ points) 
in the 3D BZ \cite{HsiehFu}, which gives rise
to a nontrivial mirror Chern number $n_{\mathcal{M}} = -2$, while its
$Z_2$ invariant is trivial, (0;000). They further predicted that the
surface state on the \{001\} surface should present a peculiar
double-Dirac-cone structure near the $\bar{X}$ points, because two
$L$ points are projected onto the same $\bar{X}$ point
[Fig. \ref{fig:TCI}(a)] \cite{HsiehFu}. 

It is interesting that the experimental verification of this TCI
prediction showcases how severe the competitions in the TI research are. The
prediction \cite{HsiehFu} 
was posted to the preprint server in February 2012. In June,
three preprints reporting experimental discovery of a TCI were posted to
the preprint server in about two weeks, and two of them were published
on the same day in September 2012 in different journals (Nature Physics 
\cite{TanakaTCI}
and Nature Materials \cite{Dziawa}), while the other one \cite{XuTCI1}
reported by a group at
Princeton University was not published; the Princeton group performed
new measurements, and a paper containing new data were submitted in
August and published in November \cite{XuTCI2}. 

Among the first two published papers, the one by Tanaka {\it et al.}
\cite{TanakaTCI}
reported straightforward confirmation of the prediction in SnTe, and the
observed surface state with the double Dirac-cone structure
[Fig. \ref{fig:TCI}] was in good qualitative agreement with
the theory; they also showed that the cousin material PbTe does not
present any surface state. The other one by Dziawa {\it et al.} 
\cite{Dziawa} reported
the TCI phase in Pb$_{0.77}$Sn$_{0.23}$Se, which shows a transition to a
trivial phase upon increasing the temperature. The work by Xu {\it
et al.} published later \cite{XuTCI2} 
reported a TCI phase in Pb$_{0.6}$Sn$_{0.4}$Te
and a trivial phase in Pb$_{0.8}$Sn$_{0.2}$Te, together with
spin-resolved ARPES data showing helical polarization on each of the
double Dirac cones. A more recent paper by Tanaka {\it et al.} 
\cite{TanakaTCI2} nailed down that the topological phase 
transition in Pb$_{1-x}$Sn$_x$Te occurs
at $x_c \simeq$ 0.25; furthermore, they found that the separation
between the two Dirac cones near the $\bar{X}$ points
systematically narrows when $x$ is reduced toward $x_c$, but they
never merge before the transition eliminates them.

In passing, the mirror Chern number $n_{\mathcal{M}}$ can also 
be used for TR-invariant 3D TIs to further classify them \cite{Teo}. For example,
Bi$_{1-x}$Sb$_x$ is a TI with $Z_2$ invariant (1;111), and it can have
$n_{\mathcal{M}} = \pm 1$. The sign of $n_{\mathcal{M}}$ is called {\it
mirror chirality}, which is related to the sign of the $g$ factor. The
first experimental work that addressed this additional topological
property in a TI was the spin-resolved ARPES done by Nishide {\it et
al.} \cite{Nishide}, who elucidated that the mirror chirality is $-1$ in
Bi$_{1-x}$Sb$_x$.

The discovery of TCIs significantly widened the scope of topological
materials. Already, detailed topological classification schemes for all
point-group symmetries have been proposed \cite{Slager}, 
and also the mirror topology
has been expanded to superconductors
\cite{TCSC1, TCSC2, TCSC3}. Experimentally, elucidating the
interplay between $Z_2$ topology and mirror topology in materials like
SnTe under uniaxial strain would be an interesting issue.

\section{How to Confirm TI Materials}

In this section, I briefly summarize the possible experimental 
procedures to confirm whether a material
is a TI or not. In the case of 2D TIs, one needs to probe the existence
of helical 1D edge state, which is possible only through quantum
transport experiments using nano-fabricated device structures. The
existence of the edge state can be seen through conductance quantization
in the insulating regime \cite{Konig2007}. 
Also, the helical spin polarization of the
edge state may be detected by transport experiments using spin Hall
effect \cite{Brune2012}.

For 3D TIs, the simplest and the most convincing is to observe the Dirac
cone by ARPES experiments. To firm up the identification of a TI, one
should employ spin-resolved ARPES to confirm that the Dirac cone is
non-degenerate and is helically spin polarized \cite{Hsieh2009,Nishide}.

Unfortunately, not all materials are suitable for ARPES, which requires
clean and flat surface that is usually obtained by cleaving single
crystals. When single crystals are not available or the material does
not cleave well, APRES becomes difficult. In such a case, one may rely on
transport experiments. Ideally, if the bulk is sufficiently insulating
and the surface carriers have high enough mobility, one would be able to
confirm that the transport is occurring through the surface and that the
surface carriers are Dirac fermions. The former can be done by looking
at the sample-size dependence of the conductance \cite{TaskinBSTS, 
SmB6-1, SmB6-2, SmB6-3}, 
and the latter may be
accomplished by elucidating the $\pi$ Berry phase in the quantum 
oscillations from the surface state \cite{RenBTS, QuBi2Te3, TaskinBSTS, 
AnalytisNP, Sacepe, XiongBerry, TaskinMBE}. 
(Detailed discussions on the identification of the $\pi$ Berry phase
will be given in Sec. 8.3.)
It should be emphasized that confirming the Dirac-fermion
nature of the surface carriers is important, because trivial
accumulation layer or inversion layer that may form on the surface of an
insulator may also give rise to surface-dominated transport \cite{KlitzingTe}. 

The Dirac-fermion nature may also be confirmed by STS experiments in
magnetic fields, because massless Dirac fermions present peculiar
Landau quantization in which the level spacing changes as $\sqrt{N}$ 
and the zero-energy Landau level is pinned to the Dirac point (see detailed
discussions in Sec. 8.2); 
by looking at the bias-voltage dependence of
the Landau quantization peaks, one can identify Dirac fermions
\cite{Sb2Te3,XueSTS, Hanaguri}.
Similarly, magneto-optics experiments to detect the Landau level
transitions can be used for detecting the peculiar quantization scheme
to identify Dirac fermions on the surface \cite{Basov}.

One should note that in reality, it is often very difficult to obtain
sufficiently bulk-insulating samples of a candidate material. In that
case, transport measurements are not very useful. If it is possible to detect
the equilibrium spin current (which is carried only by the helical Dirac
fermions and hence is not bothered by bulk carriers) by some
electromagnetic means \cite{SunSpin}, 
it would become a very useful tool for
identifying a TI. However, feasibility of such an experiment is not 
clear at the moment.

\section{Syntheses of TI Materials}

\subsection{Bulk single crystals}

Except for Bi$_{1-x}$Sb$_x$, all the confirmed 3D TI materials are
chalcogenides (i.e. compounds containing chalcogen atoms S, Se, and Te).
Since chalcogen atoms are volatile, the syntheses of chalcogenides are
done in sealed evacuated quartz-glass tubes, which can sustain
temperatures up to 1000$^{\circ}$C. Such a necessity of containment
limits the range of applicable growth techniques, and one usually uses
the Bridgman method. In this method, the temperature of the melt is gradually
reduced while keeping a temperature gradient in the tube, so that the
crystallization starts at the cold end and the crystal grows as the
solidification proceeds from this end. Most of the popular TI materials
including Bi$_2$Se$_3$, Bi$_2$Te$_3$, and Bi$_2$Te$_2$Se are 
grown by the Bridgman method. 

Another possible crystal growth technique is the vapor transport. In
this technique, one puts a chunk of polycrystalline material on one end
of a sealed quartz-glass tube. The tube is kept for a long time in a
furnace with a certain temperature gradient, in which the
polycrystalline material is on the hotter end. The temperature gradient
is chosen so that the material sublimates at the hotter end and crystalizes
at somewhere in a colder part. When some reagent such as I$_2$ is
used as a transporter, the technique is called chemical vapor transport;
if the material easily sublimates, one does not need a transporter and
the technique is called physical vapor transport. Single crystals of TCI
materials, SnTe, (Pb,Sn)Se, and (Pb,Sn)Te are usually grown with a
vapor transport technique.

Since the tetradymite compounds cleave easily, they can be made into a
very thin (down to only a few nm thick) flakes by employing an
exfoliation technique similar to that used for making graphene samples 
using Scotch tapes. Such thin flakes are particularly useful for experiments
involving gating to electrostatically control the surface chemical
potential \cite{Steinberg2010,Checkelsky2011}.

\subsection{Thin films}

For the growth of high-quality epitaxial thin films of TI materials,
molecular beam epitaxy (MBE) technique is usually employed
\cite{XueReview}. So far, reports of MBE growths have been made for
Bi$_{1-x}$Sb$_x$ \cite{Hirahara2010} or relatively simple tetradymite
compounds Bi$_2$Se$_3$ \cite{TaskinMBE, XueSTS, Bi2Se3Si1, Bi2Se3Si2,
Bi2Se3Si3, K.L.WangSdH, XueNP, STO-WAL, Bi2Se3GaAs, Bi2Se3Sapphire,
TaskinAM, Bi2Se3CdS, Bi2Se3InP}, Bi$_2$Te$_3$ \cite{XueBi2Te3}, and
Sb$_2$Te$_3$ \cite{Sb2Te3}, as well as their solid-solutions such as
(Bi,Sb)$_2$Te$_3$ \cite{XueBST}. Those materials can be grown by
co-evaporating the constituent elements with suitable flux ratios.
Chemical vapor deposition (CVD) technique has also been applied to the
growth of Bi$_2$Se$_3$ \cite{Bi2Se3CVD}, but the reported film quality
has not been as good as the best MBE-grown films. 

For epitaxial growths of thin films, the lattice matching between the
substrate and the grown material is usually very important. However, in
the case of tetradymite TI materials, thanks to the existence of the van
der Waals gap between the QLs, the lattice matching with the substrate
is not crucial and the epitaxial growth proceeds in the so-called van
der Waals epitaxy mode \cite{Koma}, in which the substrate and the films
are only weakly bonded with the van der Waals force and hence the
lattice matching condition is greatly relaxed. In particular, epitaxial
growths of Bi$_2$Se$_3$ have been reported for various substrates
including Si(111) \cite{Bi2Se3Si1, Bi2Se3Si2, Bi2Se3Si3, K.L.WangSdH},
graphene-terminated 6H-SiC(0001) \cite{XueSTS, XueNP}, SrTiO$_3$(111)
\cite{STO-WAL}, GaAs(111) \cite{Bi2Se3GaAs}, sapphire(0001)
\cite{TaskinMBE, Bi2Se3Sapphire, TaskinAM}, CdS(0001) \cite{Bi2Se3CdS},
and InP(111) \cite{Bi2Se3InP}.

It turned out that the control of the substrate temperature is the most
crucial ingredient for obtaining high-quality films of Bi$_2$Se$_3$
with a large area of atomically flat terraces \cite{TaskinAM}. In particular, a two-step
method in which the depositions of the first layer and subsequent layers
are done at different temperatures, to promote initial adhesion and
crystallization separately, has been reported to yield best quality
films \cite{TaskinMBE, Bi2Se3Si2, Bi2Se3Si3, Bi2Se3Sapphire, TaskinAM}.

For the TCI material SnTe, a technique called hot-wall epitaxy has been
used in the past \cite{SnTeFilm}, 
yielding good quality samples with a reasonably high mobility 
($\sim$2700 cm$^2$/Vs).
For SnTe which is a cubic material with rock-salt structure, good
lattice matching is crucial for epitaxial growth. BaF$_2$ has been
traditionally used as a substrate \cite{SnTeFilm}, but recently, 
using Bi$_2$Te$_3$ as a
buffer layer between SnTe and sapphire was reported to yield high
quality films that present surface SdH oscillations \cite{TaskinSnTe}.

\subsection{Nanoribbons and nanoplates}

Mesoscopic transport experiments of Bi$_2$Se$_3$ \cite{Peng2010}, 
Bi$_2$Te$_3$ \cite{Kong2010, K.L.WangNatNano}, and 
Bi$_2$Te$_2$Se \cite{Gehring} have
been performed using nanoribbons and nanoplates. 
Bi$_2$Se$_3$ nanoribbons are usually
synthesized by gold-catalyzed vapor liquid solid (VLS) technique \cite{Peng2010}.
Typically, Bi$_2$Se$_3$ powder is placed in the
center of a tube furnace through which Ar gas flows and transports
evaporated Bi$_2$Se$_3$; a silicon substrate decorated with Au
nanoparticles are placed downstream in the tube furnace, and the growth
of nanoribbons proceeds as Au nanoparticles absorb Bi$_2$Se$_3$ vapor and
leave crystallized Bi$_2$Se$_3$ nanoribbons beneath them. Naturally, the
diameter of grown nanoribbons is roughly determined by the size of the Au
nanoparticles (typically $\sim$20-nm diameter). When there are no Au
nanoparticles to work as catalyst, nanoplates (typically a few nm thick
and a few $\mu$m wide) are obtained instead of nanoribbons
\cite{Kong2010, Gehring}.

\section{Defect Chemistry and Engineering}

An important theme in the research of TI materials is to reduce
unintentionally-doped bulk carriers that hinder observations of surface
transport properties. Only in the case of HgTe thin films, which can be
grown in very high purity using an MBE technique, unwanted bulk carriers
are not an issue. In most other materials, one needs to find suitable
ways to reduce bulk carriers. In this section, I will use the
Kr\"oger-Vink notation to describe defects in crystals; for example,
$V^{\bullet\bullet}_{\rm Se}$ means a selenium vacancy with double
positive charge, and Bi$^{'}_{\rm Te}$ means a bismuth ion sitting on
the tellurium lattice site with a single negative charge.

Naturally-grown Bi$_2$Se$_3$ crystals are always electron doped, with
the typical bulk carrier density $n_{\rm 3D}$ of 10$^{19}$ cm$^{-3}$
\cite{Butch, AnalytisPRB, Eto},
because of the thermodynamically inevitable Se vacancies
($V^{\bullet\bullet}_{\rm Se}$) or Se anti-site defects
(Se$^{\bullet}_{\rm Bi}$) that have low formation energies
\cite{Scanlon, Johnson}. Doping Ca$^{2+}$ to the Bi$^{3+}$ site has been
reported to be useful for reducing $n$-type carriers and eventually
achieving $n$-to-$p$ type transition \cite{CheckelskyCa}. 
However, Ca doping obviously
introduces strong scattering centers and the electron mobility becomes
low in Ca-doped Bi$_2$Se$_3$ \cite{CheckelskyCa}. 
Optimization of the growth condition
\cite{Butch} and isovalent Sb doping to the Bi site
\cite{AnalytisNP, AnalytisPRB} are both reported to be useful for 
reducing $n_{\rm
3D}$ down to 10$^{16}$ cm$^{-3}$ while keeping a high electron mobility
that allowed observation of surface SdH oscillations
\cite{AnalytisNP}, although $n$-to-$p$ type transition was not
achieved with these routes. Recently, Ren {\it et al.} \cite{RenCd} succeeded in
growing high-mobility $p$-type crystals of Bi$_2$Se$_3$ by combining Cd
doping and a Se-rich growth condition; furthermore, subsequent annealing
of such $p$-type crystals to compensate for the $p$-type carriers by the
electrons coming from Se vacancies made it possible to obtain both $n$-
and $p$-type samples with low bulk carrier density, in which surface SdH
oscillations were observable \cite{RenCd}.

\begin{figure}[t]
\begin{center}
\includegraphics[width=7cm]{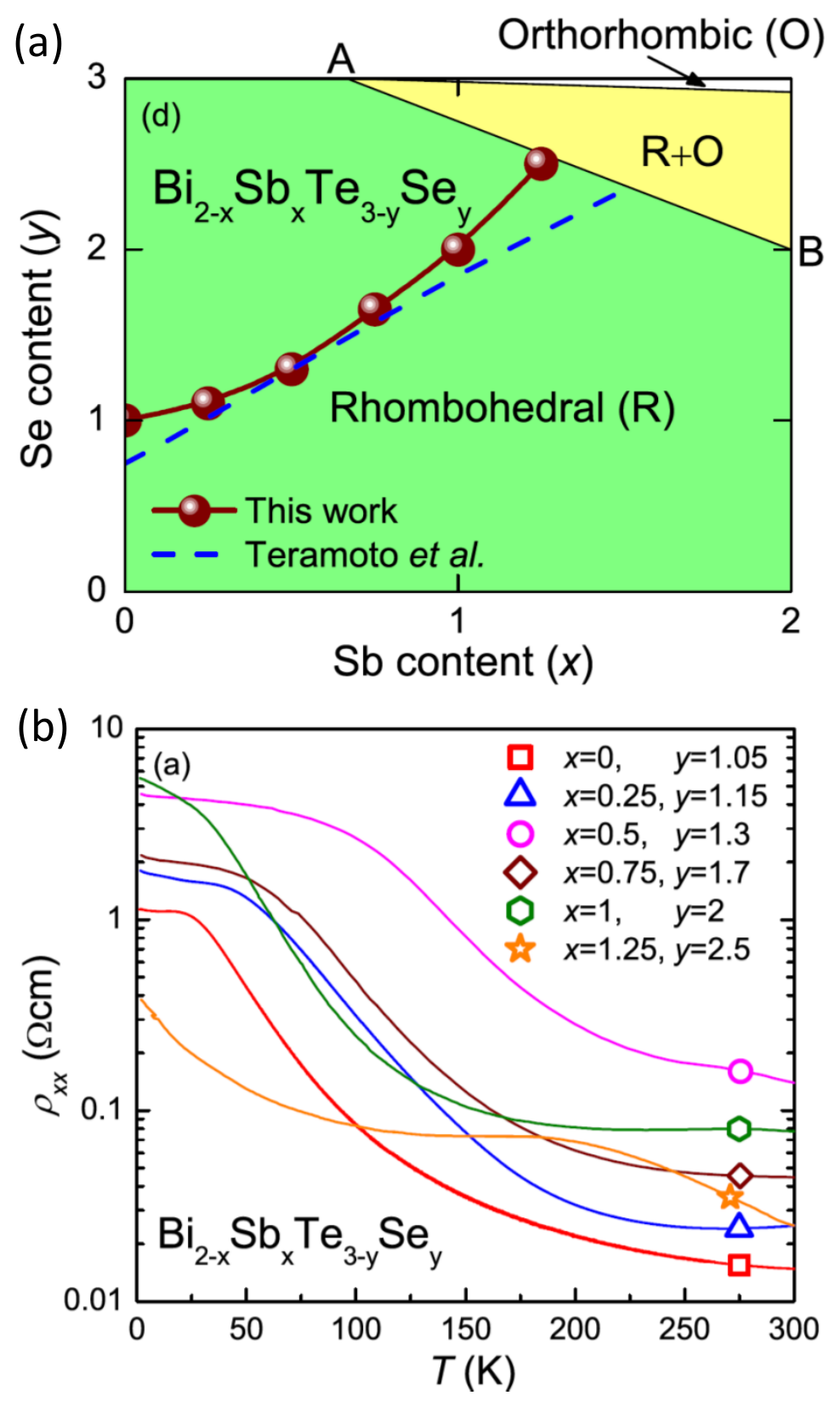}
\caption{(Color online) 
(a) Composition-structural phase diagram of the
Bi$_{2-x}$Sb$_{x}$Te$_{3-y}$Se$_{y}$ system. 
The circles denote the
compositions where the compensation between $n$- and $p$-type carriers 
are maximally achieved. The
dashed line denotes the insulating composition previously suggested 
by Teramoto and Takayanagi [J. Phys. Chem. Solids {\bf 19} (1961) 124].
(b) Temperature dependencies of $\rho_{xx}$ for the series of BSTS
samples at optimized compositions. Note that the vertical axis is in a
logarithmic scale. Taken from Ref. \citen{RenBSTS}; copyright
American Physical Society (2011).}
\label{fig:BSTS}
\end{center}
\end{figure}

\begin{figure}[t]
\begin{center}
\includegraphics[width=8.1cm]{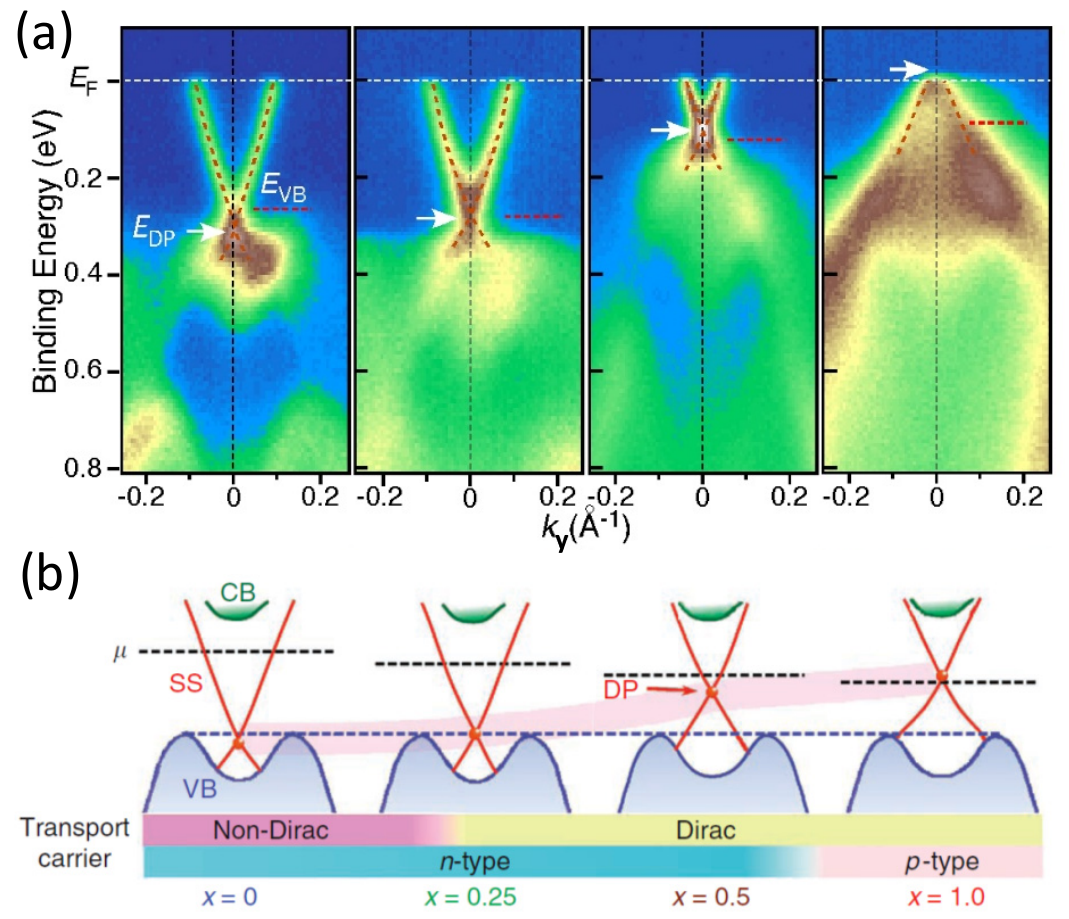}
\caption{(Color online) 
(a) ARPES intensities of Bi$_{2-x}$Sb$_{x}$Te$_{3-y}$Se$_{y}$
for, from left to right, $(x, y)$ = (0, 1), (0.25, 1.15), (0.5, 1.3), and (1, 2), 
measured with $h\nu$ = 58 eV along the $k_y$ axis 
($\bar{\Gamma}-\bar{K}$ cut) at $T$ = 30 K.
White arrows and thick (red) dashed lines indicate the energy positions of the 
Dirac point ($E_{\rm DP}$) and the valence band top ($E_{\rm VB}$), 
respectively.
(b) Schematic band diagrams for Bi$_{2-x}$Sb$_{x}$Te$_{3-y}$Se$_{y}$ 
at the insulating compositions, derived from the data shown in (a). 
Taken from Ref. \citen{Arakane}. }
\label{fig:DiracConeTuning}
\end{center}
\end{figure}

In contrast to Bi$_2$Se$_3$, Bi$_2$Te$_3$ can be grown in both $n$- and
$p$-types \cite{QuBi2Te3}. 
This is because Te anti-site defects (Te$^{\bullet}_{\rm
Bi}$) or Bi anti-site defects (Bi$^{'}_{\rm Te}$) are formed in Te-rich
or Bi-rich conditions, respectively, and Te$^{\bullet}_{\rm Bi}$ is a
donor while Bi$^{'}_{\rm Te}$ is an acceptor \cite{Scanlon, Johnson}. 
Nevertheless, it is
difficult to obtain Bi$_2$Te$_3$ samples with low $n_{\rm 3D}$. By
growing a boule of crystal with a compositional gradient, Qu {\it et
al.} were able to pick up samples with $n_{\rm 3D} \approx $ 10$^{16}$
cm$^{-3}$ in which they observed surface SdH oscillations \cite{QuBi2Te3}.

The Bi$_2$Te$_2$Se compound can be much more insulating than
Bi$_2$Se$_3$ or Bi$_2$Te$_3$ if grown in a slightly Se-rich condition
\cite{RenBTS}.
This is because Bi$_2$Te$_2$Se naturally crystallize in a
chalcogen-ordered structure with Te-Bi-Se-Bi-Te QL units 
(Fig. \ref{fig:tetradymite}) \cite{Sokolov}. This peculiar
structure solves the problems of Se-vacancies in Bi$_2$Se$_3$ and Bi/Te
antisite defects in Bi$_2$Te$_3$ at the same time: First, because Se
is concealed in the middle of the QL, Se vacancy is much more difficult
to occur than in Bi$_2$Se$_3$. Second, since the electronegativity of Se
is stronger than that of Te, Bi is more strongly bonded to Se than to
Te, which discourages the occurrence of Bi/Te antisite defects. As a
result, Bi$_2$Te$_2$Se can be reasonably bulk-insulating. In addition, its
ordered structure ensures a high mobility, and consequently the
topological surface state can be easily studied through transport
properties of this material \cite{RenBTS,XiongBTS}. 
In passing, materials with similar 
chalcogen-ordered structure, Bi$_2$Te$_{1.6}$S$_{1.4}$ 
\cite{JiBTS}, Bi$_{1.1}$Sb$_{0.9}$Te$_2$S \cite{JiBTS}, 
and Bi$_2$(Te,Se)$_2$(Se,S) (natural Kawazulite mineral) 
\cite{Kawazulite} have been recently reported, and the latter two 
are found to show moderately bulk-insulating behavior. 

Importantly, even higher bulk-insulating properties than in Bi$_2$Te$_2$Se
have been achieved by utilizing a solid solution
Bi$_{2-x}$Sb$_x$Te$_{3-y}$Se$_y$ ($y \ge 1$) \cite{RenBSTS, TaskinBSTS}. 
In this material, the middle
of the QL is preferentially occupied by Se, hence retaining the
essential virtue of Bi$_2$Te$_2$Se. In addition, the Bi/Sb ratio in the
cation layers and the Te/Se ratio in the outer layers affect the levels
of acceptors and donors, respectively, making it possible to achieve a
maximally compensated situation. Indeed, it was found that a reasonably
bulk-insulating behavior is observed for a series of compositions
$(x,y)$ which forms a curved line in the $x$ vs $y$ phase diagram (Fig.
\ref{fig:BSTS}). Furthermore, the surface Dirac-cone structure of
Bi$_{2-x}$Sb$_x$Te$_{3-y}$Se$_y$ was found to present a systematic
change along the insulating composition line 
(Fig. \ref{fig:DiracConeTuning}), which makes it possible to
{\it tune} the properties of the surface Dirac fermions \cite{Arakane}. 
Such a tuning of the Dirac-carrier properties has been 
dubbed ``Dirac-cone engineering". 
In particular, one can obtain both $p$- and $n$-type
Dirac carriers by tuning $(x,y)$ along the insulating line, which is
useful for designing a $p-n$ junction of the topological surface state.
In this respect,
similar band engineering has also been achieved in (Bi,Sb)$_2$Te$_3$
\cite{XueBST}.  

In fact, another useful approach to reducing bulk carriers is to make a solid
solution of Bi$_2$Te$_3$ and Sb$_2$Te$_3$. Growths of thin films
\cite{XueBST} and nanoplates \cite{KongBiSbTe} of (Bi,Sb)$_2$Te$_3$
as well as nanoribbons \cite{Hong-Cui} of (Bi,Sb)$_2$Se$_3$
have been reported, and in those samples the chemical potential was
successfully tuned to be close to the Dirac point in the middle of the
band gap.

\section{Properties of 3D TI Materials}

\subsection{Surface state and helical spin polarization}

The most prominent property of a TI is the existence of a gapless
surface state. The gapless nature is protected by TRS in
$Z_2$ topological insulators. What makes this surface state distinct
from ordinary surface states (including accumulation and inversion
layers) is its helical spin polarization (Fig. \ref{fig:HelicalSS}), which is also called
spin-momentum locking; namely, the surface state is spin non-degenerate
and the direction of the spin is perpendicular to the momentum vector
and is primarily confined in the surface plane. In fact, if a band has such a
peculiar spin polarization and the system preserves TRS, there
must be a Kramers partner for each eigenstate and Kramers theorem
dictates that the two eigenstates cross each other at TRIMs, which guarantees
the gapless nature of the surface state [see Fig. \ref{fig:TSS}(b)].

The helical spin polarization of the surface
state means that a dissipationless spin current exists on the surface in
equilibrium, because there is no net charge flow but the spin angular
momentum flows in the direction perpendicular to the spin direction. 
The spin helicity of the surface state (i.e. whether the up spin is associated
with $+k$ or $-k$) determines the spin current direction. In all the cases
tested so far, the spin helicity has been found to be left-handed [i.e.
the spin points to $-y$ direction for $\mathbf{k}=(+k,0)$] when $E_F$
is above the Dirac point, and it becomes right-handed for $E_F$ below 
the Dirac point [see Fig. \ref{fig:HelicalSS}(d)].

The spin-momentum locking naturally gives rise to various interesting
spin-related physics. For example, charge fluctuations are naturally
accompanied by spin fluctuations, leading to novel spin-plasmon
excitations \cite{SpinPlasmon}. 
The characteristic energy scale of such excitations is
predicted to be a few meV, and those excitations may have been observed
in ultra-high-resolution ARPES experiments \cite{KondoMode}. Also,
shining a circularly-polarized light to the surface state selectively
excites electrons with a particular spin polarization \cite{Hseih2011}, 
which means that electrons with a particular momentum direction is 
photo-excited in the TI surface state, and as
a result, a photocurrent flows in a direction dictated by
the light polarization. Such a peculiar photocurrent has actually been
experimentally observed \cite{McIver}.

However, detection of a spin-polarized current on the surface of a TI
turns out to be extremely difficult. This is because the spin-momentum
locking makes the charge and spin scattering times to be the same, and
therefore the spin diffusion length is equal to the electron mean free
path in the TI surface state \cite{Appelbaum}. 
In such a situation, the spin polarization
is significantly diminished in a diffusive transport 
(to the order of $\Delta k/k_F$, where $\Delta k$ is the shift of the 
Fermi surface induced by the applied electric field) \cite{Edelstein}, 
and the experiment must be done in a ballistic transport regime to 
detect a spin-polarized current \cite{BurkovSpin}. No such
experiment has been reported for 3D TIs, but in the CdTe/HgTe/CdTe
quantum well in the 2D TI regime, spin polarization of the edge current
has been confirmed \cite{Brune2012}, thanks to the long electron 
mean free path achievable in HgTe quantum wells.

\subsection{Dirac fermion physics}

When two spin-non-degenerate eigenstates forming a Kramers pair cross each
other at a TRIM and the energy dispersion near this crossing point can be
approximated by a linear dispersion, the low energy physics is
described by massless Dirac equation. This means that a Kramers pair of
surface states are actually forming a single 2D Dirac cone on which the
spin degeneracy is lifted [Fig. \ref{fig:HelicalSS}(d)]. 

The Dirac equation for a free particle with mass $m$ is written as
\begin{eqnarray}
E\psi(\mathbf{r}) = c \left( \begin{array}{cc} 0 & \boldsymbol{\sigma} \\ 
\boldsymbol{\sigma} & 0 \end{array} \right)
\cdot \mathbf{\hat{p}}\psi(\mathbf{r}) + mc^2 
\left( \begin{array}{cc} I & 0 \\ 0 & -I \end{array} \right) \psi(\mathbf{r}) 
& & \nonumber \\
= c \left( \begin{array}{cccc} mc & 0 & \hat{p}_z & \hat{p}_x - i\hat{p}_y \\
0 & mc & \hat{p}_x + i\hat{p}_y & - \hat{p}_z \\
\hat{p}_z & \hat{p}_x - i\hat{p}_y & -mc & 0 \\
\hat{p}_x + i\hat{p}_y & - \hat{p}_z & 0 & -mc \end{array} \right) \psi(\mathbf{r}) , & & 
\end{eqnarray}
where $\boldsymbol{\sigma}$ is the vector of Pauli matrices and 
$\mathbf{\hat{p}}$ is the momentum operator. The energy eigenvalue
of this equation is 
\begin{equation}
E = \pm c \sqrt{\mathbf{p}^2 + m^2c^2},
\end{equation}
and hence the Dirac equation always has positive and negative energy states.
One can easily see that this energy eigenvalue has a gap for a finite mass $m$,
but it becomes gapless when $m$ = 0. This is the reason why a gapless system
obeying the Dirac equation is called ``massless"; it does not mean that the 
effective mass of electrons becomes zero. In fact, when the dispersion is 
linear, its second derivative is zero and the effective mass $m^* =
\hbar^2(\partial^2E/\partial k^2)^{-1}$ diverges.

Graphene has emerged as a prototypical material to host 2D Dirac
fermions \cite{graphene}. 
Although both graphene and TIs are Dirac systems, there is an 
important difference. Namely, 
the former has both spin and valley ($K$ and $K'$ points in the BZ) 
degeneracies, while the latter is non-degenerate,
and hence the Dirac fermion physics is simpler in TIs. 
This difference in degeneracy also means that the fermion 
degrees of freedom is 1/4 in TIs compared to that in graphene.

A prominent property of Dirac fermions is that they carry the Berry
phase of $\pi$, as was initially noted by Ando {\it et al} \cite{T.Ando}. 
To see this,
let us consider 2D massless Dirac fermions with the Fermi velocity 
$v_F$, for which the $4 \times 4$ matrix equation reduces to 
$2 \times 2$ and the Dirac equation is written as 
\begin{equation}
E\psi(\mathbf{r}) = \hbar v_F \boldsymbol{\sigma} \cdot 
\mathbf{\hat{k}}\, \psi(\mathbf{r}) 
= -i \hbar v_F \boldsymbol{\sigma} \cdot \nabla \psi(\mathbf{r}) .
\end{equation}
The eigenfunctions of this equation are
\begin{equation}
\psi_{\pm}(\mathbf{r}) = \frac{1}{\sqrt{2}} \left( 
\begin{array}{c} e^{-i\theta(\mathbf{k})/2} \\
\pm e^{i\theta(\mathbf{k})/2} \end{array} \right) 
e^{i \mathbf{k} \cdot \mathbf{r}}
\equiv u_{\pm}(\mathbf{k}) e^{i \mathbf{k} \cdot \mathbf{r}},
\end{equation}
where $\theta(\mathbf{k}) = \arctan(k_y/k_x)$, and the 
energy eigenvalues are 
\begin{equation}
E_{\pm} = \pm \hbar v_{F} k.
\end{equation}
When the wave vector $\mathbf{k}$ is adiabatically rotated 
anticlockwise along a closed path $C$ to encircle the origin, 
the Berry phase $\gamma$ acquired during this adiabatic cycle is 
\begin{equation}
\gamma 
= \oint_C d\mathbf{k} \cdot  i \langle u_{\pm}(\mathbf{k}) \,| 
\nabla_k |\, u_{\pm}(\mathbf{k}) \rangle = \pi .
\end{equation}
Because of this $\pi$ Berry phase, the time-reversed scattering paths,
which in ordinary metals interfere constructively to 
cause the weak localization effect \cite{WL}, now destructively
interfere each other, leading to the weak anti-localization effect \cite{T.Ando}.

\begin{figure}[t]
\begin{center}
\includegraphics[width=7.2cm]{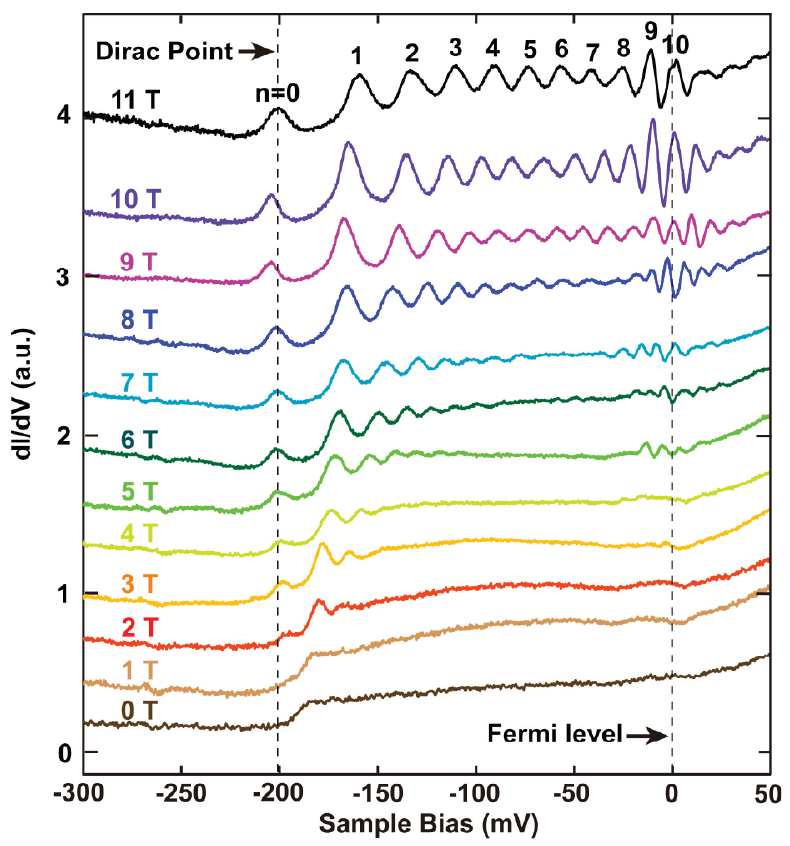}
\caption{(Color online) 
Landau quantization of the topological surface state observed
in STS experiment on a 50-nm-thick Bi$_2$Se$_3$ film.
The differential tunneling spectra were
acquired for various magnetic fields from 0 to 11 T applied
perpendicular to the surface. Taken from Ref. \citen{XueSTS}; 
copyright American Physical Society (2010).
}
\label{fig:Cheng}
\end{center}
\end{figure}

Another prominent property of Dirac fermions is their peculiar
Landau quantization of the energy states in magnetic fields.
It has been shown \cite{McClure} that the quantization occurs
in the manner 
\begin{equation}
E_{\pm}(N) = \pm \sqrt{(2e\hbar v_F^2 B/c) N},
\label{eq:DiracLandau}
\end{equation}
where $N$ = 0, 1, 2, ...
This means that in Dirac systems when the Fermi level is varied, 
the Landau-level (LL) spacing 
is not a constant but changes as $\sqrt{N}$, as opposed
to ordinary metals in which the LL spacing is
simply $\hbar \omega_c$ ($=e\hbar B/m_c c$ where $m_c$ is the 
cyclotron mass) and is independent of 
the Fermi level. Also, Eq.
(\ref{eq:DiracLandau}) indicates that there is the zeroth
LL with $N$ = 0, which is pinned to the charge
neutrality point (Dirac point).
Therefore, the Landau quantization of massless Dirac fermions 
is characterized by the occurrence of the 
zero-energy state and the symmetrical appearance of 
$\sqrt{N}$ states on both the positive and negative energy 
sides of the Dirac point. 
Such a peculiar Landau quantization of 
the surface state has indeed been observed in TIs by STS
experiments in magnetic fields (Fig. \ref{fig:Cheng})
\cite{Sb2Te3, XueSTS, Hanaguri}.

\begin{figure}[t]
\begin{center}
\includegraphics[width=7.1cm]{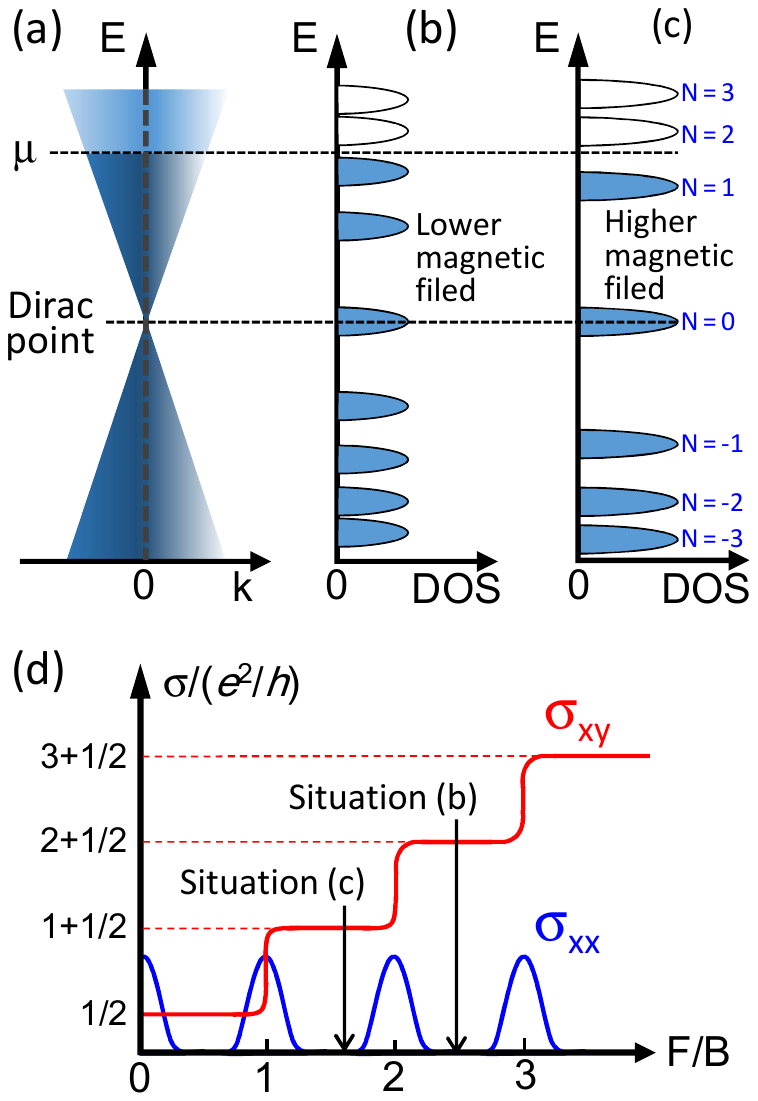}
\caption{(Color online) 
(a) Partially-filled 2D Dirac cone; $\mu$ is the Fermi level.
(b) Landau quantization of the Dirac cone; LLs below $\mu$ are
filled with electrons. Note that the spacing between LLs change as
$\sqrt{N}$ and the $N$ = 0 LL is pinned to the Dirac point.
(c) In a higher magnetic field, the spacing between LLs increases
as $\sqrt{B}$, and fewer LLs are filled.
(d) Schematic behavior of $\sigma_{xx}$ and $\sigma_{xy}$
in the quantum Hall regime of 2D Dirac fermions; the situations
depicted in panels (b) and (c) are marked by arrows. Notice
that the minima in $\sigma_{xx}$ marks a complete filling
of up to $N$th LLs. }
\label{fig:LLs}
\end{center}
\end{figure}

When the Landau quantization of the form Eq. (\ref{eq:DiracLandau})
takes place, the associated quantum Hall effect becomes
unusual, and the Hall plateau between the $N$th
and $(N+1)$th LLs is quantized to
\begin{equation}
\sigma_{xy} = - \frac{e^2}{h} \left(N+\frac{1}{2}\right) ,
\end{equation}
which is called half-integer quantization [see Fig. \ref{fig:LLs}] 
\cite{graphene}. 
This unusual quantization can be understood to be a 
result of the existence of the $N$ = 0 LL at the Dirac point, 
which dictates that the first Hall plateaus on the positive and 
negative energy sides must appear antisymmetrically, 
because $\sigma_{xy}$ is an odd function of energy.
The half-integer quantization
can also be understood to be a result of the $\pi$ Berry
phase \cite{Geim, P.Kim}, which is most easy to see in the Laughlin's
gauge argument \cite{Laughlin1981} for the quantization of $\sigma_{xy}$.

\subsection{Quantum oscillations}

When the Landau quantization of the energy states occurs in 
crystalline solids in magnetic fields, the density of states (DOS)
becomes periodically modulated as a function of magnetic
field, which leads to various sorts of oscillation phenomena
that are generally called quantum oscillations \cite{Shoenberg}. 
In particular, the oscillations
occurring in conductivity are called Shubnikov-de Haas (SdH) 
oscillations, while the oscillations occurring in magnetic
susceptibility are called de Haas-van Alphen (dHvA) 
oscillations. 

SdH oscillations play particularly important roles
in the studies of 3D TIs for two reasons. First, they provide a 
means to selectively and quantitatively characterize the 
2D surface states that coexist with 3D bulk states.
Second, the phase factor of the oscillations directly reflects 
the Berry phase of the system, which allows us to elucidate
whether the electrons showing the SdH oscillations are
Dirac fermions or not.

In SdH oscillations, conductivity oscillates periodically as a function 
of $1/B$ and the oscillatory part of the longitudinal conductivity 
$\sigma_{xx}$ follows
\begin{equation}
\Delta \sigma_{xx} \sim \cos\left[2\pi \left(\frac{F}{B}
-\frac{1}{2}+\beta\right)\right],
\label{eq:SdH}
\end{equation}
where $F$ is the oscillation frequency and $\beta$ accounts for a phase shift
($0 \le \beta < 1$). 
In fact, the same phase factor appears in the Onsager's semiclassical quantization 
condition \cite{Shoenberg}
\begin{equation}
A_{N}=\frac{2\pi e}{\hbar c}B\left(N+\frac{1}{2}-\beta\right),
\end{equation}
which is satisfied when the $N$th LL is crossing the Fermi energy $E_F$.
Here, $A_{N}$ is the area enclosed by electrons in the $k$-space with their 
cyclotron orbits on the Fermi surface.
The parameter $\beta$ is simply the Berry phase $\gamma$ divided by $2\pi$.
For spinless fermions,
it is known \cite{Shoenberg, Mikitik1999} that the Berry phase is zero for 
a parabolic energy dispersion ($\beta$ = 0) and, as already
noted, $\pi$ for Dirac fermions possessing a linear 
energy dispersion ($\beta$ = $\frac{1}{2}$).
In real TI materials, the Dirac dispersion is not strictly
linear but contain a parabolic component. 
Nevertheless, theoretically it has been elucidated that $\beta$ is robustly $\frac{1}{2}$ 
at least at large $N$ \cite{TaskinBerry, Mikitik2012, Wright}.

When $E_F$ lies at the center of a LL (which is usually
broadened due to thermal fluctuations and disorder), the DOS takes a
maximum; on the contrary, the DOS takes a minimum when $E_F$ lies in 
between two neighboring LLs [Fig. \ref{fig:LLs}(b,c)]. In the latter situation, 
a certain number of LLs are completely
filled and the next LL is empty. Therefore, a 
minimum in $\sigma_{xx}$, which occurs when DOS takes a minimum, 
signifies a complete filling of some $N$ LLs,
and one can assign an integer index $N$ to that minimum. 
In ordinary metals, this LL index $N$ corresponds to the filling factor $\nu$.
This can be easily understood by remembering the situation in the 
ordinary quantum Hall
effect, in which $\sigma_{xy}$ is quantized to $\nu e^2/h$ and 
$\sigma_{xx}$ becomes zero when the chemical potential lies between the
$\nu$th and $(\nu+1)$th LLs. (In the quantum Hall effect, the vanishing
$\sigma_{xx}$ is a reflection of the bulk gap opening at the Fermi level.)
On the other hand, in the case of Dirac 
fermions, the filling factor $\nu$ is not $N$ but is $N+\frac{1}{2}$ because
of the half-integer quantization which essentially stems from the existence 
of the zeroth LL [Fig. \ref{fig:LLs}(d)].

The phase factor $\beta$ in the SdH oscillations can be experimentally
determined from an analysis of the so-called LL fan diagram, 
in which the sequence of the
values of $1/B_N$ corresponding to the $N$th minimum in $\sigma_{xx}$ 
are plotted versus $N$. 
From Eq. (\ref{eq:SdH}), one can see that the $N$th minimum 
occurs when the argument of the cosine equals $(2N-1)\pi$, i.e.
\begin{equation}
2\pi \left(\frac{F}{B_N}-\frac{1}{2}+\beta\right) =(2N-1)\pi .
\label{eq:N_min}
\end{equation}
Therefore, the plot of $1/B_{N}$ vs $N$ makes a straight 
line with a slope $F$ corresponding to the oscillation frequency. 
Note that $F/B_1=1-\beta$ holds for the 1st minimum. Also,
it follows from Eq. (\ref{eq:N_min}) that, when a linear fit to the 
LL fan diagram is extrapolated to $1/B_N \rightarrow 0$, 
the intercept on the $N$-index axis gives the phase factor $\beta$.
When the $\beta$ value thus obtained is $\frac{1}{2}$, one may
conclude that the SdH oscillations come from Dirac fermions. 

In the past, there have been confusions about the proper way to construct
the LL fan diagram; namely, whether to assign an integer index to a 
minimum in $\rho_{xx}$ or to a minimum in $\sigma_{xx}$
\cite{RenBTS, QuBi2Te3, TaskinBSTS, AnalytisNP, Sacepe, K.L.WangSdH, 
YongChen, K.L.WangSdH2, Veldhorst}.
Since it is important to clarify this confusion, let us discuss this issue 
in some detail. In solids, the resistivity tensor is an inverse of the
conductivity tensor, and in the isotropic case their relation is 
\begin{eqnarray}
\left( \begin{array}{cc} \rho_{xx} & \rho_{xy} \\
\rho_{yx} & \rho_{xx} \end{array} \right)
&=& \left( \begin{array}{cc} \sigma_{xx} & \sigma_{xy} \\
-\sigma_{xy} & \sigma_{xx} \end{array} \right)^{-1} \nonumber \\
&=& \frac{1}{\sigma_{xx}^2+\sigma_{xy}^2}
\left( \begin{array}{cc} \sigma_{xx} & -\sigma_{xy} \\
\sigma_{xy} & \sigma_{xx} \end{array} \right) . \ \ \  
\end{eqnarray}
Therefore, when the condition $\sigma_{xx} \ll \sigma_{xy}$
is satisfied (which is usually the case with low-carrier-density semiconductors), 
$\rho_{xx} \simeq \sigma_{xx}/\sigma_{xy}^2$
and the minima in $\rho_{xx}$ coincide with those in $\sigma_{xx}$.
This is the reason why the LL fan diagram constructed from the 
$\rho_{xx}$ data can give the correct phase factor in graphene
\cite{Geim, P.Kim}.
However, in the case of TIs, due to the presence of the bulk transport
channel, often the condition $\sigma_{xx} \ll \sigma_{xy}$ does not 
strictly hold. In the extreme case, when $\sigma_{xx} \gg \sigma_{xy}$
(which is usually the case with metals), $\rho_{xx} \simeq \sigma_{xx}^{-1}$
and the minima in $\rho_{xx}$ now coincide with the {\it maxima} 
in $\sigma_{xx}$. Therefore, unless $\sigma_{xx} \ll \sigma_{xy}$
is satisfied, performing the LL fan diagram analysis using the resistivity data
is dangerous. Ideally, one should measure both $\rho_{xx}(B)$ and 
$\rho_{yx}(B)$ at the same time and convert them into $\sigma_{xx}(B)$
and $\sigma_{xy}(B)$ to perform reliable LL fan diagram analysis to elucidate
the correct Berry phase. 

\begin{figure}[b]
\begin{center}
\includegraphics[width=6.3cm]{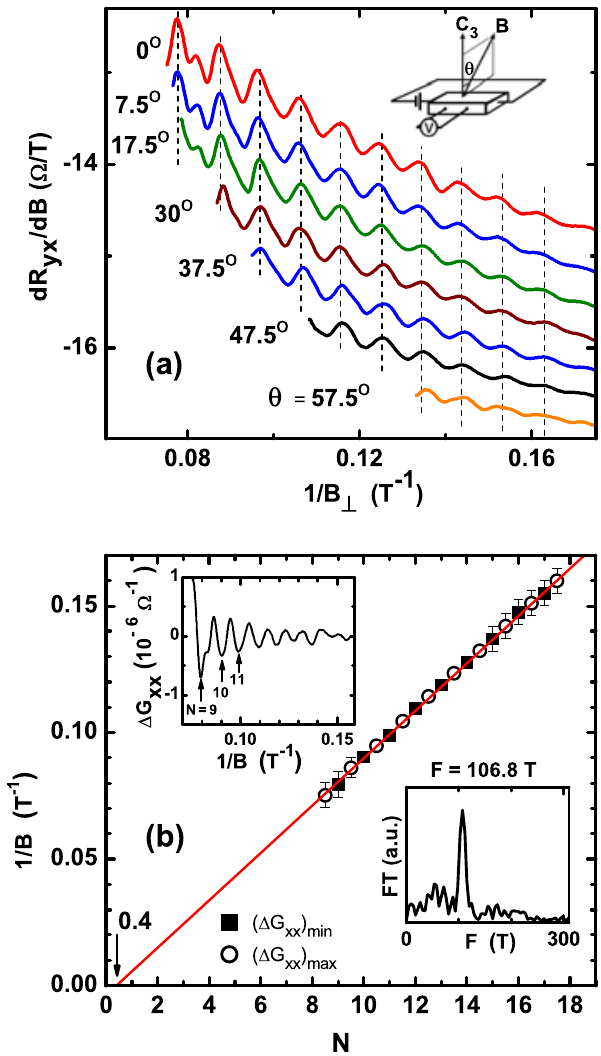}
\caption{(Color online) 
(a) SdH oscillations observed in a 10-nm-thick 
MBE-grown Bi$_2$Se$_3$ film; to enhance the visibility of
the oscillations, $dR_{yx}/dB$ is plotted vs 
$1/B_{\perp}$ ($=1/B\cos\theta$) here. Notice that the
positions of maxima and minima do not change with $1/B_{\perp}$,
which gives evidence for the 2D origin of the oscillations.
(b) LL fan diagram constructed from the analysis of the 
$\sigma_{xx}$ data measured at 1.6 K and $\theta$ = 0$^{\circ}$;
here, integer index $N$ are assigned to the minima (upper inset). 
Upon making a linear fitting to the data,
the slope is fixed at the frequency $F$ obtained from the 
Fourier analysis of the data shown in the lower inset;
the straight-line fitting extrapolates to 0.40 $\pm$ 0.04 on the 
$N$-index axis, which is close to the value 1/2 expected for
Dirac fermions. Based on the data from Ref. \citen{TaskinMBE}.}
\label{fig:FanDiagram}
\end{center}
\end{figure}

In this regard, in the early stage of the TI research, the LL fan diagram 
analyses of the SdH oscillations observed in TIs were influenced 
by the case with graphene and used the minima in $\rho_{xx}$ for
indexing integer $N$ 
\cite{RenBTS, QuBi2Te3, TaskinBSTS, AnalytisNP, Sacepe, K.L.WangSdH,
YongChen, K.L.WangSdH2, Veldhorst}. 
Therefore, the conclusions regarding the Berry
phase in those early publications should be taken with care. It was 
Xiong {\it et al.} who first switched to correctly using the minima 
in $\sigma_{xx}$ for the LL fan diagram analysis \cite{XiongBTS},
and some of the recent works performed reliable analyses and 
unambiguously demonstrated the Dirac nature of the surface state 
through determination of the Berry phase 
\cite{XiongBTS, RenSnBTS, XiongBerry, TaskinMBE}. 
In addition, to determine the Berry phase from
the LL fan diagram in a most reliable way, one should fix the slope
of the linear fitting by using the frequency $F$ obtained from the 
Fourier analysis of the data; this way, the intercept on the 
$N$-index axis is obtained with little ambiguity \cite{RenSnBTS, TaskinMBE}.
An example of the LL fan diagram obtained for an MBE-grown 
Bi$_2$Se$_3$ film is shown in Fig. \ref{fig:FanDiagram}.

\begin{figure}[[b]
\begin{center}
\includegraphics[width=6.3cm]{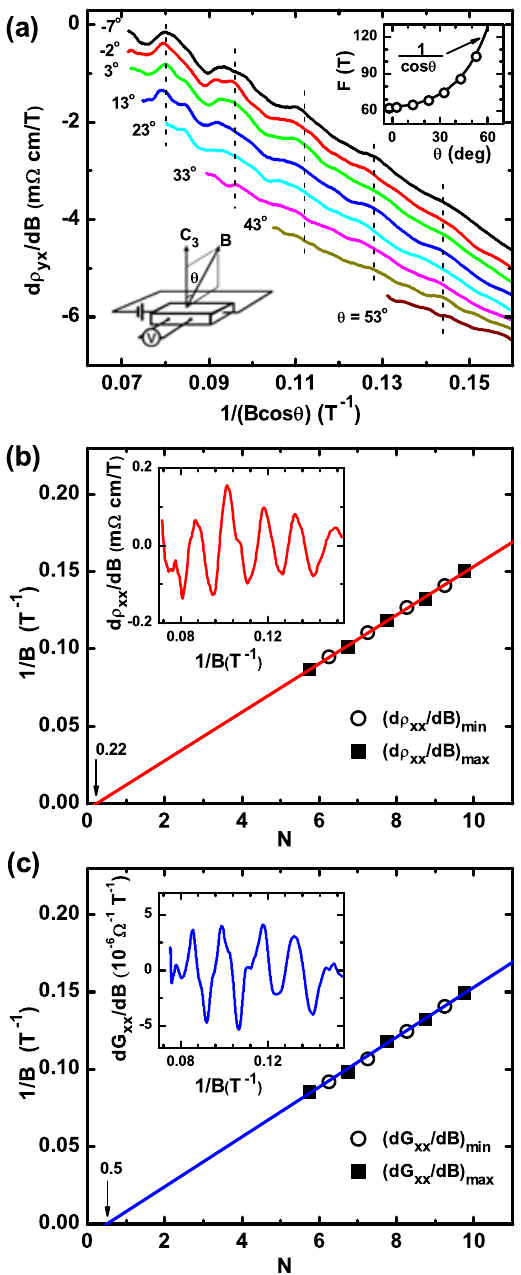}
\caption{(Color online) 
(a) SdH oscillations in $d\rho_{yx}/dB$ observed in a 260-$\mu$m-thick 
Bi$_2$Te$_2$Se single crystal; inset shows the magnetic-field 
angle dependence of the oscillation frequency $F$. 
(b) LL fan diagram based on the 
$d\rho_{xx}/dB$ data at 1.6 K and $\theta$ = 0$^{\circ}$
shown in the inset. Assuming that integer indices $N$ are to be
assigned to the minima in $\rho_{xx}$, the minima and maxima
in $d\rho_{xx}/dB$ correspond to $N + \frac{1}{4}$
and $N + \frac{3}{4}$, respectively. Linear fitting 
to the data extrapolates to 0.22 $\pm$ 0.12.
(c) LL fan diagram based on the
$d\sigma_{xx}/dB$ data shown in the inset; here, assuming that 
integer indices $N$ are to be
assigned to the minima in $\sigma_{xx}$, $N + \frac{1}{4}$
and $N + \frac{3}{4}$ are assigned to the minima and maxima
in $d\sigma_{xx}/dB$, respectively. Linear fitting 
to the data (in which the slope is constrained from the Fourier analysis) 
extrapolates to 0.5. Based on the data from Ref. \citen{RenBTS}.}
\label{fig:BTSdata}
\end{center}
\end{figure}

For the phase analysis of the SdH oscillations, an interesting lesson
can be learned from the data for Bi$_2$Te$_2$Se published in 2010
by Ren {\it et al.} \cite{RenBTS}. In their original paper, they assigned 
integer index $N$ to the minima in $\rho_{xx}$ and obtained an
intermediate phase factor $\beta$ of 0.22 $\pm$ 0.12 [see
Fig. \ref{fig:BTSdata}(b)] \cite{RenBTS}. 
However, if the data are re-analyzed by calculating $\sigma_{xx}$
from $\sigma_{xx} = \rho_{xx}/(\rho_{xx}^2 + \rho_{yx}^2)$ and 
identifying its minima to signify integer $N$, the same data set
now gives $\beta$ = 0.5 [Fig. \ref{fig:BTSdata}(c)], 
which reaffirms that Ren {\it et al.} were indeed observing SdH 
oscillations from Dirac fermions.

Besides the Berry phase, the SdH oscillations contain various useful
information. 
The Onsager's relation \cite{Shoenberg} gives $F$ in terms of the 
Fermi wave vector $k_{F}$ as
\begin{equation}
F = (\hbar c/ 2\pi e)\pi k_{F}^{2} 
\end{equation}
for a circular extremal cross section of FS, and therefore one can calculate
the averaged $k_{F}$, and hence the carrier density, from the frequency of the 
SdH oscillations. To be explicit, for the topological 2D surface state
the surface carrier density $n_s = \frac{1}{(2\pi)^2}\pi k_F^2$ 
is directly obtained 
from $F$. If the surface state is of non-topological origin (such as 
accumulation or inversion layer), $n_s$ should be multiplied by 2
to account for the spin degeneracy. Also, if the SdH oscillations
come from a 3D bulk state and the bulk Fermi surface is an ellipsoid,
measurements of the SdH frequencies in magnetic fields along 
the three principal axes give the lengths of the three semiaxes, $k_F^a$,
$k_F^b$, and $k_F^c$; the bulk carrier density $n_{\rm 3D}$ 
for this Fermi surface is obtained as $n_{\rm 3D} = 
[2/(2\pi)^3](4\pi/3) k_F^{a} k_F^{b} k_F^{c}$,
where the spin degeneracy is taken into account \cite{Eto}.

The 2D nature of the SdH oscillations can be tested by taking the 
dependence of $F$ on the angle $\theta$
between the magnetic field direction and the surface plane normal
[Figs. \ref{fig:FanDiagram}(a) and \ref{fig:BTSdata}(a)]; 
if the measured $F$ changes as $\sim 1/\cos \theta$, it is a strong 
indication that the SdH oscillations come from a 2D system
\cite{Taskin2010, RenBTS, QuBi2Te3, TaskinBSTS, AnalytisNP, TaskinMBE}.
One should note, however, that a reasonably wide range 
of $\theta$ (e.g., up to 50$^{\circ}$) had better be measured, because
SdH oscillations from an elongated 3D Fermi surface can present an
approximate $ 1/\cos \theta$ scaling for a limited range of $\theta$
\cite{QuBi2Te3}. Also, disappearance of the SdH oscillations for 
$\theta = 90^{\circ}$ gives additional support to the 2D nature.

The cyclotron mass $m_c$ of the carriers can be determined from the
analysis of the temperature dependence of the amplitude of the
SdH oscillations. This is because the quantum oscillations are expressed in the 
Lifshitz-Kosevich theory \cite{Shoenberg} as
\begin{equation}
\Delta \sigma_{xx}=
A_0 R_{T} R_{D} R_{S} \cos \left[ 2\pi 
\left(\frac{F}{B}- \frac{1}{2} + \beta \right) \right] ,
\label{eq:SdH2}
\end{equation} 
where $A_0$ is a constant and the three coefficients, 
$R_T = 2\pi^2 (k_B T/\hbar \omega_c) / \sinh[2\pi^2 (k_B T/\hbar \omega_c)]$, 
$R_D = \exp[-2\pi^2 (k_B T_D/\hbar \omega_c)]$, and
$R_S = \cos (\frac{1}{2}\pi g m_e/m_c)$ 
are called temperature, Dingle, and spin damping factors, respectively \cite{Shoenberg}, 
with $T_D$ the Dingle temperature ($g$ is the electron $g$-factor and 
$m_e$ is the free electron mass).
In a fixed magnetic field, $R_D$ does not change and the temperature
dependence shows up only through $R_T$; therefore, a fitting of the temperature
dependence of the oscillation amplitude allows one to determine $\omega_c$,
which in turn gives $m_c = e B/(c \omega_c)$. 
Remember, $m_c$ is defined as 
\begin{equation}
m_c = \frac{\hbar^2}{2\pi}\left( \frac{\partial A(E)}{\partial E} \right)_{E=E_F},
\end{equation}
where $A(E)$ is the area enclosed by the cyclotron orbit in the $k$-space.
Since the cyclotron orbit is confined on the Fermi surface in the $k$-space, 
the enclosed area changes with $E_F$. In the case of 2D Dirac fermions
with the energy dispersion $E(k) = \hbar v_F k$, one obtains $A(E_F) =
\pi k_F^2 = \pi E_F^2/(\hbar v_F)^2$ and hence $m_c = E_F/v_F^2
= \hbar k_F/v_F$. Thus, once $m_c$ is determined from the temperature
dependence of the SdH amplitude, one can calculate $v_F = \hbar k_F/m_c$
and compare it with the slope of the Dirac dispersion known from the ARPES
measurement to judge if the obtained SdH result is consistent with the
known dispersion.

After $m_c$ is determined, one can further determine the Dingle temperature 
$T_D$ ($= \hbar/2\pi k_B \tau$) from the magnetic-field dependence of the 
amplitude of the SdH 
oscillations at fixed temperature and obtain the quantum scattering time $\tau$. 
This is usually done by making the Dingle analysis, in which one plots 
$\ln[A B \sinh(\alpha T/B)]$ against $1/B$ ($A$ is the 
observed magnetic-field-dependent amplitude 
of the oscillations and $\alpha = 14.7(m_c/m_e)$ [T/K]). 
Such a plot makes a straight line and its slope gives $T_D$
because of the relation
$A = A_0 R_T R_D R_S \sim [(2\pi^2 k_B T/\hbar \omega_c) / 
\sinh(2\pi^2 k_B T/\hbar \omega_c)]
\exp(-2\pi^2 k_B T_D/\hbar \omega_c)$.
The obtained $\tau$ is used for calculating the mean free path
$\ell^{\rm SdH} = v_F \tau$, which in turn gives an estimate of the 
surface carrier mobility
$\mu^{\rm SdH}_s = e\tau/m_c = e\ell^{\rm SdH}/\hbar k_F$.  

Historically, the first observation of quantum oscillations 
coming from the 2D state of a 3D TI was made by Taskin and Ando 
in 2009 for Bi$_{1-x}$Sb$_x$, in which both dHvA
and SdH oscillations were clearly observed \cite{Taskin2009}. 
In their experiment, up to five frequencies were resolved in the Fourier transform
of the oscillations, and detailed angular dependence measurements of those
frequencies in all three basal planes elucidated the existence of one circular 2D Fermi
surface on the $C_1$ plane and three ellipsoidal electron pockets located at the 
$L$ points in the 3D BZ. Furthermore, it was possible to 
determine $m_c = 0.0057m_e$ and $T_D$ = 6.7 K for the 2D channel 
from the analyses of the 2D component in the data.
The 2D carrier density obtained from the oscillation
frequency was only $n_s$ = 1.4 $\times$ 10$^{10}$ cm$^{-2}$. 
The extremely small $m_c$ at very low $n_s$ is a
characteristic feature of massless Dirac fermions, in which 
$m_c = E_F/v_F^2$; furthermore, the phase of the 2D component of the 
oscillations indicated the Berry phase 
of $\pi$ \cite{note}, and hence the observed 2D carriers were most likely 
of topological origin. Nevertheless, it turned out that the 2D component
of the oscillations is too strong if one assumes that the 2D carriers 
reside only on the outer surface of the sample, because the total number of 
bulk electrons was $\sim$10$^5$ times larger than the total number
of 2D carriers, and yet, the dHvA oscillations from the 2D carriers 
were as strong as those from bulk carriers. Although the actual
situation in Bi$_{1-x}$Sb$_x$ is still not clear, what is probably
happening is
that Bi$_{1-x}$Sb$_x$ crystals contain a high density of planar defects 
along the crystallographic $C_1$ plane, and topological 2D states
reside on such internal ``surfaces". 

The first observations of SdH oscillations coming really from the outer surface
of 3D TIs were made for Bi$_2$Te$_3$ and Bi$_2$Se$_3$ by
Qu {\it et al.} \cite{QuBi2Te3} and by Analytis {\it et al.} \cite{AnalytisNP},
respectively, at nearly the same time in 2010.
The 2D nature of the 
SdH oscillations were confirmed by taking the angular dependence
of the oscillation frequency, as was the case with Bi$_{1-x}$Sb$_x$
\cite{Taskin2009, Taskin2010}. 
Since the surface band structures were already known from ARPES 
for those materials \cite{Hasan_Bi2Se3, Chen_Bi2Te3, Hasan_Bi2Te3}, 
the agreement of $v_F$ estimated from the 
SdH data with that known from ARPES gave confidence in the origin
of the observed 2D SdH oscillations.
Those observations were made possible by minimizing the 
naturally-doped bulk carriers in Bi$_2$Te$_3$ and Bi$_2$Se$_3$,
but still, the surface transport accounted for only less than 0.3\% of
the total conductance.

\subsection{Two-band analysis}

In real TI materials, the bulk transport channel is usually not negligible
and one should consider parallel conductions through surface and bulk. 
Such a situation can be treated by a two-band model, in which 
the composite Hall resistivity is described as \cite{RenBTS}
\begin{equation}
\rho_{yx} = \frac{(R_{s}\rho_{b}^{2}+R_{b}\rho_{s}^{2})B
+R_{s}R_{b}(R_{s}+R_{b})B^{3}}
{(\rho_{s}+\rho_{b})^{2}+(R_{s}+R_{b})^{2}B^{2}},
\end{equation}
where $R_{b}$ and $\rho_{b}$ are the Hall coefficient and
resistivity of the bulk channel, $R_{s} = t/(en_{s})$, and
$\rho_{s}=\rho_{\square}t$, with $t$ the sample thickness
and $\rho_{\square}$ the surface sheet resistance. 
By fitting the $\rho_{yx}(B)$ data with this two-band model, one can
obtain $n_{\rm 3D}$ ($= 1/eR_b$), $\rho_b$, $n_s$, 
and $\rho_{\square}$,
from which the mobilities for the bulk and surface channels are also 
obtained.

One should note that this analysis involves as many as four parameters.  
Nevertheless, the parameters must be consistent with the $\rho_{xx}$ 
value in zero field, which gives one constraint. If SdH oscillations are
observed in the surface transport, the SdH frequency $F$ fixes
the value of $n_s$, which is an additional constraint. 
With those two constraints, the 
two-band analysis becomes reasonably reliable. 

\begin{figure}[b]
\begin{center}
\includegraphics[width=5.5cm]{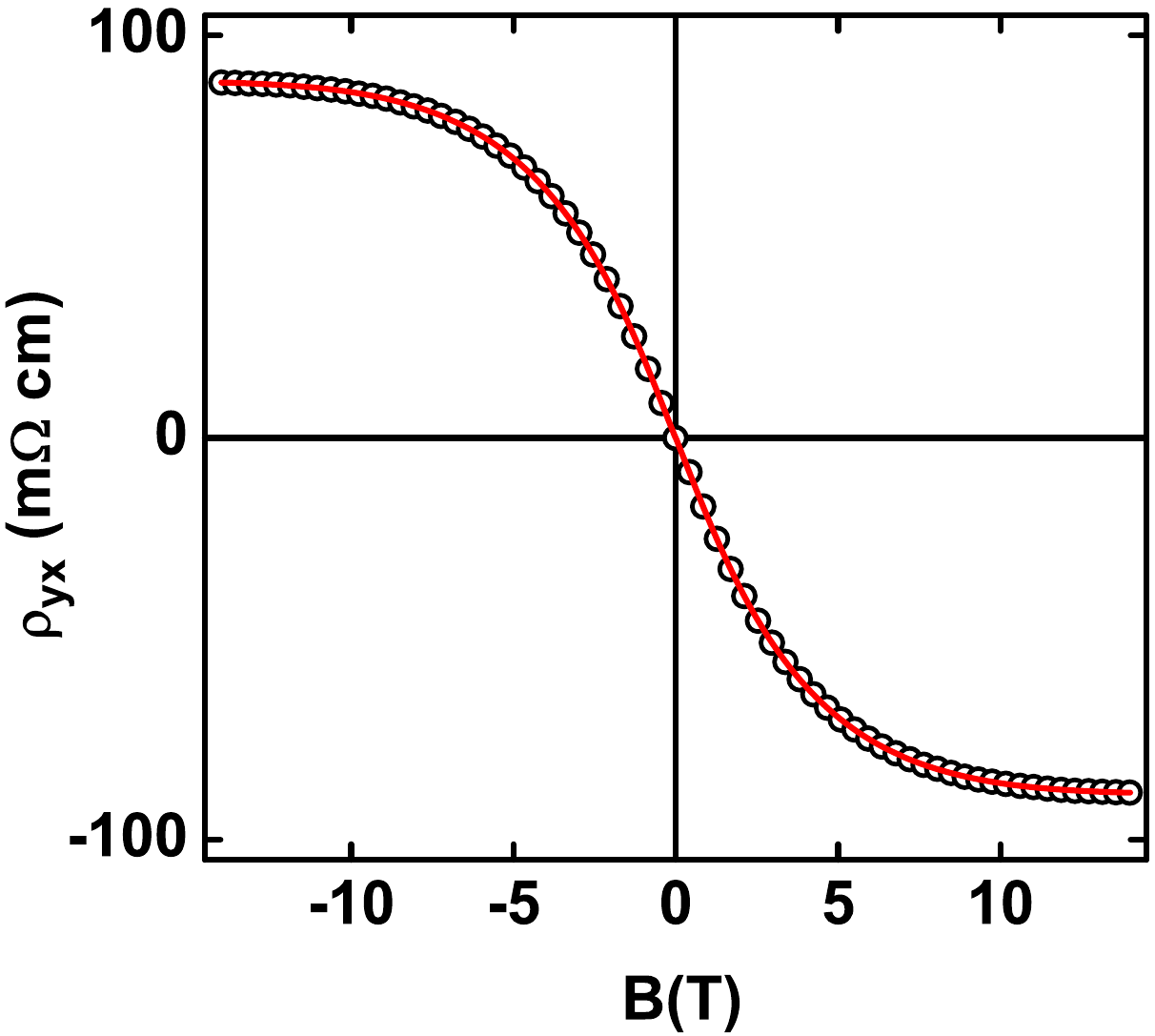}
\caption{(Color online) 
Two-band analysis of the non-linear $\rho_{yx}(B)$ data
observed in Bi$_2$Te$_2$Se at 1.6 K. Open circles are the data 
and the solid line is the fitting result. In this fitting, the surface
carrier density $n_s$ is not a free parameter but is constrained
from the SdH frequency. Based on the data from Ref. \citen{RenBTS}.}
\label{fig:BTSdata2}
\end{center}
\end{figure}

For example, in the transport study of a 260-$\mu$m-thick Bi$_2$Te$_2$Se 
crystal reported by Ren {\it et al.} \cite{RenBTS}, observation of 
surface SdH oscillations (Fig. \ref{fig:BTSdata}) allowed them to fix 
$n_s$, and the two-band analysis 
of $\rho_{yx}(B)$ (Fig. \ref{fig:BTSdata2}) gave consistent estimates 
of all the transport parameters: surface
mobility $\mu_{s}^{\rm tr}$ = 1450 cm$^{2}$/Vs, bulk mobility 
$\mu_{b}$ =
11 cm$^{2}$/Vs, $n_s$ = 1.5 $\times$ 10$^{12}$ cm$^{-2}$, and
$n_{\rm 3D}$ = 2.4 $\times$ 10$^{17}$ cm$^{-3}$. 
These values are used for calculating the fraction of the surface
contribution in the total conductance, which, in this example, was
$\sim$6\%.

However, if the SdH oscillations are {\it not} observed, one should 
take the result of the two-band analysis with a grain of salt. 
This is because the model
does not consider any magnetic-field dependence of the scattering rates,
even though the magnetoresistance $\rho_{xx}(B)$ is usually large and 
complicated in TIs. In fact, it is almost always the case that the 
combination of parameters obtained from the two-band analyses
of $\rho_{yx}(B)$ does not correctly reproduce the $\rho_{xx}(B)$
behavior.

It is often observed in TIs that the surface mobility 
$\mu^{\rm tr}_s$ estimated from the two-band analysis 
is higher than $\mu^{\rm SdH}_s$ estimated from SdH oscillations.
This discrepancy is most likely intrinsic and is a manifestation of the peculiar
helical spin polarization. 
This is because $\tau$ obtained from SdH oscillations
reflects scattering events in all directions equally, 
whereas in the transport coefficients
such as $\rho_{yx}$ the backward scattering, which is discouraged in the
TI surface states, plays the most important role.
More specifically, $\tau^{\rm tr}$ to govern the resistivity
acquires the additional factor $1/(1-\cos\phi)$ upon
spatial averaging ($\phi$ is the scattering angle), while $\tau$ to
govern the dephasing in the quantum oscillations is given by a simple
spatial averaging without such a factor \cite{DasSarma}. 
Hence, if the small-angle scattering becomes dominant 
(which is often the case at low
temperature), $\tau^{\rm tr}$ can be much larger than $\tau$.

\subsection{Weak anti-localization effect}

As briefly mentioned in conjunction with the Dirac fermion physics, the
$\pi$ Berry phase associated with charge carriers leads to the
weak anti-localization effect \cite{T.Ando}.
In the ordinary weak localization effect \cite{WL},
electrons in metals are localized due to constructive interference
of the electron wave functions between two time-reversed paths,
which enhances the probability of those electrons to be found at
a certain spatial location and reduces their ability to transport currents. 
This interference can be destroyed by breaking TRS 
with applied magnetic field, which shifts the phase factor 
of the wave functions of the two time-reversed paths differently.
Therefore, when the weak localization is taking place, application 
of a magnetic field leads to a {\it negative} magnetoresistance,
i.e., an enhancement of the conductivity. The anti-localization effect
is just the opposite of this weak localization effect \cite{T.Ando}; namely, 
because of the phase shift of $\pi$ which occurs when an electron
travels along a closed path, the two time-reversed paths interfere
destructively, reducing the probability of electrons to localize. 
Naturally, application of a magnetic field tends to lift this destructive
interference, leading to a {\it positive} magnetoresistance.

It should be noted that this weak anti-localization effect also occurs in 
a system with strong SOC \cite{HLN}, which causes a 
spin rotation whenever an electron is scattered off an impurity,
and the resulting phase change of the electron wave function 
leads to the destructive interference similar to the case of the 
$\pi$ Berry phase. Therefore, the observation of the weak anti-localization
effect alone does not give evidence for the existence of Dirac fermions.

The magnetic-field dependence of the conductivity for the weak
anti-localization effect in 2D systems was calculated by Hikami, 
Larkin, and Nagaoka as \cite{HLN}
\begin{equation}
\Delta \sigma_{xx}(B) = \alpha
\frac{e^{2}}{\pi h} \left [  \Psi \left( \frac{\hbar c}{4 e L_{\phi}^2 B} + 
\frac{1}{2} \right) 
- {\rm ln} \left( \frac{\hbar c}{4 e L_{\phi}^2 B} \right)  \right ],
\label{eq:HLN}
\end{equation}
where $\Psi$ is the digamma function and $L_{\phi}$ is the phase
coherence length. The prefactor $\alpha$ should be $-\frac{1}{2}$ for
each transport channel that either carries the $\pi$ Berry phase
or bears a strong SOC. 

\begin{figure}[t]
\begin{center}
\includegraphics[width=8.1cm]{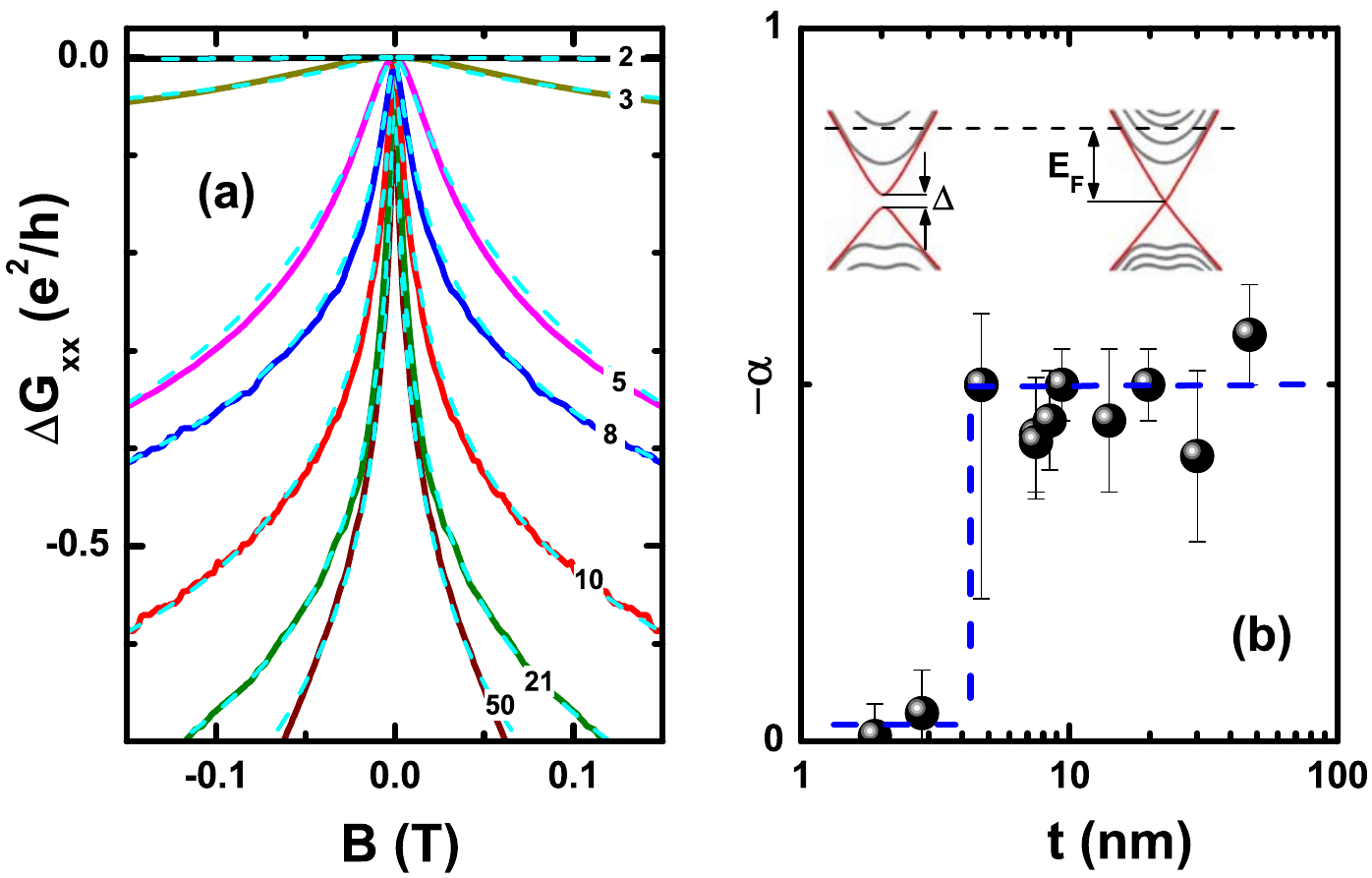}
\caption{(Color online) 
(a) Weak anti-localization behavior observed at 1.6 K in $\sigma_{xx}$ of
a series of MBE-grown Bi$_2$Se$_3$ thin films with various 
thickness shown in QL unit (1 QL is 0.95-nm thick); 
dashed lines are the fittings using the Hikami-Larkin-Nagaoka formula 
[Eq. (\ref{eq:HLN})].
(b) Thickness dependence of the parameter $\alpha$ in Eq. (\ref{eq:HLN}). 
Inset shows schematic energy bands above and below the critical thickness
at which hybridization of top and bottom surface states starts to develop. 
Taken from Ref. \citen{TaskinMBE}; copyright American Physical Society (2012).}
\label{fig:WAL}
\end{center}
\end{figure}

In Bi$_{2}$Se$_{3}$ thin films, this weak anti-localization effect is
frequently observed and $\alpha \approx -\frac{1}{2}$ is usually found.
This is because the top and bottom surfaces are connected though the
bulk channel to form a single diffusive transport channel
\cite{TaskinMBE, STO-WAL, J.Chen2011, Steinberg2011, Y.S.Kim2011}.
Remember, as long as the thickness is shorter than $L_{\phi}$ which is
usually of the order of 100--1000 nm in TIs, electrons can diffusively
travel from the bottom surface state to the top surface state though the
bulk state using three different Fermi surfaces without losing the phase
memory. Figure \ref{fig:WAL} shows an example of the weak
anti-localization behavior observed in a series of MBE-grown
Bi$_{2}$Se$_{3}$ thin films with varying thickness, in which $\alpha
\approx -\frac{1}{2}$ was consistently observed unless the films are 
too thin. 

Intriguingly, in films with low bulk-carrier density, by applying a
gate voltage
one can decouple one of the surface states from the rest through 
creation of an intervening gapped layer, and in such a 
situation $\alpha$ has been found to become $-1$, signifying
two independent diffusive transport channels 
\cite{J.Chen2011,Steinberg2011}.

\subsection{Topological protection of the surface state}

An important property of the topological surface state of a TI is its topological
protection. There are three aspects in the notion of topological protection. 
One is a consequence of the fundamental $Z_2$ topology 
which guarantees the existence of a
gapless surface states as long as TRS is preserved.
Another is a consequence of the helical spin polarization,
which makes the spin eigenvalues of the states with $\mathbf{k}$ and 
$\mathbf{-k}$ to be exactly opposite; as a result, electrons with 
momentum $\mathbf{k}$ cannot be backscattered into the $\mathbf{-k}$
state because of the spin mismatch [Fig. \ref{fig:HelicalSS}(d)], 
and hence the surface state is protected from backscattering.
(Note, however, that the scattering probability from $\mathbf{k}$ 
to $\mathbf{-k}+\mathbf{\delta}$ is non-zero, and hence the 
surface state is not dissipationless.) 
The third is the $\pi$ Berry phase associated
with massless Dirac fermions \cite{T.Ando}, which protects them from
weak localization through destructive interference of time-reversed
paths. Those three effects collectively protect the topological surface
state.

However, when TIs are thinned to the extent that the wave functions of the 
top and bottom surface
states overlap, their hybridization leads to a gap opening
at the Dirac point and results in a degenerate, massive Dirac
dispersion \cite{XueNP, Linder, OscCros, UltraThin, Sakamoto2010, 
VirginiaTech2010, Austin2011}. 
For Bi$_{2}$Se$_{3}$, it was shown by ARPES experiments \cite{XueNP,
Sakamoto2010}
that this hybridization gap opens in ultrathin films with the thickness of 
6 nm or less.

Intriguingly, it was found by Taskin {\it et al.} in transport measurements of a series 
of high-quality Bi$_{2}$Se$_{3}$ thin films that the opening of the 
hybridization gap in the Dirac dispersion leads to a drastic diminishment 
of the metallic surface transport, which helps us understand the importance of 
topological protection in the transport properties of TIs \cite{TaskinMBE}. 
Namely, in Fig. \ref{fig:WAL} one can see that the weak anti-localization
behavior is quickly diminished in ultrathin films with thickness of less than 
5 nm, and this is because in such ultrathin films the opening of the 
hybridization gap causes 
the Berry phase $\gamma$ to be reduced from $\pi$. In the simplest case,  
$\gamma$ depends on the hybridization gap $\Delta$ as\cite{S.Q.ShenPRL}
\begin{equation}
\gamma = 
\pi \left( 1- \frac{\Delta}{E_{F}} \right) .
\end{equation}
This change in the Berry phase diminishes the weak anti-localization
effect and the ordinary weak localization effect takes over. Indeed,  
it was demonstrated in Ref. \citen{TaskinMBE} that
$\rho_{xx}(T)$ of the 2- and 3-nm-thick films presented a localization
behavior at low 
temperature, whereas all the films with thickness larger than 6 nm
preserved a high surface mobility and presented pronounced 
SdH oscillations of Dirac nature. Therefore, 
the change in the transport properties in the hybridized regime 
signifies the consequences of the loss of topological protection 
\cite{TaskinMBE}.

\subsection{Magnetic topological insulator}

Magnetic-ion doping to 3D TIs has been studied
to see the effect of breaking TRS.
Mn-doping to Bi$_2$Te$_3$ was reported to induce bulk 
ferromagnetism with Curie temperature of up to 12 K
with 9\% Mn doping \cite{HorMnBiTe}.
For Bi$_2$Se$_3$, it was reported that Fe and Mn dopings do 
not induce bulk ferromagnetism, but they were found to open 
a small gap at the Dirac point in the surface Dirac cone
\cite{ChenFeMnBiSe}, possibly 
because of a ferromagnetic order that develops only on the 
surface.
Such a Dirac-fermion-mediated ferromagnetism was confirmed
in Mn-doped Bi$_2$(Se,Te)$_3$ thin flakes, in which the chemical
potential was successfully tuned into the bulk band gap so that 
the anomalous Hall effect coming only from the surface electrons can be
measured to probe their ferromagnetic order \cite{CheckelskyMn}.
Recently, in Cr-doped (Bi,Sb)$_2$Te$_3$ thin films, similar surface
ferromagnetism was observed and, furthermore, when the chemical 
potential was tuned to be close to the Dirac point where a gap
opens due to the ferromagnetic order, the {\it quantum anomalous
Hall effect} with $\sigma_{xy}$ quantized to $e^2/h$ was observed
at 30 mK, which signifies the appearance of a 1D chiral edge state
as a consequence of the TRS breaking in the 2D topological surface
state \cite{QAHE}.

\section{Prospects}

There have been great improvements in the materials properties
of TIs in the past few years to make them suitable for fundamental 
research. Now there are 
a couple of choices for bulk-insulating TI materials in which 
the surface transport
dominates over residual bulk transport at low temperature in bulk single
crystals \cite{RenBSTS, RenSnBTS}. 
High-quality thin film samples presenting pronounced
SdH oscillations can also be grown \cite{TaskinMBE,K.L.WangSdH}, 
which are useful for making devices 
to investigate various fundamental phenomena. 
In future, discoveries of new TI materials possessing bulk band gaps 
of more than 0.6 eV would be desirable for
room-temperature applications, but existing materials such as 
Sn-doped Bi$_2$Te$_2$Se \cite{RenSnBTS}
are good enough for studying the 
topological surface transport at temperatures below $\sim$100 K.
It is expected that the focus of the TI research will gradually 
move to actually realizing novel topological phenomena using 
high-quality samples that have become available.
In this regard, there are several major themes for future investigations,
on which I elaborate below.

\subsection{Topological magnetoelectric effects}

The topological field theory for $Z_2$ TIs \cite{QHZ} leads to 
the appearance of an unusual 
$\mathbf{E}\cdot\mathbf{B}$ term in the Lagrangian
\begin{equation}
\mathcal{L} = \frac{1}{8\pi}\left(\epsilon \mathbf{E}^2 
-\frac{1}{\mu}\mathbf{B}^2 \right) 
+ \left( \frac{\alpha}{4\pi^2} \right) \theta \, \mathbf{E}\cdot\mathbf{B},
\end{equation}
where $\epsilon$ is dielectric constant, $\mu$ is magnetic permeability, 
$\alpha = e^2/\hbar c$ is the fine structure constant, and $\theta$ = 0 or $\pi$ 
(mod $2\pi$) is a topological invariant, which takes the value $\pi$ in $Z_2$ TIs.
When $\theta$ is allowed to take any value, the second term (called $\theta$-term) 
describes the electrodynamics of an exotic fictitious particle {\it axion}.
Due to the existence of the $\theta$-term, the constituent equations of a TI
become
\begin{eqnarray}
\mathbf{D} &=& \mathbf{E} + 4\pi \mathbf{P} 
- \frac{\alpha \theta}{\pi} \mathbf{B} \\
\mathbf{H} &=& \mathbf{B} - 4\pi \mathbf{M} 
+ \frac{\alpha \theta}{\pi} \mathbf{E},
\end{eqnarray}
where $\mathbf{D}$ is electric induction, $\mathbf{P}$ is
electric polarization, and $\mathbf{M}$ is magnetization.

The most important consequence of this peculiar electromagnetism is that
in TIs where $\theta = \pi$, electric field $\mathbf{E}$ induces 
magnetization $4\pi \mathbf{M} = \alpha \mathbf{E}$
and the proportionality coefficient is a universal constant $\alpha$.
Similarly, magnetic field $\mathbf{B}$ induces electric polarization
$4\pi \mathbf{P} = \alpha \mathbf{B}$. 
This is the topological magnetoelectric effect to characterize the 
nontrivial $Z_2$ topology of a TI in its electromagnetic properties. 
It should be noted, however,
that the metallic surface states electrically short-circuit the 
whole surface, which makes it impossible to maintain 
electric field or polarization necessary for any magnetoelectric effects; 
therefore, to observe the topological magnetoelectric
effect, one should gap out the surface states to avoid the short-circuiting
effect \cite{QHZ,Nomura}. 
This can, in principle, be done by depositing a ferromagnetic insulator
whose magnetization is kept perpendicular to the surface. 
However, because of the practical difficulty to achieve such a gapping of the surface
state, the topological magnetoelectric effect remains to be experimentally
discovered. Even without opening a gap, magneto-optical properties of 
the surface state may show topological magnetoelectric effects in terms
of a Faraday rotation quantized in integer multiples of $\alpha$ and a 
gigantic Kerr rotation of $\pi/2$ \cite{MacDonald}. 
Another intriguing consequence of the $\theta$-term to be 
tested in future is 
the appearance of an image magnetic monopole \cite{monopole}; 
namely, if the surface of
a TI is gapped out by some means and a point charge is placed near the
surface, the response of the TI looks as if there is a magnetic 
monopole in the TI. 

\subsection{New types of topologies}

Widening the scope of topological materials is an important theme.
Since superconductors have a superconducting gap at the Fermi level,
they are in a way similar to insulators and one can conceive {\it topological
superconductors} characterized by a topological invariant that is protected
by the existence of a gap \cite{QiZhang, Tanaka_JPSJ}.
So far, topological classifications of insulators and superconductors
based on three discrete symmetries (TR, particle-hole, and chiral) 
have been established \cite{Schnyder, Ryu_NJP}. 
Recently, as mentioned in Sec. 4.5, new topological classifications
based on point-group symmetry of the crystal lattice is attracting
significant interest \cite{Slager, TCSC1, TCSC2, TCSC3}, 
particularly after the new type of topological
materials called topological crystalline insulators \cite{HsiehFu,FuTCI} have been 
experimentally discovered \cite{TanakaTCI, Dziawa, XuTCI2}. 
Also, whereas it was thought to be
necessary for topological materials to have a fully-gapped 
energy spectrum for a topological invariant to be well defined
\cite{Schnyder},
it is becoming possible to conceive a nontrivial topology
for gapless systems 
\cite{Savrasov, TurnerWeyl, NodalTSC1, Sato2006, 
SatoFujimoto, NodalTSC2, Sasaki2011}. 
It is an interesting time of topological
expansion. Naturally, experimental discoveries of concrete 
materials that are nontrivial with respect to new topologies
will continue to be crucially important.

\subsection{Majorana fermions}

Majorana fermions are exotic charge-neutral particles that are their
own antiparticles \cite{Wilczek, Alicea, Beenakker, Stanescu}. 
While their existence in nature as elementary particles
has not been confirmed since its prediction in 1937, recently its realizations
in condensed matter as quasiparticles are attracting significant attention
because of their fundamental novelty as well as their potential for 
being used as a qubit of fault-tolerant topological quantum computing \cite{Wilczek}.
To realize Majorana fermions in condensed matter, one must achieve
particle-hole symmetry in a spin-non-degenerate system
\cite{Alicea, Beenakker, Stanescu}. The former can 
be easily achieved in superconductors in which low-energy quasiparticles
obey Bogoliubov-de Gennes equation characterized by inherent 
particle-hole symmetry; however, achieving a spin-non-degenerate 
superconductivity is a difficult task \cite{Read}. 

In this context, Fu and Kane realized
that, since the topological surface state of TIs is spin non-degenerate, 
if superconductivity could be induced in the surface state by using
proximity effect from an attached BCS superconductor, the resulting 
superconducting state harbors Majorana fermions \cite{FuKaneMajorana}. 
It is useful to note that such a proximity-induced superconducting state
on the surface of a TI has singlet Cooper pairs, but nonetheless it is 
topologically nontrivial due to the $\pi$ Berry phase born by the 
surface Dirac electrons; therefore, such a surface can be considered a 
2D topological superconductor.
There have been a number of experimental reports to confirm the 
superconducting proximity effect in the topological surface states
\cite{Sacepe, Veldhorst, Williams, HgTeSC, Burch, SamarthSC, LiLu1, 
LiLu2, LiLu3, Koren, Mason},
but the existence of Majorana fermions has not been elucidated. 

Majorana fermions may also be found in a doped TI 
that becomes a superconductor. 
In this case, surface Dirac fermions
may obtain superconductivity due to the proximity effect from the
bulk and become a spin-non-degenerate topological superconductor.
Another 
possibility is that the bulk superconductivity in a doped TI is itself
topologically nontrivial (i.e. it is a bulk topological superconductor) 
due to strong SOC  \cite{FuBerg}, and its surface harbors 
dispersing, massless Majorana fermions. In this regard, 
the superconductor Cu$_x$Bi$_2$Se$_3$ \cite{Hor},
which is a doped 3D TI, has attracted a lot of attention
\cite{FuBerg, Wray, Kriener, Sasaki2011, HsiehFuSC, Yamakage, HaoLee, 
Kirzhner, Georgia, Levy, Kriener2, Bay, Kriener3}.
In fact, Cu$_x$Bi$_2$Se$_3$ was recently found to present
signatures of unconventional superconductivity in its
point-contact spectra \cite{Sasaki2011}, and an unconventional 
superconductivity in this material is necessarily topological for
symmetry reasons \cite{Sasaki2011}. 

It should be mentioned that the experimental situation for 
Cu$_x$Bi$_2$Se$_3$ is currently rather controversial. 
While some of the follow-up point-contact 
measurements \cite{Kirzhner, Georgia} supported unconventional 
superconductivity originally reported by Sasaki {\it et al.} \cite{Sasaki2011}, 
a recent STS study \cite{Levy} reported spectra 
that are consistent with conventional BCS superconductivity. 
This confusion essentially stems from the fact that available samples of 
superconducting Cu$_x$Bi$_2$Se$_3$ are inhomogeneous 
\cite{Kriener2} and 
the superconducting volume fraction never exceeds 70\%
\cite{Kriener, Kriener2}; furthermore, the author's group has found
that one of the impurity phases is CuSe$_2$, which is a 
conventional superconductor with the transition temperature of 2.4 K.
Obviously, improvements in the sample quality are desirable for local-probe
measurements. Nevertheless, bulk superconducting properties like the 
temperature dependence of 
the upper critical field \cite{Bay} and the disorder dependence of the 
superfluid density \cite{Kriener3}
supported unconventional superconductivity in this material. 

More recently, a superconducting 
doped TCI, Sn$_{1-x}$In$_x$Te, was also found to present similar signatures
of unconventional superconductivity in the point-contact spectroscopy 
\cite{Sasaki2012}, and here again,
an unconventional superconductivity is necessarily 
topological \cite{Sasaki2012}.
Those bulk superconducting materials are intriguing candidates of 
topological superconductors to host massless Majorana fermions.

\subsection{Spintronics device applications}

The dissipationless spin current that exists in the topological surface
state in equilibrium is expected to be useful for low energy
consumption spintronic devices. Nevertheless, it is not clear how
to exploit the dissipationless spin current for spintronic operations.
This is obviously an important theme for both theory and experiment.
When one breaks the equilibrium and push current through the
surface state, the resulting current is spin polarized; however, as
already discussed in Sec. 8.1, the expected spin polarization is 
extremely small in the diffusive regime and the experiment should
be done in the ballistic transport regime \cite{BurkovSpin}, 
which has been difficult
in practice. If successful, a useful feature of the current-induced
spin polarization would be that the polarization direction can be
easily switched by tuning the chemical potential across the Dirac
point with gating. 

\subsection{Hybrid structures based on TIs}

Once ideal samples of TIs  become available, one can fabricate various
types of hybrid structures involving TIs to exploit their novel 
properties. For example, unusual magneto-transport properties 
stemming from the peculiar spin-momentum locking on the 
surface of TIs have been theoretically predicted for TI-Ferromagnet 
hybrid structures \cite{FM-TI1,
FM-TI2, FM-TI3}. In this respect, recent report on the successful growth 
of the ferromagnetic insulator EuS on Bi$_2$Se$_3$ is an encouraging
progress \cite{EuS}. 
Also, combinations of a TI, a superconductor, and 
a ferromagnetic insulator would allow one to create and manipulate 
Majorana fermions \cite{FM-TI-SC1, FM-TI-SC2, FM-TI-SC3, FM-TI-SC4,
FM-TI-SC5, FM-TI-SC6}.

\subsection{Electron interactions}

An important future direction is the merger of 
strong electron correlations and topology. 
There are already a number of theoretical works to elucidate
the effect of electron correlations in TIs \cite{Correlation1, 
Correlation2, Correlation3, Correlation4, Correlation5, Correlation6}.
As the experiments on TIs become refined, the electron correlation
physics will gradually be elucidated.
Perhaps more importantly, if a material 
that becomes both insulating and topologically nontrivial 
{\it because of} strong electron correlations is discovered, 
such a materials would be called a topological Mott insulator
\cite{TMI} and will create an entirely new and rich 
field of research.
There are theoretical discussions that oxides containing 5$d$ 
transition-metal elements might realize such a novel state
of matter \cite{Presin}, and discoveries along this line would 
be extremely interesting.


\begin{flushleft}
{\bf Acknowledgments}\\
\end{flushleft}

The author would like to thank T. Ando, L. Fu, H. Fukuyama, 
Y. Fuseya, A. Kapitulnik, D. Loss, N. Nagaosa, K. Nomura, S. Murakami, N. P. Ong, 
M. Sato, Y. Tanaka, and S.C. Zhang,
for useful discussions and comments. Also, the author greatly acknowledges 
the contributions of his collaborators, K. Segawa, A. Taskin, Z. Ren, S. Sasaki, K. Eto,
M. Kriener, T. Sato, S. Souma, T. Takahashi, I. Matsuda, F. Komori, D. N. Basov, 
T. Kondo, S. Shin, T. Tsuda, and S. Kuwabata.
The author acknowledges the support by JSPS (NEXT Program and KAKENHI 25220708), 
MEXT (Innovative Area ``Topological Quantum Phenomena"), and AFOSR (AOARD 124038).


\profile{Yoichi Ando}{was born in Tokyo, Japan in 1964.
He obtained B.Sc. (1987), M.Sc. (1989), and Ph.D. (1994) degrees from
the University of Tokyo. He was a research scientist at the Central Research
Institute of Electric Power Industry (CRIEPI, 1989-1991) and at the 
Superconductivity Research Laboratory, International Superconductivity
Technology Center (SRL-ISTEC, 1991-1994), where he worked on his
own toward his Ph.D. thesis. He did his postdoc at Bell Laboratories 
in the United States (1994-1996), and then returned to CRIEPI, 
where he led a research group as a senior research scientist (1996-2007)
and also served as a department head (2004-2005). 
Since 2007, he has been a professor at the Institute of 
Scientific and Industrial Research, Osaka University. 
His research is aimed at discovery and understanding of novel
quantum materials. To this end, he synthesizes new materials, grows
high-quality single crystals, and performs top-notch measurements of
various fundamental properties. 
He has made numerous contributions in the field of
high-$T_c$ superconductivity by taking advantage of high-quality single
crystals grown in his laboratory, but in addition, he has made a number
of well-cited contributions to a thermoelectric material, a
giant-magnetoresistance material, and a solid-oxide fuel cell material. 
Recently, he is most interested in revealing new physics in topological 
insulators and topological superconductors, and he has already contributed
significantly to new materials discoveries of those materials.
He has received the prestigious Japan Society for the Promotion of Science
(JSPS) Prize in 2006, and also received the Superconductivity Science 
and Technology Award in 2003 and in 2013.
}

%
%

\end{document}